\documentclass[aps,pra,10pt,twocolumn,superscriptaddress]{revtex4-2} % On govt laptop
\usepackage{amsmath}
\usepackage{amssymb}
\usepackage{graphicx}
\usepackage{color}
\usepackage{url}
\usepackage{xfrac}
\usepackage{listings}
\usepackage{lipsum}
\newcommand{\be}[1]{\begin{equation}\label{#1}}
\newcommand{\ee}{\end{equation}}
\newcommand{\bea}[1]{\begin{eqnarray}\label{#1}}
\newcommand{\eea}{\end{eqnarray}}
\newcommand{\no}{\nonumber \\}
\newcommand{\Fig}[1]{Fig.(\ref{#1})}
\newcommand{\Tbl}[1]{Table \ref{#1}}
\newcommand{\Eq}[1]{Eq.(\ref{#1})}
\newcommand{\App}[1]{Appendix~\ref{#1}}
\newcommand{\Sec}[1]{Section~\ref{#1}}
\newcommand{\Lst}[1]{Listing~\ref{#1}}
\newcommand{\bsub}{\begin{subequations}}
\newcommand{\esub}{\end{subequations}}
\newcommand{\bwt}{\begin{widetext}}
\newcommand{\ewt}{\end{widetext}}

\newcommand{\baa}[1]{\begin{alignat}{#1}} % THIS WORKS, BUT DOES NOT RECOGNIZE THE \label{#2}
\usepackage{color}

\def\trm#1{\textrm{#1}}
\def\tit#1{\textit{#1}}
\def\tbf#1{\textbf{#1}}

\def\a0{{\alpha_0}}

\def\da0{{\dot{\alpha}_0}}

\def\myoverDefn#1#2{\hbox{\space \raise-2mm\hbox{$\textstyle{#1} \atop \scriptstyle{#2}$} }}
\def\G{{\Gamma}}

\def\a{{\alpha}}

\def\dag{\dagger}

\def\rp{r_{P}}
\def\rp2{r_{p}^{2}}
\def\Tr{\textrm{Tr}}

\def\Xbar{\bar{X}}
\def\Xstar{X^*}
\def\Xbarmax{\bar{X}^{max}}
\def\varphibar{\bar{\varphi}}
\def\varphibarmin{\bar{\varphi}^{min}}
\def\varphibarmax{\bar{\varphi}^{max}}

\def\H{\mathcal{H}}
\def\rbarmax{\bar{r}^{max}}
\def\N{{\mathcal{N}}}

\newcommand{\half}{\frac{1}{2}}

\newcommand{\ket}[1]{|#1\rangle}
\newcommand{\bra}[1]{\langle #1|}

\def\rhod{\rho_d}
\def\lambdavecE{\vec{\lambda}^{(4)}_E}
\def\PkN#1{\vec{P}^{(#1,4)}_E}
\def\PKN#1#2{\vec{P}^{(#1,#2)}_E}
\def\e#1{\hat{e}_{#1}}
\def\bfe{\mathbf{e}}
\def\bfeT{\mathbf{e}^{T}}
\def\pNvecE#1{\vec{p}_E^{\,(#1)}}
\def\pNvece#1{\vec{p}_e^{\,(#1)}}
\def\rmu{r_{\mu_4}}

\def\Max{\textrm{Max}}
\def\Min{\textrm{Min}}

\usepackage{fancyhdr}
\begin{document}
%%======== fancyhdr: puts the page number back in on the [R]ight vs the [L]eft =====
\fancyhead[R]{\ifnum\value{page}<2\relax\else\thepage\fi}
%%================================================================

%============================================
%\flushleft{\large DRAFT: rr\_losses\_paper\_v3.tex}
%\vspace{-1em}\flushleft{(\textit{v1: original rough draft; v2: references added; v3: (somewhat) reduced appendices})}
%\vspace{-1em}\flushleft{({\textit{\color{red} Conclusion still needed})}
%% _v4.tex PMA version to all authors 18July2016
%% _v5.tex AMS edits 21July2016
%% _v6.tex PMA edits and final proof read - still need to check figure captions
%============================================
\title{The distribution of density matrices at fixed purity for arbitrary dimensions}
\author{Paul M. Alsing}\email{corresponding author: paul.alsing@us.af.mil}
%\author{\tit{2nd author}}\author{\tit{3rd author}}
%\affiliation{Air Force Research Laboratory, Information Directorate, 525 Brooks Rd, Rome, NY, 13411}
%\author{$\cdots$}
\affiliation{Air Force Research Laboratory, Information Directorate, 525 Brooks Rd, Rome, NY, 13411}
\author{Christopher C. Tison}\author{James Schneeloch}
\affiliation{Air Force Research Laboratory, Information Directorate, 525 Brooks Rd, Rome, NY, 13411}
\author{Richard J. Birrittella}
\affiliation{Air Force Research Laboratory, Information Directorate, 525 Brooks Rd, Rome, NY, 13411}
\affiliation{National Research Council Postdoctoral Fellow, National Academy of Sciences, 500 Fifth St., N.W.
Washington, D.C. 20001}
\author{Michael L. Fanto}
\affiliation{Air Force Research Laboratory, Information Directorate, 525 Brooks Rd, Rome, NY, 13411}
%============================================

\date{\today}
%\maketitle

%====================================
\begin{abstract}
%====================================
We present marginal cumulative distribution functions (CDF) for density matrices $\rho$ of fixed purity 
$\tfrac{1}{N}\le\mu_N(\rho)=\Tr[\rho^2]\le 1$ for arbitrary dimension $N$. 
We provide closed-form analytic formulas  for the cases $N=2$ (trivial), $N=3$ and $N=4$,
 and present a prescription for CDFs of higher arbitrary dimensions. 
These formulas allow one to uniformly sample density matrices at a user-selected, fixed constant purity, and also detail how these density matrices are distributed nonlinearly in the range $\mu_N(\rho)\in[\tfrac{1}{N}, 1]$.
As an illustration of these formulas, we compare the logarithmic negativity and quantum discord 
to the (Wootter's) concurrence 
spanning a range of fixed purity values in
$\mu_4(\rho)\in[\tfrac{1}{4}, 1]$ for the case of $N=4$ (two qubits). 
%
%Additionally, we
%investigate the efficacy of a proposed entanglement witness based on the difference 
%of the purities of the composite bipartite state with its reduced single qubit state, 
%attempting to generalize the linear entropy for pure states, to mixed states.
%
We also investigate the distribution of eigenvalues of a reduced $N$-dimensional density matrices obtained by
tracing out the reservoir of its higher-dimensional purification.
Lastly, we numerically investigate a recently proposed complementary-quantum correlation conjecture which lower bounds the quantum mutual information of a bipartite system by the sum of classical mutual informations obtained from two pairs of mutually unbiased measurements.
Finally, numerical implementation issues for the computation of the CDFs and inverse CDFs necessary for uniform sampling $\rho$ for fixed purity at very high dimension are briefly discussed.
\end{abstract}
\maketitle
%=================
% needed for fancyhdr
%%=================
\thispagestyle{fancy}
%%=================

%==================================================
%\clearpage
%\newpage
%==================================================
%\tableofcontents
%\newpage
%==================================================

%==================================================
\section{Introduction}\label{sec:Intro}
%==================================================
The random generation of density matrices uniformly distributed according to the Haar measure 
is well known \cite{Zyczkowski:2011,Mezzadri:2007}, 
and constitutes an extremely powerful tool in quantum information science 
from the exploration of measures of entanglement \cite{ Zyczkowski_2ndEd:2020}
to quantum data locking protocols \cite{Lloyd:2013, Lum:2016}.
While the numerical algorithm to generate uniformly random density matrices $\rho_N$ of arbitrary dimension $N$ is straightforward, the samples are biased towards the lower values of purity $\mu_N(\rho)=\Tr[\rho^2]\in[\tfrac{1}{N}, 1]$ (nearer the maximally mixed state (MMS) $\rho_{MMS}=\boldmath{I_{N\times N}}/N$ with $\mu(\rho_{MMS})=\tfrac{1}{N}$), with ever increasing rarity to  sample pure state $\rho_N$ ($\mu_N(\rho)\to1$). The implication is that one has to generate an extremely large set random density matrices to obtain a statistically relevant number of samples as the purity nears unity. This problem is only exacerbated as the dimension $N$ of the quantum state increases.

In this work, we derive the cumulative distribution functions (CDFs) of density matrices $\rho_N$ for fixed purity, that will allow us to uniformly generate random density matrices, but now at a user-chosen fixed purity, across the complete 
range $\mu_N(\rho)\in[\tfrac{1}{N}, 1]$. These formulas will reveal the distribution of the eigenvalues of  $\rho_N$ on the sphere $S^{(N-2)}$ of radius $r_N\equiv \sqrt{\mu_N-1/N}\in [0,1-\tfrac{1}{N}]$ centered on the MMS, as a function of the purity $\mu_N$.

While most witnesses or measures of entanglement are based on quantities derived from properties the reduced density matrix of a higher dimensional composite quantum state (pure or mixed), having the ability to sample the later at fixed purity provides a surgical tool to numerically explore the  entanglement relationship between the subsystems derived from the composite state. As an example, for the $N=d^2$-dimensional Werner state, given by 
the convex combination of the $d$-dimensional maximally entangled bipartite Bell state 
$\ket{\Psi}_{ab}=\tfrac{1}{\sqrt{d}}\sum_{0}^{d-1}\ket{n,n}_{ab}$ with the MMS, written as 
$\rho^{(W, N)} = p\,\ket{\Psi}_{ab}\bra{\Psi} + (1~-~p)\,\boldmath{I}_{N\times N}/N$ with $p\in[0,1]$, one can compute logarithmic negativity ($LN$) analytically (see \App{app:Werner:state}) as 
$LN = \log_2(1+2\,\N)$, where the negativity (sum of the absolute values of the negative eigenvalues of the partial transpose (PT) of the composite state), is given by 
$\N=\tfrac{1}{2} \tfrac{d-1}{d} \big((d+1)p-1\big)$.
Further, one can then show that the 
probability $p$ and the purity 
$\mu_{(W, d^2)}\in[\tfrac{1}{d^2},1]$ are related by 
$p=\tfrac{d^2\,\mu_{(W, d^2)}-1}{d^2-1}$. 
Thus, the $LN$ entanglement measure is directly seen as an analytic function of the purity of the composite state.
The CDFs derived in this work allow the exploration, for example, of the $LN$ to the purity of the 
composite random state $\rho_N$ for arbitrary dimension $N$. For the case of $N=4$, i.e. two-qubits, we will numerically compare the relationship of the $LN$ to the purity of the composite state with the Wootter's concurrence \cite{Wootters:1998}, the only entanglement measure valid for both pure and mixed states (applicable as well to $N=6$, a qubit-qutrit system).

This paper is outlined as follows:
In \Sec{sec:uniform:U:rho} we review the generation of random unitary matrices $U_N\in U(N)$, 
which can then be used generate random density matrices via the prescription $\rho_N = U_N\,\rho_N^{(diag)}\, U_N^\dag$ where $\rho_N^{(diag)}$ is a random diagonal density matrix of dimension $N$, whose  eigenvalues are unchanged by the similarity transformation generating $\rho_N$. We review several easily implementable methods for the uniform generation of $\rho_N^{(diag)}$.
In \Sec{sec:WC} we provide a parameterization of $\rho_N^{(diag)}$ as a vector $\vec{p}_E^{\,(N)}\in \mathbb{R}^N$  in the Weyl chamber (WC) (see Chapter 8.5 of \cite{Zyczkowski_2ndEd:2020}) 
%
% PMA wording
%
%as a convex combination of the pure state (unit eigenvalues) and the MMS (all equal eigenvalues of $\sfrac{1}{N}$)
%of the $N-1$ simplex $\Delta_{N-1}$ of the eigenvalues of $\rho_N^{(diag)}$.
%
%James' wording
%
whose components are the probability eigenvalues of $\rho_{N}^{(diag)}$. Where the convex hull (i.e., set of all convex combinations) of all probability vectors $\vec{p}_{E}^{(N)}\in \mathbb{R}^{N}$ defines an $N-1$-dimensional simplex in 
$\mathbb{R}^{N}$ (denoted as $\Delta_{N-1}$), the Weyl chamber can be defined as the convex hull of all truncated maximally mixed states (i.e., 
maximally mixed, but truncated to a fixed number of nonzero entries. See \Eq{PkN} for example), and this convex hull is
also and $N-1$-dimension simplex (here denote as $\tilde{\Delta}_{N-1}$). 
We then transform $\vec{p}_E^{\,(N)}$ to a vector $\vec{p}_e^{\,(N)}$
 centered about the MMS (and whose first component is always given by $\sqrt{\mu_N} = \sqrt{\sfrac{1}{N}}$),
and associate the remaining $N-1$ components with a vector lying on a sphere $S^{(N-2)}$ of radius 
$r_N = \sqrt{\mu_N-1/N}$, for fixed value of the purity $\mu_N$.
We then analytically solve for the boundary ranges of the $N-3$ spherical polar angles.
We provide a formulation of these bounds for arbitrary dimension $N$.
In \Sec{sec:N:2:3:4} we illustrate the previous formulas for the analytically tractable cases of $N=2$ (which is trivial), $N=3$ (a qutrit) and $N=4$ (a pair of qubits). We provide closed form analytic solutions for the cumulative distribution functions (CDF) of both the (spherical polar) angles and the purity radius vector $r_N$ for $N=\{3,4\}$. The CDFs for the radial purity vector allows us to examine the probability of obtaining a density matrix of a given purity \cite{CDF:note} if one were to generate density matrices by the standard method employing the uniform distribution of $\rho_N$ according to the Haar measure.
We compare our results to that of previous numerical simulations employing the uniform Haar measure generation method.
In \Sec{sec:Applications} we explore applications of the above formulas for  $N=4$.
For the case of two-qubits $N=4$, we compare the logarithmic negativity to the Wootter's concurrence as a function of the composite state purity $\mu_4$. 
In addition, we compare these measures to a ``baseline" entanglement witness based on the difference 
of the purities of the composite bipartite state with its reduced single qubit state,
extending the concept the linear entropy for pure states, to mixed states.
%%
%investigate the efficacy of a proposed entanglement witness based on the difference 
%of the purities of the composite bipartite state with its reduced single qubit state, 
%attempting to generalize the linear entropy for pure states, to mixed states.
%%
We also investigate the distribution of eigenvalues of a reduced $N$-dimensional density matrix obtained by
tracing out the reservoir of its higher-dimensional purification.
Lastly, we numerically investigate a recently proposed complementary-quantum correlation conjecture \cite{Scheeloch:2014}
which lower bounds the quantum mutual information of a bipartite system by the sum of classical mutual informations obtained from two pairs of mutually unbiased measurements.
In \Sec{sec:N5:and:beyond} we discuss the case of $N=5$ which serves as an instructive example for how these formulas are applied for all higher dimensions $N>4$. The formulation ceases to be purely analytically tractable due the presence of powers of the sine of the angles in the integration measure. At this point one must turn to numerical simulation employing the non-trivial analytically specified boundary regions for each polar angle.
We further numerically explore the case of $N=6$,  a qubit-qutrit system, and provide a simple, implementable few-line numerical code (which easily generalizes to higher dimension) to generate its radial CDF. We discuss issues of numerical accuracy, and avenues for approximation for larger $N$ values.
%
%In \Sec{sec:numerical:implementation:issues} we discuss numerical implementation issues for computing CDFs 
%and  their inverses, and approximations applicable to  large $N$.
%
In \Sec{sec:Discussion} we summarize our results and discuss avenues for future research directions.

%=======================================================================
\section{Generation unitary and density matrices distributed with respect to the Haar measure}\label{sec:uniform:U:rho}
%=======================================================================
There are many well-known schemes for  uniformly generating density matrices according to the Haar measure.
One of the oldest constructive approaches
dating back to Hurwitz in 1887 (see \cite{Zyczkowski_Kus:1994} and references therein) relies on the $N^2$ angle parameterization of $U(N)=SU(N)\times U(1)$ by all possible products of $2\times 2$ $SU(2)$ rotation matrices 
\be{Eij}
E^{(i,j)}=
\left(\begin{array}{cc}
\cos\phi_{ij}\,e^{i\,\psi_{ij}} & \sin\phi_{ij}\,e^{i\,\chi_{ij}} \\
 -\sin\phi_{ij}\,e^{-i\,\chi_{ij}}  & \cos\phi_{ij}\,e^{-i\,\psi_{ij}}
\end{array}\right)
\ee
embedded within $N\times N$ matrices, with the $\cos$ terms at position $(i,i)$ and $(j,j)$ with $1\le i<j\le N$, and the $\pm\sin$ terms at positions $(i,j)$ and $(j,i)$ respectively, with the remaining elements unity on the diagonal $k\ne(i,j)$, and zeros on the off diagonals (see \cite{Zyczkowski_Kus:1994}, and \App{app:uniform:U:rho}).
There is an overall $U(1)$ phase $e^{i\,\alpha}$. The angles are taken from the intervals 
$0\le\phi_{r\,s}\le\pi/2$, $0\le\psi_{r\,s}, \chi_{1\,s} \le\pi/2$ and $0\le\alpha\le 2\,\pi$ 
uniformly with respect to the Haar measure with probability measure
$P_U(dU)=\sqrt{N! 2^{N(N-1)}}\,
\Pi_{1\le r\le s\le N}(\tfrac{1}{2\,r})d\left[(\sin\phi_{rs})^{2r}\right]\,
\Pi_{1\le s\le N}\,d\chi_{1s}$. One then samples $\alpha, \phi_{rs}, \chi_{1s}$ uniformly on the interval $[0,2\,\pi)$.
Additionally, one takes  $\phi_{rs} = \sin^{-1}(\xi^{1/2r})$ with $\xi$ drawn uniformly in $[0,1)$, for 
$r\in\{1,2,\ldots,N-1\}$.

With today's fast and accurate computer linear algebra subroutines, a simple (requiring only a few lines of code) routine centers around the $QR$ decomposition \cite{NumRecC2ndEd:1992} of an invertible $N\times N$ complex matrix $Z$ with entries $z_{jk}$ (the Ginibre ensemble) (see Mezzadri \cite{Mezzadri:2007}, and \App{app:uniform:U:rho}).
%=======================================
% From LogNegApprox_17Feb2022.tex, Appendix A
%=======================================
Such matrices have matrix elements that are independent and identically distributed (i.i.d.) standard normal complex random variables with probability distribution $p(z_{jk})=\tfrac{1}{\pi} e^{-|z_{jk}|^2}$.
The joint probability distribution for the matrix elements (also statistically independent) is given by
$P(Z) = \tfrac{1}{\pi^{N^2}} \prod_{j,k=1}^N e^{-|z_{jk}|^2} = 
 \tfrac{1}{\pi^{N^2}} \exp\left[\sum_{j,k=1}^N e^{-|z_{ij}|^2}\right]$ $= \tfrac{1}{\pi^{N^2}} \exp\left(-\Tr[Z^\dag\,Z]\right)$.
$P(Z)$ is normalized to unity via $\int_{\mathbb{C}^{N^2}} P(Z)\, dZ=1$ where
$dZ=\prod_{j,k=1}^N dx_{jk}\,dy_{jk}$ and $z_{jk} = x_{jk} + i\,y_{jk}$. The integration measure on the Ginibre ensemble $\mathbb{C}^{N\times N}\cong \mathbb{C}^{N^2}$ is $d\mu_G(Z) = P(Z)\,dZ$ which can be thought of as an infinitesimal volume in $\mathbb{C}^{N^2}$. 
The crucial point is that $d\mu_G(Z)$ is invariant under left and right multiplications of $Z$ by arbitrary unitary matrices i.e. $d\mu_G(U\,Z)=d\mu_G(Z\,V)=d\mu_G(Z)$ for  $U, V\in U(N)$. The proof follows trivially from the property of the trace in the definition of $P(Z)$ since 
$\Tr[ (U\,Z)^\dag\,(U\,Z)]= \Tr[ (Z\,V)^\dag\,(Z\,V)]=\Tr[Z^\dag\,Z]$, and hence $P(U\,Z)=P(Z\,V)=P(Z)$.
This Haar measure is the matrix analogue of a uniform probability distribution in one dimension 
$p(\theta) = \tfrac{1}{2\,\pi}$.

The above algorithm to generate random unitary matrices distributed with the Haar measure uses the $QR$ decomposition of $Z$ (vs the less stable, at higher dimensions, Gram-Schmidt orthonormalization routine).  
Here $Q$ is a unitary matrix and $R$ is an upper-triangular matrix.
One caveat exits though. The $QR$ decomposition is not unique, since if $Z=Q\,R$ then so is
$Z=Q'\,R'\equiv (Q\,\Lambda) \, (\Lambda^{-1}\,R)$, with $\Lambda$ a diagonal matrix so that $Q'$ and $R'$ are again unitary and upper-triangular matrices. To make the decomposition unique, the solution (see Mezzadri \cite{Mezzadri:2007} for details) is to choose $\Lambda_{ii} = R_{ii}/|R_{ii}|$ where $R_{ii}$ are the diagonal elements of $R$. Thus, the diagonal matrix elements of $R'=(\Lambda^{-1}\,R)$ are real and strictly positive, which renders 
the matrix $Q'=Q\,\Lambda$ unique and uniformly distributed with the Haar measure. This is the procedure implemented in the \App{app:uniform:U:rho} Mathematica codes. 

Given a code to generate random unitary matrices $U$, one can subsequently  uniformly generate density matrices
$\rho$ (positive semi-definite Hermitian matrices of unit trace)
 by the simple decomposition $\rho = U\,\rho_{N}^{(diag)}\, U^\dag >0$. To generate the diagonal density matrix 
$\rho_{N}^{(diag)}$, an element of the Weyl chamber (see section 8.5 of \cite{Zyczkowski_2ndEd:2020}) such that
$\sum_{i=1}^N \big(\rho_{N}^{(diag)}\big)_{ii}=1$, one simply generates another random unitary $U'$ and takes the absolute square of a random row (or column) as the the diagonal entries. 
The above is the method used in this work to generate uniformly random unitary matrices. 
These codes are easily implementable in commonly used coding languages, such as Python, 
or Mathematica (see \App{app:uniform:U:rho}).
%=======================================

Lastly, another insightful method to generate an element $\rho_{N}^{(diag)} = \trm{diagonal}\{\lambda_1, \lambda_2,\ldots,\lambda_{N-1}\}$ of the WC (the subset of the $N-1$ simplex $\Delta_{(N-1)}$ defined by the eigenvalues arranged in decreasing order $\{\lambda_1\ge\lambda_2\ldots\ge\lambda_{N-1}\}$ 
with $\lambda_N\equiv 1-\sum_k^{N-1}\,\lambda_k$) is to choose the eigenvalues 
according to the formula (see Appendix A of Zyczkowski \tit{et al.} \cite{Zyczkowski:1998}, and \App{app:uniform:U:rho}):
\bsub
\bea{Zykowski:lambdas}
\hspace{-0.15in}
\lambda_k &=& \left[1- \xi_1^{1/(N-k)}\right]\,\left(1-\sum_{i=1}^{k-1} \lambda_i\right), \; 
\lambda_1 = 1- \xi_1^{1/(N-1)},\qquad \label{Zykowski:lambdas:line1}\\
\hspace{-0.15in}
\lambda_2 &=& \left[1- \xi_2^{1/(N-2)}\right]\,(1-\lambda_1),\;
\quad \lambda_N\equiv 1-\sum_{k=1}^{N-1} \lambda_k, \label{Zykowski:lambdas:line2}
\eea
\esub
where the $\{\xi_k\}$ for $k\in\{1,2,\ldots,N-1\}$ are uniform deviates in $[0,1]$.
For $N=2$ we have trivially $\lambda_1 = 1-\xi_1$, so that $\lambda_1$ is uniform
in $[0,1]$. Since $\rho_{(N=2)}=\tfrac{1}{2}\left(\mathbb{I} + \vec{r}\cdot\vec{\sigma} \right)$
with eigenvalues $\lambda_\pm = \tfrac{1}{2} (1\pm r)$ with $r=\sqrt{2}\,r_2=\sqrt{2\,\mu_2-1}$,
we see that $\lambda_1=\lambda_+$ and hence $r_2$ ($\mu_2$) is distributed uniformly (quadratically)
in $[0,\tfrac{1}{\sqrt{2}}]$ ($[\tfrac{1}{2},1]$).
For $N=3$ we have $\lambda_1= 1-\sqrt{\xi_1}$, $\lambda_2 = (1-\xi_2)\sqrt{\xi_1}$ with
$\lambda_3 = \xi_2\,\sqrt{\xi_1}$. Note the term $\sqrt{\xi_1}$ appears with alternating signs in
$\lambda_1$ and $\lambda_2$, while the same is true for the term $\xi_2\,\sqrt{\xi_1}$
in $\lambda_2$ and $\lambda_3$. These terms cancel in successive sums of eigenvalues (in decreasing order) in order to ensure the unit trace condition $\Tr[\rho_{diagonal}]= \sum_{k=1}^{N} \lambda_k=1$. 

The formulas in \Eq{Zykowski:lambdas:line1} and \Eq{Zykowski:lambdas:line2}
%are presented in \cite{Zyczkowski:1998} without derivation, the later of which is given
are derived
%in \App{app:uniform:U:rho}, 
 by assuming a generic form of $\lambda_1=1-z_1$ and 
 $\lambda_k = (1-z_k)\Pi_{i=1}^{k-1}\,z_i$ for $k=\{2,\ldots,N-1\}$ and $\lambda_N=1-\sum_{k=1}^{N-1} \lambda_k$
 which by construction, trivially enforces the unit trace condition. Then by assuming $z_k=(\xi_k)^{\alpha_k}$, 
 and requiring that the Jacobian $\det\left[\tfrac{\partial \lambda_i(\xi)}{\partial \xi_j}\right]$ 
 of the transformation between the $\lambda$ and $\xi$ 
 variables is a constant (so that the $\xi$ variables are uniformly distributed \cite{NumRecC2ndEd:1992}), 
 the powers $\alpha_k = \tfrac{1}{N-k}$ appearing in  
  \Eq{Zykowski:lambdas:line1} and \Eq{Zykowski:lambdas:line2} are obtained.
 
%=======================================================================
\section{Spherical polar representation of diagonal density matrices in the Weyl chamber}\label{sec:WC}
The previous section discussed the uniform generation of density matrices satisfying 
the unit trace (linear eigenvalue) constraint $\Tr[\rho_N]=1$. 
In this section we now turn our attention to the main focus of this work, namely, 
the uniform generation of density matrices with the additional implementation of
the quadratic constraint of fixed (chosen) purity $\mu_N=\Tr[\rho_N^2]$.
%============================
\begin{figure}[ht]
%\begin{center}
%\begin{tabular}{cc}
%\hspace{-0.15in}
\includegraphics[width=3.75in,height=2.85in]{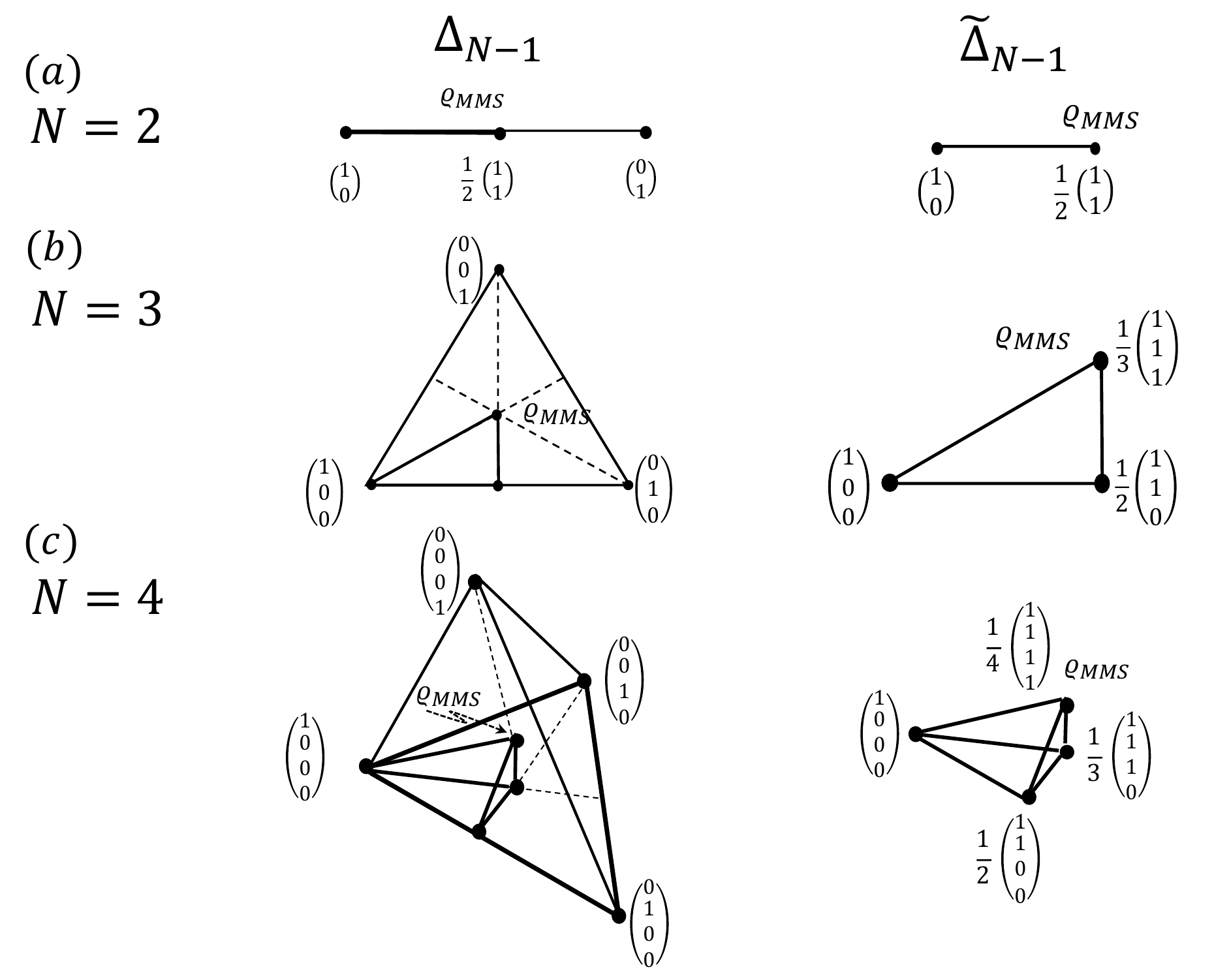}%&
%\includegraphics[width=3.75in,height=2.85in]{fig_WC_N_2_3_4_11Mar2022}%&
%\includegraphics[width=3.5in,height=2.5in]{PS2N2K2_4_8_16_32_64_8Mar2021} 
%\end{tabular}
%\end{center}
\caption{(left) Eigenvalue simplex $\Delta_{N-1}$ and (right) Weyl chamber (WC) $\tilde{\Delta}_{N-1}$ for 
(a) $N=2$,
(b) $N=3$,
(c) $N=4$.
}\label{fig:Simplex:WC}
\end{figure}
%============================
Before we begin we first introduce several different coordinate systems 
that will be useful in our derivation. We will illustrate this with examples for $N=4$.

\subsection{Bases}
\subsubsection{E-basis in $\mathbb{R}^N$}
Let us call the orthogonal ``E-basis" the canonical basis in $\mathbb{R}^N$ with basis vectors 
$E_i~=~\{0,\ldots,1,\ldots,0\}^T~=~\delta_{i,j}$ for $i,j\in\{1,\ldots,N\}$.
In this basis we define the $N=4$ diagonal density matrix (ddm) $\rho_d$ as the $N=4$ vector~$\lambdavecE$
\be{lambda4vecE}
\lambdavecE = 
\left(\begin{array}{c}\lambda_1 \\ \lambda_2  \\ \lambda_3  \\ \lambda_4 \end{array}\right), 
\qquad \sum_{i=1}^4 \lambda_i =1, 
\qquad \sum_{i=1}^4 \lambda^2_i \equiv\mu_4.
\ee

%\subsubsection{The (non-orthogonal) P-basis: Vertices of the $(N-1)$-Simplex in dimension $N$}
\subsubsection{The (non-orthogonal) P-basis: Vertices of the Weyl Chamber $(N-1)$-Simplex in dimension $N$}
We now define the vertices of the  $(N-1)$-Simplex of the Weyl Chamber (denoted) $\tilde{\Delta}_{N-1}$ in dimension $N$ 
\Fig{fig:Simplex:WC}(c)(right)
as the vectors  $\vec{P}_E^{(k,N)}$ as (here illustrated for $N=4$)
\bea{PkN}
\PkN{1} &=& \left(\begin{array}{c}1 \\ 0  \\ 0 \\ 0 \end{array}\right), \quad
\quad\PkN{2} = \dfrac{1}{2}\,\left(\begin{array}{c}1 \\ 1  \\ 0 \\ 0 \end{array}\right),\quad \no
\PkN{3} &=& \dfrac{1}{3}\,\left(\begin{array}{c}1 \\ 1  \\ 1 \\ 0 \end{array}\right),\quad
\PkN{4} = \dfrac{1}{4}\,\left(\begin{array}{c}1 \\ 1  \\ 1 \\ 1 \end{array}\right),
\eea
which should be read as \tit{the maximally mixed state (MMS) of dimension $k$ embedded in dimension $N=4$}.
That is, for $N=1$ we can regard the 0-Simplex as the  1D ``vector" $\PKN{1}{1}\equiv(1)$.
For $N=2$ we embed 
$\PKN{1}{1}\equiv(1)\mapsto \PKN{1}{2} \equiv\tiny{\left(\begin{array}{c}1 \\ 0  \end{array}\right)}\in\mathbb{R}^2$, and then ``add" the MMS in dimension $N=2$ as  
$\PKN{2}{2}= \tiny{\frac{1}{2}\,\left(\begin{array}{c}1 \\ 1  \end{array}\right)}$, which are the vertices of the (one dimensional) 1-Simplex $\Delta_{1}$. For $N=3$ we continue this process by embedding 
$\PKN{1}{2}\mapsto\PKN{1}{3}$ and  $\PKN{2}{2}\mapsto\PKN{2}{3}$ in $\mathbb{R}^3$, and then ``adding" the MMS in dimension $N=3$ as  $\PKN{3}{3}=  \tfrac{1}{3}\,\tiny{\left(\begin{array}{c}1 \\ 1 \\ 1 \end{array}\right)}$. Continuing this process for $N=4$ produces \Eq{PkN}, which can be trivially extended iteratively to arbitrary dimension $N$.
Note that in dimension $N$, the vector $\PKN{1}{N}$ is a \tit{pure} state, while $\PKN{N}{N}$ is the MMS for that dimension.

In dimension $N$, there are $N!$ ways to order the eigenvalues $\{\lambda_k\}$, $k\in\{1,\ldots,N\}$ of the ddm $\rhod$  
(in the E-basis). 
We define the  \tit{Weyl Chamber} (WC) as the particular \tit{descending order} 
$\{\lambda_1\ge\lambda_2\ge\ldots\ge\lambda_N\}$, 
which is denoted as the $(N-1)$-Simplex $\tilde{\Delta}_{N-1}$.
%===============================
%In dimension $N$, there are $N!$ ways to order the eigenvalues $\{\lambda_k\}$, $k\in\{1,\ldots,N\}$ of the ddm $\rhod$  (in the E-basis). 
%We define the \tit{Weyl Chamber} (WC) as the particular \tit{descending order} $\{\lambda_1\ge\lambda_2\ge\ldots\ge\lambda_N\}$, 
%which is denoted as the $(N-1)$-Simplex $\tilde{\Delta}_{N-1}$.
%===============================

\subsubsection{The (orthogonal) e-basis; the ``in-WC"  basis}
We now form the \tit{e-basis}, or  the \tit{in-WC}  basis formed from normalized successive differences of the 
$\tilde{\Delta}_{N-1}$ vertex vectors $\PKN{k}{N}$ for $k\in\{0,\ldots,N-2\}$,
\be{e_basis}
\hspace{-0.15in}
\e{1} = \dfrac{\PKN{N}{N}}{\left|\PKN{N}{N}\right|} , 
\;\; \e{k+2} = \dfrac{\PKN{N-(k+1)}{N} - \PKN{N-k)}{N}}{\left|\PKN{N-(k-1)}{N} - \PKN{N-k)}{N}\right|}, 
\ee
For $N=4$ these are the the normalized vectors
$ \{\PkN{4}, \PkN{3}-\PkN{4}, \PkN{2}-\PkN{3},\PkN{1}-~\PkN{2}\} \mapsto \{\e{1},\e{2},\e{3},\e{4}\}$
Here, $\e{1}$ is a unit vector pointing from the origin to the MMS  $\PKN{N}{N}$, and lies ``outside" the WC.
The vectors $\e{k+2,\,(k\in[0,N-2])}$ 
are then unit vectors that lie within the WC and point from the embedded $(N-k)$-dimensional MMS  to the embedded $(N-(k+1))$-dimensional MMS.
We will designate the \tit{orthogonal} matrix whose rows are the unit vectors  $\e{k}$ 
as $\bfe$, and its transpose as $\bfeT$.
For $N=\{3, 4\}$ we explicitly have 
\be{eMatrix_eMatrixT}
\hspace{-0.20in}
%%%%%%%%%% e: N = e %%%%%%%%%%%%%%%
\bfe_{N=3} = 
\left(
\begin{array}{ccc}
\frac{1}{\sqrt{3}} & \frac{1}{\sqrt{3}}  & \frac{1}{\sqrt{3}}  \\
 \frac{1}{\sqrt{6}} & \frac{1}{\sqrt{6}}  & -\sqrt{\frac{2}{3}} \\
\frac{1}{\sqrt{2}} & -\frac{1}{\sqrt{2}} & 0 \\
\end{array}
\right),
\;
%%%%%%%%%% e: N = 4 %%%%%%%%%%%%%%%
\bfe_{N=4} = 
\left(
\begin{array}{cccc}
\frac{1}{2} & \frac{1}{2}  & \frac{1}{2}             & \frac{1}{2} \\
 \frac{1}{2\sqrt{3}} &  \frac{1}{2\sqrt{3}} &  \frac{1}{2\sqrt{3}}            & - \frac{\sqrt{3}}{2} \\
\frac{1}{\sqrt{6}} & \frac{1}{\sqrt{6}} &  -\sqrt{\frac{2}{3}} & 0 \\
\frac{1}{\sqrt{2}} &  -\frac{1}{\sqrt{2}}  & 0                                   & 0
\end{array}
\right),
%
%\;
%%%%%%%%%% e^T: N = 4 %%%%%%%%%%%%%%%
%%\bfeT = 
%%\left(
%%\begin{array}{cccc}
%%\frac{1}{2} & \frac{1}{2\sqrt{3}}  & \frac{1}{\sqrt{6}}             & \frac{1}{\sqrt{2}} \\
%%\frac{1}{2} &  \frac{1}{2\sqrt{3}} &  \frac{1}{\sqrt{6}}            & - \frac{1}{\sqrt{2}} \\
%%\frac{1}{2} &  \frac{1}{2\sqrt{3}} &  -\frac{\sqrt{2}}{\sqrt{3}} & 0 \\
%%\frac{1}{2} &  -\frac{3}{\sqrt{2}}  & 0                                   & 0
%%\end{array}
%%\right), 
%%
%\qquad
%\bfe\cdot\bfeT = \mathbf{1}.
\ee
where 
$\bfe_N\cdot\bfeT_N = \mathbf{1}$.
These matrices will useful for transforming (column) vectors $\vec{p}$ between the E-basis $\in\mathbb{R}^N$ and the 
MMS-centered e-basis:
\be{vector_transformations}
\vec{p}^{(N)}_{e-basis} =  \bfe\cdot\vec{p}^{(N)}_{E-basis}, \quad  \vec{p}^{(N)}_{E-basis} =  \bfeT\cdot\vec{p}^{(N)}_{e-basis}.
\ee
%(Note that for some perverse reason I actually implemented this in the Mathematica codes via an operation on row vectors $\vec{p}^{\;T}$ via
%$ \vec{p}^{\,(N)\;T}_{e-basis} = \vec{p}^{\,(N)\;T}_{E-basis} \cdot   \bfeT$; either way works satisfactorily).

The utility of the particular formulation will be the simplicity (regularity) of the  vector $\vec{\lambda}^{(N)}_E$
when transformed to the e-basis.
 For $N=4$ we have
  $\vec{\lambda}^{(N)}_e =  \bfe\cdot\vec{\lambda}^{(N)}_E$
  %
 % \bsub
  \bea{lambda4vecebasis}
  \vec{\lambda}^{(N)}_e &=&  
 %\bfe\cdot\vec{\lambda}^{(N)}_E 
%  &=& 
%  \left(\begin{array}{c} 
% \frac{1}{2}( \lambda_1+\lambda_2+\lambda_3+\lambda_4) \\ 
%  \frac{1}{2\sqrt{3}}( \lambda_1+\lambda_2+\lambda_3-3\lambda_4) \\
%   \frac{1}{\sqrt{2}\sqrt{3}}( \lambda_1+\lambda_2-2\lambda_3) \\
%   \frac{1}{\sqrt{2}}( \lambda_1-\lambda_2)\end{array}\right), \label{lambda4vecebasis:1}\\
%   %
%   &=&
   %
  \left(
  \begin{array}{c} 
 \frac{1}{2} \\ 
  \frac{1}{2\sqrt{3}}[ (\lambda_1-\lambda_4)+(\lambda_2-\lambda_4)+(\lambda_3-\lambda_4)] \\
   \frac{1}{\sqrt{2}\sqrt{3}}[(\lambda_1-\lambda_3)+(\lambda_2-\lambda_3)] \\
   \frac{1}{\sqrt{2}}( \lambda_1-\lambda_2)
   \end{array}
   \right)  \label{lambda4vecebasis:2}, \no
   &\equiv&
   \left(
  \begin{array}{c} 
 \frac{1}{2} \\ 
 \rmu \sin\varphi_3\cos\varphi_2 \\
  \rmu \sin\varphi_3\sin\varphi_2 \\
  \rmu \cos\varphi_3\\
 \end{array}
   \right),
  \eea
 % \esub
  where we have defined the spherical coordinates $(\rmu,\varphi_2,\varphi_3)$ on $S^2$, 
  which maps the WC tetrahedron to the  ball with purity $\mu_4\in[1/4,1]$ for $N=4$. 
  (Note: the reason for the ordering of the components and the numbering of the angles starting from 
  $\varphi_2$ will be made clear shortly).
  In fact, the first component of $ \vec{\lambda}^{(4)}_e$ is actually $\sqrt{(\mu_{4,min}=1/4)} = 1/2$. For arbitrary $N$ this generalizes to the first component being  $\sqrt{(\mu_{N,min}=1/N)}~=~1/\sqrt{N}$.
  
  Note that the first matrix in \Eq{lambda4vecebasis:2} is written in such a way, that in the WC each term in parenthesis is positive (greater or equal to zero). This implies that the tetrahedron for $N=4$ has been mapped to the positive octant of $S^2$ (this is generalized for arbitrary $N$).
  Given a randomly generated $\vec{\lambda}^{(4)}_E\mapsto\vec{\lambda}^{(4)}_e$
 we can back out the spherical coordinates as follows. 
 Since $\mu_4 = \vec{\lambda}^{(4)}_E\cdot  \vec{\lambda}^{(4)}_E = \vec{\lambda}^{(4)}_e \cdot \vec{\lambda}^{(4)}_e = \frac{1}{4} + \rmu^2$ we have that $\rmu=\sqrt{\mu_4 -1/4}$. Comparing the terms in \Eq{lambda4vecebasis:2} yields for $N=4$,
\bsub
 \bea{lambda:to:spherical:coords}
 \rmu&=& \sqrt{\mu_4-1/4},  \label{lambda:to:spherical:coords:1} \\
 \cos\varphi_3 &=&  \frac{( \lambda_1-\lambda_2) }{\sqrt{2}\,r}, \label{lambda:to:spherical:coords:2} \\
 \tan\varphi_2 &=&   \sqrt{2}\,\frac{(\lambda_1-\lambda_3)+(\lambda_2-\lambda_3) }{(\lambda_1-\lambda_4)+(\lambda_2-\lambda_4)+(\lambda_3-\lambda_4)}, \label{lambda:to:spherical:coords:3} %\\
% %
% &=&  \sqrt{2}\,\frac{(\lambda_1-\lambda_2)+2 (\lambda_2-\lambda_3) }{(\lambda_1-\lambda_2)+2(\lambda_2-\lambda_3)+3(\lambda_3-\lambda_4)}, \no
% %
% &=&
% \sqrt{2}\,\frac{(\sqrt{2}\,r \cos\theta + 2 (\lambda_2-\lambda_3) }{\sqrt{2}\,r \cos\theta+2(\lambda_2-\lambda_3)+3(\lambda_3-\lambda_4)}, \label{lambda:to:spherical:coords:4}\\
% %
% &\equiv&
%  \sqrt{2}\,\frac{(\sqrt{2} \cos\theta + 2 (\tilde{\lambda}_2-\tilde{\lambda}_3) }{\sqrt{2} \cos\theta+2(\tilde{\lambda}_2-\tilde{\lambda}_3)+3(\tilde{\lambda}_3-\tilde{\lambda}_4)}, \label{lambda:to:spherical:coords:5}
 \eea
 \esub 
%where in the very last line we have removed the radial dependence by defining $\tilde{\lambda}_k = r\,\lambda_k$.
% The last line shows that $\phi = \phi(\theta,(\tilde{\lambda}_2-\tilde{\lambda}_3),(\tilde{\lambda}_3-\tilde{\lambda}_4))$.
% Thus, in the desired case of generating (hopefully, uniformly) randomly generated ddm given a fixed purity $\mu$ (equivalently, fixed $r$)
% %
% we can clearly state that from    \Eq{lambda:to:spherical:coords:2},  $0\le\cos\theta\le\frac{1}{\sqrt{2}\,r}$ (where the lower limit comes from cases where $\lambda_1=\lambda_2$ and the upper limit comes from the pure state $\lambda_1=1, \lambda_2=0= \lambda_3= \lambda_4$.
% This key issue is to enumerate the upper and lower limits of $\phi = \phi(\theta)$ for a fixed $\mu$ from \Eq{lambda:to:spherical:coords:2} -  \Eq{lambda:to:spherical:coords:5}.
showing that the spherical polar angles $(\varphi_2, \varphi_3)$ on $S^{2}$ depend on the values of the eigenvalues selected.

In the next section we present a different parameterization of the WC, leading again to a mapping on the sphere $S^{N-2}$ (via a transformation of an WC eigenvalue vector in the E-basis, to the e-basis, via $\bfe$) analogous to \Eq{lambda4vecebasis:2}, but with a ``nicer" (more amenable) structure.

%=======================================================================
\subsection{Parameterization of the WC in terms of its vertices $\boldsymbol{\PKN{k}{N}}$: purity coordinates}
 In this section we present a systematic parameterization of the WC in terms of its vertices $\PKN{k}{N}$.
 We will build this up recursively from $N=2$ to a general $N$.
 %but will focus the numerics on the case $N=4$ where we can plot ddm on $S^2$.
 
 \subsubsection{Derivation of the purity coordinates $\mu_k$}
 For $N=2$ the WC $\tilde{\Delta}_1$ is a 1-dimensional line with a single coordinate interpolating between the pure state $\PKN{1}{2}$ and the MMS of dimension $N=2$, $\PKN{2}{2}$ (i.e. a convex combination of the 2D vertices of the 1D line). 
 Let us write a point $\vec{p}^{\,(N=2)_E}$  on this line in the E-basis as
\bsub
 \bea{N2:param}
 \hspace{-0.5in}
 \pNvecE{2}(x_2) &=& x_2\,\PKN{1}{2} + (1-x_2)\,\PKN{2}{2}, \qquad Tr[\pNvecE{2}]=1, \quad\\
  \hspace{-0.5in}
 1/2\le\mu_2 &=&  \pNvecE{2}\cdot \pNvecE{2}\le1, \qquad \pNvecE{2}(x_2) = \pNvecE{2}(\mu_2).
 \eea
 \esub
 Note that $ Tr[\pNvecE{2}]=1$ follows since (i)  $Tr[\PKN{k}{N}]=1$ holds for every $k$-dimensional MMS embedded in dimension $N$,  and (ii) by construction of \Eq{N2:param} we have made a convex combination of the vertices.
 Note that we have also defined the $N=2$ purity $\mu_2$ as  $1/2\le\mu_2 =  \pNvecE{2}\cdot \pNvecE{2}\le1$.

 To extend this from the 1D-line to the 2D-triangle we need to add a new ``polar angle" which we accomplish as follows
\bsub
 \bea{N3:param}
\hspace{-0.70in}
 \pNvecE{3}(x_3,x_2) &=& x_3\,\pNvecE{2}(x_2) + (1-x_3)\, \PKN{3}{3} , \quad Tr[\pNvecE{3}]=1, \qquad\; \\
\hspace{-0.70in}
 &=& x_3\Big(x_2\,\PKN{1}{3} + (1-x_2) \PKN{2}{3}\Big) + (1-x_3)\, \PKN{3}{3},\qquad \\
 \hspace{-0.70in}
 1/3\le\mu_3 &=&  \pNvecE{3}\cdot \pNvecE{3}\le1, \qquad  \pNvecE{3}(x_3,x_2) =  \pNvecE{3}(\mu_3,\mu_2).
 \eea
 \esub
 \Eq{N3:param} states that we now have a convex combination of points $\pNvecE{2}(x_2)$ on the 1D WC line, parameterized by the $x_2$, or equivalent the $N=2$ purity $\mu_2$, and the MMS in dimension $N=3$, $ \PKN{3}{3}$. The new coordinate that governs this convex combination from points on the 1D line $\tilde{\Delta}_1$ for $N=2$ to points into the 2D triangle $\tilde{\Delta}_2$ for $N=3$ is $x_3$, or equivalently the $N=3$ purity $\mu_3$.
 (Note: we will find it convenient to use both the $x_k$ and $\mu_k$ coordinates, and later demonstrate that $x_k=x_k(\mu_k,\mu_{k-1})$ for each $k$. That is, for a fixed $k\ge2$, $x_k$ \tit{only} depends on $\mu_k$ and $\mu_{k-1}$, where we \tit{define} $\mu_1\equiv 1$).
 
 We now derive an important relationship between the $x_k$ and the $\mu_k$ coordinates for $N=3$, which generalizes readily to arbitrary $N$.
 Consider the following calculation
 \bsub
 \bea{x3tomu3mu2}
  \hspace{-0.15in}
 \mu_3 &=& \pNvecE{3}\cdot \pNvecE{3}, \label{x3tomu3mu2:1} \\
  \hspace{-0.15in}
 &=&  \left( x_3\,\pNvecE{2} + (1-x_3)\, \PKN{3}{3}\right)^2, \no
   \hspace{-0.15in}
 &=& x^2_3\, \left(\pNvecE{2}\cdot\pNvecE{2}\right) + 2 \,x_3\,(1-x_3)\, \left(\pNvecE{2}\cdot\PKN{3}{3}\right)\no
  \hspace{-0.15in}
  &{}& \hspace{0.95in}+ (1-x_3)^2\, \left(\PKN{3}{3}\cdot \PKN{3}{3}\right), \no
 \hspace{-0.15in}
 &=& x^2_3\,\mu_2 + 2 \,x_3\,(1-x_3)\, \mu_3^{(min)} + (1-x_3)^2\, \mu_3^{(min)},\no
  \hspace{-0.15in}
 &=& x^2_3\,\mu_2 + (1-x^2_3)\, \mu_3^{(min)}, \no
  \hspace{-0.15in}
 \Rightarrow  x_3 &=& \sqrt{\dfrac{\mu_3-\mu_3^{(min)}}{\mu_2-\mu_3^{(min)}}}, \quad\qquad \mu_3^{(min)} =1/3,  
 \;  \mu_1\equiv 1, \label{x3tomu3mu2:2}\\
 \hspace{-0.15in}
 \Rightarrow  x_k &=& \sqrt{\dfrac{\mu_k-\mu_k^{(min)}}{\mu_{k-1}-\mu_k^{(min)}}} = \sqrt{\dfrac{k\,\mu_k-1}{k\,\mu_{k-1}-1}}, 
 \; \mu_k^{(min)} =1/k.\quad\quad\;\;  \label{x3tomu3mu2:3}
\eea
\esub
 In the above we have used that for the MMS in dimension $N$, we have $\PKN{N}{N} = \frac{1}{N}\,(1,\ldots,1)$ so that 
 $\pNvecE{k}\cdot\PKN{N}{N} = \frac{1}{N}\,Tr[\pNvecE{k}] =  \frac{1}{N}\, = \mu_N^{(min)}$.
In addition, we also have $\PKN{N}{N}\cdot \PKN{N}{N}=  \frac{1}{N}\, = \mu_N^{(min)}$.

\Eq{x3tomu3mu2:3} is the central result of this approach, which then characterizes the ddm in dimension $N$ in terms of all the purities
$\{\mu_N, \mu_{N-1},\ldots,\mu_3,\mu_2\}$ of dimensions $k\in[2,N]$ \cite{Tr:rho:cubed:note}.

For completeness, for the case of $N=4$, we have
\bsub
 \bea{N4:param}
 \hspace{-0.2in}
 \pNvecE{4}(x_4,x_3,x_2) &=& x_4\,\pNvecE{3}(x_3,x_2) + (1-x_4)\, \PKN{4}{4} , \; Tr[\pNvecE{4}]=1,\qquad\; \\
 \hspace{-0.2in}
 &=& x_4\left(x_3\Big(x_2\,\PKN{1}{3} + (1-x_2) \PKN{2}{3}\Big)\right., \no
 \hspace{-0.2in}
 &+& \left. (1-x_3)\, \PKN{3}{3}\right) +(1-x_4)\, \PKN{4}{4}, \\
 \hspace{-0.2in}
  1/4\le\mu_4 =  \pNvecE{4}&\cdot& \pNvecE{4}\le1, \;  \pNvecE{4}(x_4,x_3,x_2) =  \pNvecE{4}(\mu_4,\mu_3,\mu_2).
 \eea
\esub
 %=======================================================================
 \subsubsection{$\boldsymbol{\pNvecE{4}(x_4,x_3,x_2)\mapsto  \pNvece{4}(x_4,x_3,x_2)}$ and derivation of the angle constraints}\label{sec:constraints}
 Upon computing $ \pNvece{4}(x_4,x_3,x_2) =  \bfe\cdot\pNvecE{4}(x_4,x_3,x_2)$ we obtain the very ``nice" (regular) form
 \be{p4vece}
 %\hspace{-0.25in}
  \pNvece{4}(x_4,x_3,x_2) = 
 \left(\begin{array}{c}
 \frac{1}{2} \\
  \frac{1}{\sqrt{4\cdot 3}}\,x_4 \\
  \frac{1}{\sqrt{3\cdot 2}}\,x_4\,x_3  \\
  \frac{1}{\sqrt{2\cdot 1}}\,x_4\,x_3\,x_2
  \end{array}\right)
 \equiv
\left(
\begin{array}{c} 
 \frac{1}{2} \\ 
  \rmu\, \cos\varphi_3\\
  \rmu\, \sin\varphi_3\cos\varphi_2 \\
  \rmu\, \sin\varphi_3\sin\varphi_2 
\end{array}
\right),
 \ee
 where we have introduced the polar and azimuthal angles $\varphi_3$ and  $\varphi_2$ respectively, and the radius $\rmu = \sqrt{\mu_4-1/4}$ as before. Note that we have also permuted the order of the spherical coordinates in order to make an easier identification with the $x_k$ coordinates.
 By taking ratios of the coordinates we can easily determine that
 \bea{determining:angles}
x_4 &=& 2\,\sqrt{3}\,\rmu\,\cos\varphi_3,  \label{determining:angles:1} \\
x_3 &=& \frac{1}{\sqrt{2}}\,\tan\varphi_3\,\cos\varphi_2, \label{determining:angles:2}  \\
x_2 &=&  \frac{1}{\sqrt{3}}\,\tan\varphi_2.  \label{determining:angles:3}
 \eea
 Note that since a ddm in the WC is a valid convex combinations of WC vertices, we must require \tit{in general} that for each $k$, $0\le x_k\le 1$. 
If we now employ \Eq{x3tomu3mu2:3}, a little algebra reveals that
\bsub
\bea{tanphik}
\tan\varphi_2 &=& \sqrt{3}\,x_2 =  \sqrt{3}\,\sqrt{\dfrac{2\mu_2-1}{2\mu_1-1}} = \sqrt{3}\,\sqrt{2\mu_2-1}, \\
\tan\varphi_3 &=& \dfrac{\sqrt{2}}{\cos\varphi_2}\,x_3 = \sqrt{2}\, \left(\sqrt{2}\,\sqrt{3\,\mu_2-1}\right)\,  \sqrt{\dfrac{3\,\mu_3-1}{3\,\mu_2-1}} \no
&{}&  \hspace{0.6in}=\sqrt{4}\,\sqrt{3\mu_3-1}.
\eea
\esub
We can easily prove by induction that for arbitrary $k$ the above pattern generalizes  to
%\bea{angles:k}
%\tan\varphi_k &=& \sqrt{k+1}\,\sqrt{k\,\mu_k-1}, \quad\Rightarrow\quad 0\le\tan\varphi_k\le \sqrt{k^2-1}, \\
%\Rightarrow 
%\sin\varphi_k &=& \sqrt{\dfrac{k+1}{k}} \,\sqrt{\dfrac{k\,\mu_k-1}{(k+1)\,\mu_k-1}}, \quad\Rightarrow\quad 0\le\sin\varphi_k\le \dfrac{\sqrt{k^2-1}}{k}, \\
%%
%\Rightarrow 
%\cos\varphi_k &=& \dfrac{1}{\sqrt{k}}\,\dfrac{1}{\sqrt{(k+1)\,\mu_k-1}}, \quad\Rightarrow\quad \dfrac{1}{k}\le\sin\varphi_k\le 1,
%\eea
%
\bwt
\bsub
\begin{alignat}{3}
\tan\varphi_k &= \sqrt{k+1}\,\sqrt{k\,\mu_k-1}, \quad && \Rightarrow\quad 0\le\tan\varphi_k\le \sqrt{k^2-1}, \label{condition:tan:varphik}\\
\Rightarrow 
\sin\varphi_k &= \sqrt{\dfrac{k+1}{k}} \,\sqrt{\dfrac{k\,\mu_k-1}{(k+1)\,\mu_k-1}}, \quad && \Rightarrow\quad 0\le\sin\varphi_k\le \dfrac{\sqrt{k^2-1}}{k}, \label{condition:sin:varphik}\\
\Rightarrow 
\cos\varphi_k &= \dfrac{1}{\sqrt{k}}\,\dfrac{1}{\sqrt{(k+1)\,\mu_k-1}}, \quad && \Rightarrow\quad \dfrac{1}{k}\le\cos\varphi_k\le 1, \label{condition:cos:varphik}
\end{alignat}
\esub
\ewt
where we obtained the bounds on the above angles, by examining the  expressions on the left at $\mu_k \to \mu^{(min)}_k = 1/k$ and  $\mu_k \to \mu^{(max)}_k = 1$.

Finally, let us examine the condition $0\le x_k\le1$ (for creating convex combinations of vertices) and the restrictions on the angles as given above to obtain
\bwt
\bsub
\begin{alignat}{3}
0\le x_4 &= \sqrt{2}\,\sqrt{3}\,\rmu\,\cos\varphi_3\le 1 \quad && \Rightarrow\quad  \left(0\le\cos\varphi_3\le \dfrac{1}{\sqrt{2}\sqrt{3}\,\rmu} \right)
\cap \left(\dfrac{1}{3}\le\cos\varphi_3\le 1\right),\no
{} &{} {} && \Rightarrow\quad \dfrac{1}{3} \le X_3\equiv\cos\varphi_3 \le \Min\left[ \dfrac{1}{\sqrt{2}\sqrt{3}\,\rmu}, 1 \right], \label{Condition:x4}
\end{alignat}
and
\begin{alignat}{3}
0\le x_3 &= \dfrac{1}{\sqrt{2}} \tan\varphi_3\,\cos\varphi_2\le 1 \quad && \Rightarrow\quad 
 \left(0\le\cos\varphi_2\le \dfrac{\sqrt{2}}{\tan\varphi_3} \right)
\cap \left(\dfrac{1}{2}\le\cos\varphi_2\le 1\right),\no
{} &{} {} && \Rightarrow\quad \dfrac{1}{2} \le \cos\varphi_2 \le \Min\left[ \dfrac{\sqrt{2}}{\tan\varphi_3}, 1 \right] \label{Condition:x3:1}, \\
{} &{} {} && \Rightarrow\quad \cos^{-1}\left( \Min\left[ \dfrac{\sqrt{2}}{\tan\varphi_3}, 1 \right]\right) \le X_2=\varphi_2 \le \cos^{-1}(1/2)  = \dfrac{\pi}{3} \label{Condition:x3:2}.
\end{alignat}
\esub
\ewt
In the above we have defined the uniform deviates $X_3=\cos\varphi_3$ and $X'_2=\varphi_2$, since the volume element on 
 $\mathbb{R}^3 = \mathbb{R}\times S^2$ is given by
 \bsub
 \bea{dV}
 \hspace{-0.25in}
 dV &=& dr\,  (r\,d\varphi_3)\, (r\,\sin\varphi_3\,d\varphi_2) = dr\,r^2\, \sin\varphi_3\,d\varphi_3\,d\varphi_2,\no
  \hspace{-0.25in}
 &=& d(r^3/3)\,d(\cos\varphi_3)\,d(\varphi_2), \\
  \hspace{-0.25in}
 &\equiv& dX_4\,dX_3\,dX'_2, \quad\Rightarrow\quad (\trm{flat}) \\
  \hspace{-0.25in}
 \trm{with } X_4 &=& r^3/3, \quad X_3 = \cos\varphi_3, \quad X'_2 = \varphi_2,
 \eea
 \esub
(modulo an insignificant minus sign on $d(\cos\varphi_3)$ which simply changes the order of the upper and lower bounds; the volume comes from a Jacobian which involves an absolute value so that this sign drops out, i.e. $|-1|=1$).
The implication of the above is that the variables that constitute the \tit{uniform deviates} are $X_3=\cos\varphi_3$ and $X'_2=\varphi_2$ 
(i.e. not the angles $(\varphi_3, \varphi_2)$ directly). The variables  $X_3$ and $X'_2$ should be chosen \tit{uniformly} over the range of their upper and lower limits. Determining those limits, is the whole issue at hand here.
%=======================================================================

\subsection{Angle ranges for arbitrary $N$}
To generalize the above formulas for arbitrary $N$, it is instructive to write out the $x_k$ for $N=\{2,3,4,5,6\}$ to discern the general pattern by induction.
% see N5_and_iterated_integral_24Mar_4Aug2021.nb  
% in Desktop/random dm papers/notes/good working codes copies/sampling
\bsub
\bea{xk:N:2:3:4:5}
\hspace{-0.15in}
N=2:\, x_2 &=& \sqrt{2\cdot1}\, r_2 \cos\varphi_1 = \sqrt{2}\, r_2, \quad (\varphi_1\equiv\pi/2),  \\
\hspace{-0.15in}
N=3:\, x_2 &=& \frac{\tan\varphi_2}{\sqrt{3}},\; x_3 = \sqrt{3\cdot2} \, r_3 \cos\varphi_2, \\
\hspace{-0.15in}
N=4:\,  x_2 &=& \frac{\tan\varphi_2}{\sqrt{3}},\; x_3 = \frac{\cos\varphi_2\tan\varphi_3}{\sqrt{2}},\;
 x_4 = \sqrt{4\cdot 3}\,r_4\,\cos\varphi_3, \\
\hspace{-0.15in}
N=5:\,  x_2 &=& \frac{\tan\varphi_2}{\sqrt{3}},\; x_3 = \frac{\cos\varphi_2\tan\varphi_3}{\sqrt{2}},\; 
 x_4 = \sqrt{\frac{3}{5}}\,r_4\,\cos\varphi_3\tan\varphi_4,\; \no
\hspace{-0.15in}
            x_5 &=& \sqrt{5\cdot4}\,r_5\,\cos\varphi_5,\\
\hspace{-0.15in}
N=6:\,  x_2 &=& \frac{\tan\varphi_2}{\sqrt{3}},\; x_3 = \frac{\cos\varphi_2\tan\varphi_3}{\sqrt{2}},\; 
 x_4 = \sqrt{\frac{3}{5}}\,\cos\varphi_3\tan\varphi_4, \no
\hspace{-0.15in}
            x_5 &=& \sqrt{\frac{2}{3}}\,\cos\varphi_4\tan\varphi_5,\; x_6 = \sqrt{6\cdot5}\,r_6\,\cos\varphi_5,                        
\eea
\esub
with $r_k=\sqrt{\mu_k-1/k}$. From inspection of the above equations a general pattern is deduced.
For a given $N$, we have the $N-2$ angles $(\varphi_2; \varphi_{k\in\{3,\ldots,N-2\}}; \varphi_{N-1})$ 
which we denote as lowest angle: $\varphi_2$, middle angles: $\varphi_{k\in\{3,\ldots,N-2\}}$, and
highest angle: $\varphi_{N-1}$, which form the angles of an $(N-2)$-sphere $S^{(N-2)}$ for a fixed radius $r_k$.
Further, the condition (range) on the variables  $0\le\{x_k\}\le 1$ for $k\in\{1,2,\ldots,N\}$ used to construct the convex combinations of vertices in the WC are given for arbitrary $N$ by
\bwt
\bsub
 \bea{range:of:xs:N}
 \trm{radius:}\quad  0\le r_{\mu_N} &\equiv& \sqrt{\mu_N-1/N}\le  \sqrt{1-1/N}, \qquad 1/N\le\mu_N\equiv\Tr[\rho_N^2]\le 1, \label{range:of:xs:N:1} \\
 \trm{highest indexed angle:}\quad   0\le x_N &=& \frac{\cos\varphi_{N-1}}{\left(\frac{1}{ \sqrt{N\,(N-1)}\;r_{\mu_N}}\right)}  \equiv\frac{X_{N-1}}{\bar{X}_{N-1}}\le 1,  \label{range:of:xs:N:2}\\
 \trm{middle indexed angles:}\quad   0\le x_k &=&  \cos\varphi_{k-1}\; 
% \frac{\tan\varphi_k}{\left(\frac{b_k}{b_{k+1}}\right)} 
 \frac{\tan\varphi_k}{\sqrt{\frac{k+1}{k-1}}} \le 1,  \quad k\in\{3,\ldots,N-1\}, \qquad  \label{range:of:xs:N:3} \\
  \trm{smallest indexed angle:}\quad   0\le x_2 &=& \frac{\tan\varphi_2}{\sqrt{3}}  \le 1,  \label{range:of:xs:N:4}
 \eea
 \esub
 \ewt
 where we have defined the variable (capital) $X_k\equiv\cos\varphi_k$ which satisfies
 \be{Xk:defn}
 \hspace{-0.25in}
\frac{1}{k} \le X_k \equiv \cos\varphi_k \equiv \frac{1}{\sqrt{k}}\,\frac{1}{\sqrt{(k+1)\mu_k-1}}\le 1, \; \frac{1}{k}\le\mu_k\le 1, 
%\quad k\in\{3,\ldots,N-1\}
 \ee
 where the left and right bounds in \Eq{Xk:defn} arise from evaluating the  $\cos\varphi_k$ at $\mu_k=1$ and   $\mu_k=1/k$ respectively.

 In addition to the bounds imposed by the definition of $X_k \equiv \cos\varphi_k$, we must also impose the range restrictions \Eq{range:of:xs:N:1}-\Eq{range:of:xs:N:4}
  imposed by the  coefficients $\{x_k\}$ (little ``x") of the convex combinations of WC vertices.
  For this, we note the following
\bea{tanvarphik:defn}
\hspace{-0.25in}
0\le \tan\varphi_k &\equiv& \sqrt{k+1}\,\sqrt{k\,\mu_k-1}, \no
 &=& \frac{\sqrt{1-X^2_k}}{X_k}\le \sqrt{k^2-1}, \quad \frac{1}{k}\le\mu_k\le 1.
 \eea
 Therefore, the constraint \Eq{range:of:xs:N:4} becomes (shifting $k~\to~k+1$)
 \bsub
 \bea{X_k:constraint}
 \hspace{-0.60in}
 \frac{1}{k} &\le& X_{k} \equiv \cos\varphi_{k} \le \Min\left[ \frac{ \sqrt{\frac{k+2}{k}}\,X_{k+1}}{\sqrt{1-X^2_{k+1}}},1\right], 
  \; k\in\{2,\ldots,N-2\},\; \label{X_k:constraint:1} \qquad \\
 \hspace{-0.60in}
 &\Rightarrow& 
\cos^{-1}\left( \Min\left[ \frac{ \sqrt{\frac{k+2}{k}}\,X_{k+1}}{\sqrt{1-X^2_{k+1}}},1\right] \right) \le \varphi_k\le \cos^{-1}(1/k), 
\label{X_k:constraint:2}
 \eea
 \esub
where we have used the fact that since $\cos(\varphi)$ is monotonically decreasing for our range of interest, then
$a\le\cos(\varphi)\le b \Rightarrow \cos^{-1}(b)  \le \varphi \le \cos^{-1}(a)$.
For the lowest indexed angle $\varphi_2$, we additionally  have the constraint \Eq{range:of:xs:N:4}, $0\le\tan\varphi_2\le \sqrt{3} \Rightarrow 0\le \tan^{-1}(\sqrt{3}) = \cos^{-1}(1/2) = \pi/3$, which is self-consistent with \Eq{X_k:constraint:2} for $k=2$
The constraint for the  highest indexed angle $X_{N-1}$ for the variable $0\le x_N\le 1$ 
can be written from \Eq{X_k:constraint:2} as
\bsub
\bea{XN:constraint}
\frac{1}{N-1} &\le&  X_{N-1} = \cos\varphi_{N-1} \no
&\le& \bar{X}_{N-1} \equiv \Min\left[\frac{1}{\left( r_{\mu_N}\,\sqrt{N (N-1)}\right)},\,1\right], \qquad  \label{XN:constraint:line1}\\
&\Rightarrow & \cos^{-1}\left( \Min\left[\frac{1}{\left( r_{\mu_N}\,\sqrt{N (N-1)}\right)},\,1\right] \right)\no
&\le& \varphi_{N-1} \le  \cos^{-1}\left(\frac{1}{N-1}\right),\label{XN:constraint:line2}
\eea
\esub
where the limits on $ 0\le~r_{\mu_N}~\equiv~\sqrt{\mu_N-1/N}~\le ~\sqrt{1-1/N}, $ are given by \Eq{range:of:xs:N:1}.

The important point to note here is that $X^{(min)}_k = 1/k$, while  $X^{(max)}_k = X^{(max)}_k(X_{k+1})$ namely, the upper limit of $X_k$ is a function of the value of $X_{k+1}$. This means that in terms of nested integrals, we can give the values of the highest indexed $X_k$, which will determine the range of integration over the lower indexed $X_{k'<k}$. In terms of computing the CDFs, we should start from $X_2$ (or $\varphi_2$) and integrate ``upwards" successively in the angle variables $X_2\to X_3\to\cdots X_{N-1}$. 
After integration over these angle variables, we can then integrate over the radius variable $r_{\mu_N}$
to find the CDF for the distribution of density matrices as a function of their purity.

In applying these formulas for generating density matrices of fixed purity, we actually work ``downwards," i.e. given an
$r_N$, we then uniformly sample $X_{N-1}$ whose upper bound is determined by the former. 
The sampled value of $X_{N-1}$ then determines the upper bound of $X_{N-2}$, the next lower angle moving downwards. 
The sampled value of $X_{N-2}$ then determines the upper bound of $X_{N-3}$, and so on, until we reach $X_2$ (although for this lowest angle we will work with $\varphi_2$ directly).

In the following section we will illustrate these formulas for the case of $N=2$ (which is trivial), 
and $N=3$ (qutrit) and $N=4$ (two qubits).
We then indicate the minor changes that occur for $N=5$, which is the first dimension with proper lowest, middle and highest angles, from which the case of arbitrary $N>5$ is readily inferred. 
%=======================================================================

\subsection{Volume Element for arbitrary $\boldsymbol{N}$}
For a given value of $N$, we have defined above coordinates on an $(N-2)$-sphere $S^{N-2}$ with volume element given by
\bsub
\bea{Volume:defn:N}
\hspace{-0.15in}
dV &=& r^{N-2}_{\mu_N}\,dr_{\mu_N}\,d\varphi_2\,d\varphi_3\,\dots d\varphi_{N-1}\,\sin\varphi_3\,\sin\varphi^2_4\,\ldots\sin\varphi^{N-3}_{N-1}, \qquad\;
\label{Volume:defn:N:1} \\
\hspace{-0.15in}
&=& r^{N-2}_{\mu_N}\,dr_{\mu_N}\,\prod_{k=2}^{N-1} \sin^{k-2}\varphi_{k}\,d\varphi_k, \label{Volume:defn:N:2} 
\eea
\esub
%====================
In particular we have
\bsub
\bea{Vol:N4:N5}
\hspace{-0.15in}
N=2:\; dV_2 &=& dr_2, \\
\hspace{-0.15in}
N=3:\; dV_3 &=& dr_3\,d\varphi_2, \\
\hspace{-0.15in}
N=4:\; 
dV_4 &=& r^{2}_{\mu_4}\,dr_{\mu_4}\,(\sin\varphi_3\,d\varphi_3)\,(d\varphi_2)\no
\hspace{-0.15in}
&\equiv& r^{2}_{\mu_4}\,dr_{\mu_4}\,dX_3\,d\varphi_2, \label{Volume:defn:N:3} \\
\hspace{-0.15in}
N=5:\; 
dV_5 &=& r^{3}_{\mu_5}\,dr_{\mu_5}\,(\sin^2\varphi_4\,d\varphi_4)\,(\sin\varphi_3\,d\varphi_3)\,(d\varphi_2),\no
\hspace{-0.15in}
&\equiv& r^{3}_{\mu_5}\,dr_{\mu_5}\,\left(1-X^2_4 \right)^{1/2}\,dX_4\, dX_3\, d\varphi_2, \qquad \label{Volume:defn:N:4} \\
\hspace{-0.15in}
&{}& \no
\hspace{-0.15in}
N:\; \Rightarrow
dV_N &=& r^{N-2}_{\mu_N}\,dr_{\mu_N}\, \left(\prod_{k=3}^{N-1} \,\left(1-X^2_k \right)^{(k-3)/2}\,dX_k\right) \, d\varphi_2, \qquad\; \label{Volume:defn:N:5} \\
\hspace{-0.15in}
%\trm{where}\; X_2&\equiv& \varphi_2, \;
%(\trm{vs}\; X_2\leftarrow \cos\varphi_2), \; 
X_k &=& \cos\varphi_k, \; k\in\{2,\ldots,N-1\}.  \label{Volume:defn:N:5}
\eea
\esub
Note that while in general $X_k\leftrightarrow \cos\varphi_k$, the variable $\varphi_2$ is already a uniform deviate, 
so we have separated it out.
% and redefined $X_2\equiv \varphi_2$. 
That is, the definition of $\cos\varphi_k = [k\big((k+1)\mu_k-1\big)]^{-1/2}$ holds for all 
$k\in\{2,\ldots,N-1\}$, while the assignment $X_k\to\cos\varphi_k$ holds for $k\in\{3,\ldots,N-1\}$.
%with $X_2\equiv\varphi_2$.

Integrating out successive sets of variables, starting from the lowest angle ($k=2$), to the highest angle ($k=N-1$), and then over the radius variable ($k=N$), will produced the desire cumulative distribution functions from which we can uniformly sample the variables.

%=======================================================================
\section{Analytic CDF\lowercase{s} for $\boldsymbol{N=\{2,3,4\}}$}\label{sec:N:2:3:4}
For $N=\{2,3,4\}$ the above formulas can be computed analytically.
They are then very informative explicit examples of the type of features exhibited in the CDF for larger values of $N$.

\subsection{$\boldsymbol{N=2}$: a single qubit}
%While the particular case of $N=2$ turns out to be trivial, it worthy of consideration to show that the formulas of the previous section hold a this ``lower boundary."
In this section we give more details on the derivation of the cumulative distribution functions for $N=2$.
%in \Eq{p:of:X3:analytics:cumulative:1}- \Eq{p:of:X3:analytics:cumulative:3}. 
 This is in fact a trivial case, but it serves to illustrate the previous applicability of the general formulas at the ``boundary condition" of the smallest density matrix dimension, $N=2$. 
 For $N=2$ we have $\pNvece{2}(x_2)$ given by
\be{p4vece:N2}
\hspace{-0.5in}
N=2:\;  \pNvece{2}(x_2) = 
%  %
%  \left(\begin{array}{c}
% b_1 \\
%  b_2\,x_2
%  \end{array}\right) \leftrightarrow
%  %
 \left(\begin{array}{c}
 \frac{1}{\sqrt{2}} \\
  \frac{1}{\sqrt{2\cdot 1}}\,x_2
  \end{array}\right)
 \equiv
\left(
\begin{array}{c} 
 \frac{1}{\sqrt{2}} \\ 
  r_2\, \cos\varphi_1
\end{array}
\right),\;
r_2 = \sqrt{\mu_2-1/2}.
 \ee
 This implies $0\le x_2 = \sqrt{2}\,r_2\,\cos\varphi_1\le 1$.
However, from the general formula 
$\tfrac{1}{k}|_{\mu_k=1} \le X_k = \cos\varphi_k = \tfrac{1}{\sqrt{k}}\,\tfrac{1}{\sqrt{(k+1)\mu_k-1}} \le 1|_{\mu_k=\tfrac{1}{k}}$, we have for $k=1$ and $\mu_1\overset{\trm{def}}{=}1$ that
$1\le X_1=\cos\varphi_1\le1\Rightarrow \cos\varphi_1=\tfrac{\pi}{2}$. This implies that 
$0\le x_2 = \sqrt{2}\,r_2 = \sqrt{2\,\mu_2-1}\le 1$, which is consistent with $\mu_4^{(1)}\in[\tfrac{1}{2},1]$.
Since $x_2$ is uniformly distributed in $x_2\in[0,1]$, then so is $r_2$ in 
 $r_2^{(1)}\in[0,\tfrac{1}{\sqrt{2}}]$.
 Both of these variables act as the coordinate along the 1D simplex in \Fig{fig:Simplex:WC}(a).
We therefore have
\bsub
\bea{CDFN2r2}
\hspace{-0.25in}
F^{(N=2)}_2(r_2) &=& 
\dfrac
{
\int_{0}^{r_2\le r_2^{max}=1/\sqrt{2}} dr'_2 
}
{
\int_{0}^{r_2^{max}=1/\sqrt{2}} dr'_2 
}
= \frac{r_2}{1/\sqrt{2}} = \sqrt{2\,\mu_2-1}, \qquad\; \\
\hspace{-0.25in}
&\equiv&
\dfrac{F^{(N=2)Num}_2(r_2)}{F^{(N=2)Denom}_2} \label{CDFN2r2:line2}
\eea
\esub
where have employed the notation (to be used henceforth) $F^{(N)}_k$ to indicate the CDF for the 
variable $X_k$ for a given dimension $N$. Note that in this ``boundary condition'' case of $N=2$ 
there are no angles $\varphi_k$; only the radius vector $r_2$. 
In \Eq{CDFN2r2:line2} we have introduced the notation
$F^{(N=2)Num}_2(r_2)$ and $F^{(N=2)Denom}_2$ where the latter is the CDF normalizing denominator.
The difference in the two expressions is that the denominator $F^{(N=2)Denom}_2$ is integrated over the full range of
the variable $r'_2\in[0,r_2^{max}=\tfrac{1}{\sqrt{2}}]$ while the numerator $F^{(N=2)Num}_2$ is
integrated over  $r'_2\in[0,r_2\le r_2^{max}]$.
We will see that the denominator expressions are then reused in 
iterated integrals for the computation of CDFs for larger $N$.
For larger dimensions, and for the middle angles we will have that  
$F_k^{(N)Num}(X_k;X_{k+1})$ represents the numerator of the CDF for $X_k$, whose upper limits depends on the 
choice of the next higher up ``angle" $X_{k+1}$, while $F_k^{(N)Denom}(X_{k+1})$ represents the normalizing denominator.
This notation will be used throughout the rest of this work.
%=======================================================================

\subsection{$\boldsymbol{N=3}$: a single qutrit}
For $N=3$, the eigenvalue simplex is an equilateral triangle \Fig{fig:Simplex:WC}(b)(left) with coordinates $(\varphi_2, r_3)$. 
Here, the new polar angle $\varphi_2$ is measured relative to the $N~=~3$ MMS  \Fig{fig:Simplex:WC}(b)(right).
Since the volume element is $dV_2 = dr_3\,d\varphi_2$ we shall always use the angle $\varphi_2$ directly (vs our use of 
$X_k\equiv\cos\varphi_k$ for $k\ge3$).

For $N=3$ we have $\pNvece{3}(x_2, x_3)$ given by
\be{p3vece:N3}
%\hspace{-0.15in}
\pNvece{3}(x_2, x_3) = 
 \left(\begin{array}{c}
 \frac{1}{\sqrt{3}} \\
  \frac{1}{\sqrt{3\cdot 2}}\,x_2\\
  \frac{1}{\sqrt{2\cdot 1}}\,x_3\,x_2
  \end{array}\right)
 \equiv
\left(
\begin{array}{c} 
 \frac{1}{\sqrt{3}} \\ 
 r_3\, \cos\varphi_2\\
 r_3\, \sin\varphi_2
\end{array}
\right), \;
r_3 = \sqrt{\mu_3-1/3}.
 \ee
This implies $0\le x_2 = \sqrt{6}\,r_3\,\cos\varphi_2\le 1$.
However, from the general formula 
$\tfrac{1}{k}|_{\mu_k=1} \le X_k = \cos\varphi_k = \tfrac{1}{\sqrt{k}}\,\tfrac{1}{\sqrt{(k+1)\mu_k-1}} \le 1|_{\mu_k=\tfrac{1}{k}}$, 
we have for $k=2$  that
$\tfrac{1}{2}\le X_2=\cos\varphi_2\le1$.
This implies that 
$0\le x_2 = \sqrt{6}\,r_3\cos\varphi_2 \le 1$, or that
%=====================================
\bwt
\bsub
\bea{X2:varphi2:limits}
\tfrac{1}{2} &\le& X_2\equiv\cos\varphi_2 \le 1,  \label{X2:varphi2:limits:line1} \\
\Rightarrow 
\varphibarmin_2(r_3)
&\equiv& \cos^{-1}\left(\Min\left[ \tfrac{1}{\sqrt{3\cdot 2}\,r_3},1\right] \right)\le \varphi_2
\le
 \cos^{-1}\left(\tfrac{1}{2}\right)=\tfrac{\pi}{3}\equiv\varphibarmax_2, \label{X2:varphi2:limits:line2} \\
\trm{where}\; \varphibarmin_2 
&=& 
\left\{
\begin{array}{c}
0, \quad 
\hspace{1.3in}
r^{(1)}_3\in[0,\tfrac{1}{\sqrt{3\cdot 2}}]\leftrightarrow \mu^{(1)}_3\in[\tfrac{1}{3},\tfrac{1}{2}],\\
\cos^{-1}\left( \tfrac{1}{\sqrt{3\cdot 2}\,r_3} \right)\equiv \tilde{\varphi}_2(r_3), \quad 
r^{(2)}_3\in[\tfrac{1}{\sqrt{3\cdot 2}},\sqrt{\tfrac{2}{3}}]\leftrightarrow \mu^{(2)}_3\in[\tfrac{1}{2},1],
\end{array}
\right.  \label{X2:varphi2:limits:line3} 
\eea
\esub
\ewt
%=====================================
where for notational convenience 
we have defined $\tilde{\varphi}_2(r_3)\overset{\trm{def}}{=}\cos^{-1}\left( \tfrac{1}{\sqrt{3\cdot 2}\,r_3}\right)$.
In \Eq{X2:varphi2:limits:line2} we see that the upper limit of the angle $\varphi_2$ is always 
$\varphibarmax_2=\tfrac{\pi}{3}$, while the lower limit $\varphibarmin_2(r_3)$ depends on the selection of 
the radius $r_3=\sqrt{\mu_3-1/3}$. Since $X_2=\cos\varphi_2\le \tfrac{1}{\sqrt{6}\,r_3}$, 
$\varphibarmin_2(r_3)$ naturally breaks up into two regions depending on the magnitude of 
$\tfrac{1}{\sqrt{6}\,r_3}$ being greater or less than unity. The dividing (or ``breakpoint") occurs when 
$\tfrac{1}{\sqrt{6}\,r_3}=1$ corresponding to $\mu_3=\tfrac{1}{3}$. 

We will see that this last result is a general feature, namely, for the highest angle $\varphi_{N-1}$ ($\varphi_2$ in this case), its range of values will always be divided into two regions $[\tfrac{1}{N}, \tfrac{1}{N-1}]\cup[\tfrac{1}{N-1}, 1]$.
It will be seen that the first purity region $\mu_N^{(1)}\equiv [\tfrac{1}{N}, \tfrac{1}{N-1}]$
(where in general we define the $\mu_N$-regions $\mu_N^{(i)}\overset{\trm{def}}{=}[\tfrac{1}{N-(i-1)}, \tfrac{1}{N-i}]$ 
for $i\in\{1,\ldots,N-1\}$), is where the diagonal density matrices in the simplex are uniformly distributed. The purity value 
$\mu_N=\tfrac{1}{N-1}$ is the largest ``$\mu$-radius" for which the sphere $S^{(N-2)}$ can be inscribed within the $N-1$ simplex $\Delta^{(N-1)}$ (the ``in-sphere," see \cite{Zyczkowski_2ndEd:2020}). For the regions $\mu_N^{(i>1)}$ the sphere of fixed $\mu$-radius extends beyond the simplex $\Delta^{(N-1)}$, and intersects it in disjoint regions, which this work details for the WC. The other $N!-1$ regions of $\Delta^{(N-1)}$ are easily obtained by first randomly permuting the values of the diagonal density $\rho_{N,d}$ matrices before applying the unitary similarity transformation to obtain the density matrix $\rho_N=U\,\rho_{N,d}\,U^\dag$. 

We now compute the angle $\varphi_2$, and radius $r_3$, CDFs for $N=3$  
illustrating the complex dependency of their limits on each other. 
This will be needed for  subsequent calculations of CDFs for $N\ge4$.

%=================================
\subsubsection{CDF $F_2^{(N=3)}(\varphi_2; r_3)$}
Using the above expressions for $\varphibarmin_2(r_3)$, we then have
\bea{FN3k2:formal}
\hspace{-0.5in}
F_2^{(N=3)}(\varphi_2; r_3) &=& 
\dfrac
{
\int_{\varphibarmin_2(r_3)}^{\varphi_2\le\varphibarmax_2=\pi/3}\, d\varphi'_2
}
{
\int_{\varphibarmin_2(r_3)}^{\varphibarmax_2=\pi/3}\, d\varphi'_2
}
=
\frac
{
\varphi_2-\varphibarmin(r_3)
}
{
\pi/3-\varphibarmin(r_3)
} \no
\hspace{-0.5in}
&\equiv&
\frac
{
F^{(N=3)Num}_2(\varphi_2; r_3)
}
{
F^{(N=3)Denom}_2(r_3)
}
\eea
We compute the $r_3$-dependent denominator $F^{(N=3)Denom}_2(r_3)$ as follows
\bea{FN3k2Denom:by:region}
\hspace{-0.75in}
&{}& F_2^{(N=3)Denom}(r_3) \no
\hspace{-0.75in}
&=&
\left\{
\begin{array}{c}
\int_{\varphibarmin_2(r_3)=0}^{\varphibarmax_2=\pi/3}\, d\varphi'_2  =  \tfrac{\pi}{3}\hspace{0.35in}, 
\quad 
\hspace{0.75in} r^{(1)}_3\in\left[0, \tfrac{1}{\sqrt{3\cdot 2}}\right] \\
{}\\
\int_{\varphibarmin_2(r_3)=\tilde{\varphi}_2(r_3)}^{\varphibarmax_2=\pi/3}\, d\varphi'_2 
= \tfrac{\pi}{3}-\cos^{-1}\left( \tfrac{1}{\sqrt{3\cdot 2}\,r_3} \right), 
\; r^{(2)}_3\in\left[\tfrac{1}{\sqrt{3\cdot 2}},\sqrt{\tfrac{2}{3}}\right]. 
\end{array}
\right.
\eea
We therefore have
\be{FN3k2varphi2:r3}
\hspace{-0.5in}
F_2^{(N=3)}(\varphi_2; r_3) = 
\left\{
\begin{array}{c}
  \dfrac{\varphi_2}{\tfrac{\pi}{3}}, 
\quad 
\hspace{0.9in} r^{(1)}_3\in\left[0,\tfrac{1}{\sqrt{3\cdot 2}}\right] \\
\dfrac
{\varphi_2-\cos^{-1}\left( \tfrac{1}{\sqrt{3\cdot 2}\,r_3} \right)}
{\tfrac{\pi}{3}-\cos^{-1}\left( \tfrac{1}{\sqrt{3\cdot 2}\,r_3} \right)}, 
\quad r^{(2)}_3\in\left[\tfrac{1}{\sqrt{3\cdot 2}},\sqrt{\tfrac{2}{3}}\right]. 
\end{array}
\right. 
\ee

%=================================
\subsubsection{$F_3^{(N=3)}( r_3)$}
To compute the radial CDF $F_3^{(N=3)}( r_3)$ we use the expression above for $F_2^{(N=3)Denom}(r_3)$ 
in \Eq{FN3k2Denom:by:region} 
\bsub
\bea{CDFN3r3:formal}
\hspace{-0.5in}
F_3^{(N=3)}(r_3) &=& 
\dfrac
{
\int_{0}^{r_3\le r_3^{max}=\sqrt{2/3}} dr'_3\, r'_3\,F_2^{(N=3)Denom}(r'_3)
}
{
\int_{0}^{ r_3^{max}=\sqrt{2/3}} dr'_3\,  r'_3\, F_2^{(N=3)Denom}(r'_3)
},\\
\hspace{-0.5in}
&\equiv&
\dfrac
{
F_3^{(N=3)Num}(r_3) 
}
{
F_3^{(N=3)Denom} 
}.
\eea
\esub
Here, $F_3^{(N=3)Denom} $ is integrated over the whole range of 
$r_3\in r^{(1)}_3\cup r^{(2)}_3=\left[0,\tfrac{1}{\sqrt{3\cdot 2}}\right]\cup \left[\tfrac{1}{\sqrt{3\cdot 2}},\sqrt{\tfrac{2}{3}}\right]$, and hence is a constant.
Using \Eq{FN3k2Denom:by:region}  we have
\bea{F3N3Denom}
&{}& F_3^{(N=3)Denom} =
\int_{0}^{\tfrac{1}{\sqrt{6}}} dr'_3\,  r'_3\,   \tfrac{\pi}{3} \no
&{}&
+\int_{\tfrac{1}{\sqrt{6}}}^{ r_3^{max}=\sqrt{2/3}} dr'_3\,  r'_3\, \left[ \tfrac{\pi}{3}-\cos^{-1}\left( \tfrac{1}{\sqrt{3\cdot 2}\,r_3} \right) \right] 
=
\frac{1}{4\,\sqrt{3}}.\qquad\;
\eea
We then have 
for $r^{(1)}_3\in[0,\tfrac{1}{\sqrt{3\cdot 2}}]\leftrightarrow \mu^{(1)}_3\in[\tfrac{1}{3},\tfrac{1}{2}]$,
and 
$r^{(2)}_3\in\left[\tfrac{1}{\sqrt{3\cdot 2}},\sqrt{\tfrac{2}{3}}\right]\leftrightarrow \mu^{(2)}_3\in[\tfrac{1}{2},1]$,
%\bwt
\bsub
\bea{F3N3k3r3:by:region}
\hspace{-0.15in}
F_3^{(N=3)} (r_3) &=&
\frac{\int_{0}^{r_3\le\tfrac{1}{\sqrt{6}}} dr'_3\,  r'_3\,   \tfrac{\pi}{3} }{F_3^{(N=3)Denom}}
= \dfrac{\tfrac{\pi\,r_3^2}{6}}{\tfrac{1}{4\,\sqrt{3}}} = \frac{2\,\pi\,r_3^2}{\sqrt{3}}, 
%\quad 
%\hspace{1.25in}
%%%
%%%r^{(1)}_3\in[0,\tfrac{1}{\sqrt{3\cdot 2}}]\leftrightarrow \mu^{(1)}_3\in[\tfrac{1}{3},\tfrac{1}{2}],
\\
\hspace{-0.15in}
&=& 
\frac{
\int_{0}^{\tfrac{1}{\sqrt{6}}} dr'_3\,  r'_3\,   \tfrac{\pi}{3} 
+
\int_{\tfrac{1}{\sqrt{6}}}^{r_3\le\sqrt{\tfrac{2}{3}}} dr'_3\,  r'_3\,   
\left[
\tfrac{\pi}{3}
-\cos^{-1}\left(\tfrac{1}{\sqrt{3\cdot 2}\,r'_3}\right)
\right] 
}
{F_3^{(N=3)Denom}},
%%%
%%%\quad 
%%%r^{(2)}_3\in\left[\tfrac{1}{\sqrt{3\cdot 2}},\sqrt{\tfrac{2}{3}}\right]
%%%\leftrightarrow \mu^{(2)}_3\in[\tfrac{1}{2},1], \qquad\; 
\no
\hspace{-0.15in}
 &\equiv&  \dfrac{g_3^{(N=3)}(r_3)}{\tfrac{1}{4\,\sqrt{3}}}, \\
\hspace{-0.15in}
%\trm{where}\; 
g_3^{(N=3)}(r_3) &=& 
\frac{1}{12}
\left(
2\,\pi\,r_3^2 + \sqrt{6\,r_3^2 -1}\, -\, 6\,r_3^2\,\cos^{-1}\left( \tfrac{1}{\sqrt{6}\,r_3} \right).
\right)
\eea
\esub
%\ewt
Note that as a function of purity $\mu_3$ we simply substitute for these expressions
$r_3\to\sqrt{\mu_3-1/3}$, i.e.
 $F_3^{(N=3)} (\mu_3)~=~F_3^{(N=3)} (r_3=\sqrt{\mu_3-1/3})$.
 We will hold off plotting these CDFs until after the next section where we compute the CDFs for $N=4$.
%=================================

%=======================================================================
\subsection{$\boldsymbol{N=4}$: a pair of qubits}
For the case of $N=4$, representing a pair of qubits, 
we have the variables $(\varphi_2, X_3=\cos\varphi_3, r_4)$, with volume element
$dV_4 = r_4^2\,dr_4\, dX_3\,d\varphi_2$. In this section we analytically compute the
CDFs $F_2^{(N=4)}(\varphi_2; X_3)$, $F_3^{(N=4)}(X_3; r_4)$ and $F_4^{(N=4)}(r_4)$
where the the $F_k^{(N=4)}$ are functions of the first argument, for fixed selected value of the
second argument. For $N=4$ the lower angle is $\varphi_2$, there are no middle angles, and the 
highest angle is $X_3$. As such we now have the limiting values of $\varphi_2$ and $X_3$ given by
\bsub
\bea{N4:varphi2:X3:bounds}
\varphibarmin_2(X_3) &=& \cos^{-1}\left(\Min\left[\Xbar_2(X_3)\equiv\frac{\sqrt{2} X_3}{\sqrt{1-X_3^2}},1\right]\right),\qquad \\
&=&
\left\{
\begin{array}{c}
% \cos^{-1}\left( \frac{\sqrt{2} X_3}{\sqrt{1-X_3^2}}\right)\;   X_3\in[\tfrac{1}{3}, \tfrac{1}{\sqrt{3}}], \\
 \cos^{-1}\Big( \Xbar_2(X_3) \Big),\;   X_3\in[\tfrac{1}{3}, \tfrac{1}{\sqrt{3}}]\equiv X_3^{(I)}, \quad \\
 0,\;                                                             \hspace{0.85in} X_3\in[\tfrac{1}{\sqrt{3}}, 1]\equiv X_3^{(II)}, 
\end{array}
\right. \quad \\
X_3^{(max)}(r_4)&=&\Min\left[\Xbar_3(r_4) \equiv \frac{1}{\sqrt{4\cdot3}\; r_4},\, 1\right], \\
&=&
\left\{
\begin{array}{c}
 1,\;             \hspace{0.35in} r_4\in[0, \tfrac{1}{2\sqrt{3}}] \leftrightarrow \mu_4\in[\tfrac{1}{4},\tfrac{1}{3}], \\
 \frac{1}{2\sqrt{3}\; r_4},\;     r_4\in[\tfrac{1}{2\sqrt{3}}, 1] \leftrightarrow \mu_4\in[\tfrac{1}{3},1],
\end{array}
\right.
\eea
\esub
with $\varphibarmax_2\equiv \tfrac{\pi}{3}$, 
and $X_k^{(min)} = \tfrac{1}{k}$
for all $N$.

%===================================
\subsubsection{CDF: $F_2^{(N=4)}(\varphi_2; X_3)$}
%\subsection{CDF: $\boldsymbol{F_2^{(N=4)}(\varphi_2; X_3)}$}
For the CDF of $\varphi_2$, given a value of $X_3$, which we denote as $F_2^{(N=4)}(\varphi_2;X_3)$, we have
\bsub
\bea{F2N4:calc}
\hspace{-0.5in}
&{}& F_2^{(N=4)}(\varphi_2;X_3) 
=
\frac
{
 \int_{\varphibarmin_2(X_3)}^{\varphi_2\le\pi/3} d\varphi'_2
}
{
 \int_{\varphibarmin_2(X_3)}^{\pi/3} d\varphi'_2
}, \\
\hspace{-0.5in}
&{}&
\hspace{0.95in}
\equiv
\frac{F_2^{(N=4)Num}(\varphi_2;X_3)}{F_2^{(N=4)Denom}(X_3)}, \\
%
%%
%&=&
%\frac{
%\varphi_2-\varphibarmin_2(X_3)
%}
%{
%\pi/3-\varphibarmin_2(X_3)
%}
%=
%\frac{
%\varphi_2-\cos^{-1}\left(\Min\left[ \frac{\sqrt{2}X_3}{\sqrt{1-X^2_3}},\,1\right]\right)
%}
%{
%\pi/3-\cos^{-1}\left(\Min\left[ \frac{\sqrt{2}X_3}{\sqrt{1-X^2_3}},\,1\right]\right)
%}, \qquad \\
%
\hspace{-0.5in}
&=& \left\{
\frac{
\varphi_2-\cos^{-1}\left( \frac{\sqrt{2}X_3}{\sqrt{1-X^2_3}} \right)
}
{
\pi/3-\cos^{-1}\left( \frac{\sqrt{2}X_3}{\sqrt{1-X^2_3}} \right)
} \right. , 
 \;  X_3\in X_3^{(I)}, \\
\hspace{-0.5in}
&=&
\left\{
\frac{
\varphi_2
}
{
\pi/3
}\right. ,  
\; \hspace{1.2in}  X_3\in X_3^{(II)}, \quad
\eea
\esub
where in the last line we have used that in $X_3~\in~[\tfrac{1}{\sqrt{3}}, 1]\equiv X_3^{(II)}$ we have
$\cos^{-1}\left(\Min\left[ \frac{\sqrt{2}X_3}{\sqrt{1-X^2_3}},\,1\right]\right) = 
\cos^{-1}\left(1\right)~=~0$.
The normalizing denominator $F_2^{(N=4)Denom}(X_3)$ is given by
\bea{F2N4DenomX3}
&{}& F_2^{(N=4)Denom}(X_3)  \no
&{}& 
\hspace{0.2in}
= 
\left\{
\left.\begin{array}{c}
\pi/3-\cos^{-1}\left( \frac{\sqrt{2}X_3}{\sqrt{1-X^2_3}} \right),
\quad  X_3\in X_3^{(I)}, 
\\
\pi/3,
\quad \hspace{1.3in}  X_3\in X_3^{(II)}.
\end{array}\right.
\right.
\eea

%===========================================
\subsubsection{CDF: $F_3^{(N=4)}(X_3; r_4)$}
For the CDF of $X_3$, given a value of $r_4\leftrightarrow \mu_4$, we have
\bsub
\bea{F3N4:calc}
\hspace{-0.25in}
F_3^{(N=4)}(X_3; r_4) &=&
\frac
{
\int_{1/3}^{X_3\le\Xbarmax(r_4)}\,dX'_3 \, F_2^{(N=4)Denom}(X'_3)
}
{
\int_{1/3}^{\Xbarmax(r_4)}\,dX'_3 \, F_2^{(N=4)Denom}(X'_3)
}\no
\hspace{-0.25in}
&=& 
\frac
{
\int_{1/3}^{X_3\le\Xbarmax(r_4)}\,dX'_3 \, \int_{\varphibarmin_2(X_3')}^{\pi/3} d\varphi'_2
}
{
\int_{1/3}^{\Xbarmax(r_4)}\,dX'_3 \, \int_{\varphibarmin_2(X_3')}^{\pi/3} d\varphi'_2
}, \\
\hspace{-0.25in}
&\equiv&
\frac{F_3^{(N=4)(i)Num}(X_3; r_4)}{F_3^{(N=4)(i)Denom}(r_4)},\; \trm{for}\; \mu_4\in\mu^{(i)}_4. \qquad\;
\eea
\esub
%================================
%
Defining (for notational convenience)
$\varphibar_{2,c}^{(min)}(X_3)\equiv\cos^{-1}\left(\tfrac{\sqrt{2}\,X_3}{\sqrt{1-X_3^2}}\right)$
we have  $F_3^{(N=4)Num}(X_3; r_4)$ by $\mu_4$-region given by
%========================================================
\bsub
\bea{F3N4:r4:Num:by:region}
\hspace{-0.2in}
\trm{\underline{\tbf{Num:}}\; }\hspace{0.75in}  &{}&\no
\hspace{-0.2in}
%%%%%%% Region 1 %%%%%%%%%%%
\trm{\underline{Region 1}:\; }  
1/4\le \mu_4 &\le& 1/3,\; \Leftrightarrow\; 1_{\left(\mu_4=1/3\right)}\le \tfrac{1}{2\sqrt{3}\,\rmu}\le \infty_{\left(\mu_4=1/4\right)},   \no
\hspace{-0.15in}
F_3^{(N=4)(1)Num}(r_4)
&=& 
\left\{\begin{array}{c}
\int_{1/3}^{X_3\le1/\sqrt{3}} dX_3 \int_{ \varphibar_{2,c}^{(min)}(X_3) }^{\pi/3} d\varphi_2, \\
= f_{NL}(X_3), \quad\qquad X_3\in X_3^{(I)}, % \frac{1}{3} \le X_3 \le \frac{1}{\sqrt{3}}, 
 \end{array}\right. \label{F3N4:r4:Num:by:region:1:I}\\
\hspace{-0.15in}
&{}&\no 
\hspace{-0.15in}
&=&
\left\{\begin{array}{l}
 \int_{1/3}^{1/\sqrt{3}} dX_3 \int_{ \varphibar_{2,c}^{(min)}(X_3)}^{\pi/3} d\varphi_2 \\
 \hspace{0.25in} +  \int_{1/\sqrt{3}}^{X_3\le\Xbar_3=\tfrac{1}{2\sqrt{3}\rmu}=1} dX_3 \int_{0}^{\pi/3} d\varphi_2, \\
\dfrac{\pi}{6}\,(2\,X_3-1),  \quad\quad X_3\in X_3^{(II)}, %\;\; \frac{1}{\sqrt{3}} \le X_3 \le 1
\end{array}\right.  \qquad \label{F3N4:r4:Num:by:region:1:II} %\\
\eea
\bea{}
%
%%%%%%%%% Region 2 %%%%%%%%%%%%%%
%\hspace{-0.25in}
%&{}&\no
%%%%
\hspace{-0.15in}
\trm{\underline{Region 2:}\; }  
1/3\le \mu_4 &\le& 1/2, \,\Leftrightarrow\, 1/\sqrt{3}_{\left(\mu_4=1/2\right)}\le \tfrac{1}{2\sqrt{3}\,\rmu}\le 1_{\left(\mu_4=1/3\right)} ,   \no
\hspace{-0.15in}
 F_3^{(N=4)(2)Num}(r_4) &=& 
\left\{\begin{array}{l}
\int_{1/3}^{X_3\le\Xbar_3(r_4)\le1/\sqrt{3}} dX_3 \int_{ \varphibar_{2,c}^{(min)}(X_3)}^{\pi/3} d\varphi_2, \\
%
 %=\frac{\pi}{3}\,\left(\frac{1}{2\sqrt{3}\,\rmu}\right) - \frac{\pi}{6}, 
= f_{NL}(X_3),
 \quad\qquad  X_3\in X_3^{(I)}, % \frac{1}{3} \le X_3 \le \frac{1}{\sqrt{3}},
\end{array}\right.  \label{F3N4:r4:Num:by:region:2:I}\\
%%%%
\hspace{-0.15in}
&{}&\no
\hspace{-0.15in}
&=&
\left\{\begin{array}{l}
\int_{1/3}^{1/\sqrt{3}} dX_3 \int_{ \varphibar_{2,c}^{(min)}(X_3) }^{\pi/3} d\varphi_2 \\
\hspace{0.15in} +  \int_{1/\sqrt{3}}^{X_3\le\Xbar_3=\tfrac{1}{2\sqrt{3}\,\rmu}<1} dX_3 \int_{0}^{\pi/3} d\varphi_2, \\
%
%= \frac{\pi}{3}\,\left(\frac{1}{2\sqrt{3}\,\rmu}\right) - \frac{\pi}{6}, 
= \dfrac{\pi}{6}\,(2 X_3-1),
\quad\;  X_3 \in\,[\tfrac{1}{\sqrt{3}}, \bar{X}_3(r_4)], %\frac{1}{\sqrt{3}} \le X_3 \le 1,
\end{array}\right. \qquad \label{F3N4:r4:Num:by:region:2:II} %\\
\eea
%%%%
%&{}&\no
%\hspace{-0.25in}
%%%%%%%% Region 3 %%%%%%%%%%%%%%%%%%
\bea{}
\hspace{-0.15in}
\trm{\underline{Region 3}:\; }  
1/2\le \mu_4 &\le& 1, \,\Leftrightarrow\, 1/3_{\left(\mu_4=1\right)}\le \tfrac{1}{2\sqrt{3}\,\rmu}\le 1/\sqrt{3}_{\left(\mu_4=1/2\right)},   \no
\hspace{-0.25in}
 F_3^{(N=4)(3)Num}(r_4)
  &=& \int_{1/3}^{X_3\le\Xbar_3=\tfrac{1}{2\sqrt{3}\,\rmu}\le\tfrac{1}{\sqrt{3}}} dX_3
   \int_{ \varphibar_{2,c}^{(min)}(X_3)}^{\pi/3} d\varphi_2,
   \no
\hspace{-0.25in}
&\equiv&
f_{NL}(X_3)=
X_3\,\left(\frac{\pi}{3}-y(X_3) \right)\\  
&{}& 
%\hspace{0.15in}
+ \sin^{-1}\left(\frac{\sin y(X_3)}{\sqrt{3}}\right) - \frac{\pi}{6},\;  \;\; X_3 \in\,[\tfrac{1}{3}, \bar{X}_3(r_4)]\no
y(X_3)&=&\cos^{-1}\left(\tfrac{\sqrt{2}\,X_3}{\sqrt{1-X_3^2}}\right) \equiv \varphibar_{2,c}^{(min)}(X_3),  \label{F3N4:r4:Num:by:region:3} \no
\Xbar_3(r_4) &=& \frac{1}{2\sqrt{3}\,r_4}. \nonumber
\eea
\esub
%=========================

Repeating the same calculations as above, but now with $X_3\to X_3^{max}(r_4)$,
we have for the normalizing denominator $F_3^{(N=4)Denom}(r_4)$ by $\mu_4$-region
%========================================================
%\vspace{-.7in}
\bsub
\bea{F3N4:r4:Denom:by:region}
\hspace{-0.65in}
\trm{\underline{\tbf{Denom:}}\; }\hspace{0.7in}  &{}&\no
\hspace{-0.65in}
\trm{\underline{Region 1}:\; }  
1/4\le \mu_4 &\le& 1/3, \,\Leftrightarrow\,
 1_{\left(\mu_4=1/3\right)}\le \tfrac{1}{2\sqrt{3}\,\rmu}\le \infty_{\left(\mu_4=1/4\right)},   \no
 \hspace{-0.65in}
F_3^{(N=4)(1)Denom}(r_4)
 &=& \int_{1/3}^{1/\sqrt{3}} dX_3 \int_{ \varphibar_{2,c}^{(min)}(X_3)}^{\pi/3} d\varphi_2 \no
\hspace{-0.65in}
&+&  
\int_{1/\sqrt{3}}^{1} dX_3 \int_{0}^{\pi/3} d\varphi_2 \no
\hspace{-0.65in}
&=& \frac{\pi}{6} \equiv \frac{4\pi}{4!},%\Rightarrow\; \trm{uniform distribution over\;} S^2, 
\label{F3N4:r4:Denom:by:region:1} \\
%%%
\hspace{-0.65in}
&{}&\no
%%%
 \hspace{-0.65in}
\trm{\underline{Region 2:}\; }  
1/3\le \mu_4 &\le& 1/2, \,\Leftrightarrow\,
1/\sqrt{3}_{\left(\mu_4=1/2\right)}\le \tfrac{1}{2\sqrt{3}\,\rmu}\le 1_{\left(\mu_4=1/3\right)} ,   \no
\hspace{-0.65in}
 F_3^{(N=4)(2)Denom}(r_4) &=& \int_{1/3}^{1/\sqrt{3}} dX_3 \int_{ \varphibar_{2,c}^{(min)}(X_3)}^{\pi/3} d\varphi_2 \no
&+&  \int_{1/\sqrt{3}}^{\tfrac{1}{2\sqrt{3}\,\rmu}\le1} dX_3 \int_{0}^{\pi/3} d\varphi_2, \no
\hspace{-0.65in}
&=& \frac{\pi}{6}\,( 2\,\Xbar_3 - 1), \;
%\quad\Rightarrow A_{WC}^{(2)}(\mu_4=\tfrac{1}{3}) =  \frac{\pi}{6},\;   A_{WC}^{(2)}(\mu_4=\tfrac{1}{2}) = \frac{\pi}{6}\left(\frac{2}{\sqrt{3}}-1\right), \qquad   
\label{F3N4:r4:Denom:by:region:2}\\
%%%%
\hspace{-0.65in}
&{}&\no
%%%%
 \hspace{-0.65in}
\trm{\underline{Region 3}:\; }  
1/2\le \mu_4 &\le& 1, \,\Leftrightarrow\, 1/3_{\left(\mu_4=1\right)}\le \tfrac{1}{2\sqrt{3}\,\rmu}\le 1/\sqrt{3}_{\left(\mu_4=1/2\right)},   \no
\hspace{-0.65in}
 F_3^{(N=4)(3)Denom}(r_4)
  &=& \int_{1/3}^{\tfrac{1}{2\sqrt{3}\,\rmu}\le\tfrac{1}{\sqrt{3}}} dX_3 \int_{ \varphibar_{2,c}^{(min)}(X_3) }^{\pi/3} d\varphi_2 \label{F3N4:r4:Denom:by:region:3}
 %=  A_{WC}^{(3)}(\rmu) 
 \no
\hspace{-0.65in}
&=& f_{NL}(\Xbar_3), \qquad \Xbar_3(r_4) = \frac{1}{2\sqrt{3}\,r_4}.
%\frac{1}{2\sqrt{3}\,\rmu}\,\left(\frac{\pi}{3}-\bar{y} \right)  + \sin^{-1}\left(\frac{\sin\bar{y}}{\sqrt{3}}\right) - \frac{\pi}{6},\;\; 
%\bar{y} =\cos^{-1}\left(\tfrac{\sqrt{2}\,\bar{X}_3}{\sqrt{1-\bar{X}_3^2}}\right), \; \bar{X}_3 = \frac{1}{2\sqrt{3}\,\rmu}\; \qquad  \label{F3N4:r4:Denom:by:region:3}
\eea
\esub
%===========================================
%
%===========================================
%% page 13 onwards if invoking \clearpage\newpage
%\clearpage
%\newpage
%===========================================
Putting this altogether we finally obtain the CDF for $F_3^{(N=4)}(X_3; r_4)$ by $\mu^{(i)}_4$-region
%==========================
\bsub
\bea{p:of:X3:analytics:cumulative}
\hspace{-0.6in}
\trm{\underline{Region 1}:\; }  
1/4\le \mu_4 &\le& 1/3, \,\Leftrightarrow\, 1_{\left(\mu_4=1/3\right)}\le \tfrac{1}{2\sqrt{3}\,\rmu}\le \infty_{\left(\mu_4=1/4\right)},  \no
\hspace{-0.6in}
F_3^{(N=4)(1)}(X_3; r_4) &=& 
\left\{ 
\begin{array}{c}
\dfrac{f_{NL}(X_3)}{\pi/6}, \quad X_3\in X_3^{(I)},\\
2\,X_3-1, \qquad X_3\in X_3^{(II)},
\end{array}
\right. \label{p:of:X3:analytics:cumulative:1} \\
\hspace{-0.6in}
&{}& \no
\hspace{-0.6in}
\trm{\underline{Region 2:}\; }  
1/3\le \mu_4 &\le& 1/2, \,\Leftrightarrow\, 1/\sqrt{3}_{\left(\mu_4=1/2\right)}\le \tfrac{1}{2\sqrt{3}\,\rmu}\le 1_{\left(\mu_4=1/3\right)}   \no
%%%
\hspace{-0.6in}
F_3^{(N=4)(2)}(X_3; r_4) &=& 
\left\{ 
\begin{array}{c}
\hspace{-4em}
\dfrac{f_{NL}(X_3)}{\frac{\pi}{6}\,(2\, \bar{X}_3 -1)}, \quad X_3\in X_3^{(I)},\\
{} \\
\dfrac{2\,X_3-1}{2\,\bar{X}_3-1}, \qquad \frac{1}{\sqrt{3}} \le X_3 \le \bar{X}_3=\frac{1}{2\,\sqrt{3}\,\rmu}
\end{array}
\right. \label{p:of:X3:analytics:cumulative:2} \\
\hspace{-0.6in}
&{}& \no
%%%
\hspace{-0.6in}
\trm{\underline{Region 3}:\; }  
1/2\le \mu_4 &\le& 1, \,\Leftrightarrow\, 1/3_{\left(\mu_4=1\right)}\le \tfrac{1}{2\sqrt{3}\,\rmu}\le 1/\sqrt{3}_{\left(\mu_4=1/2\right)},   \no
\hspace{-0.6in}
F_3^{(N=4)(3)}(X_3; r_4) &=& 
\dfrac{f_{NL}(X_3)}{f_{NL}(\bar{X}_3)}, \qquad \frac{1}{3} \le X_3 \le \bar{X}_3 = \tfrac{1}{2\sqrt{3}\rmu},\;, \label{p:of:X3:analytics:cumulative:3} \\
\hspace{-0.6in}
&{}& \no
%%%
\hspace{-0.6in}
\trm{where}\; f_{NL}(X_3) &=& X_3\left(\frac{\pi}{3}-y(X_3) \right) 
\no
\hspace{-0.6in}
&+& 
\sin^{-1}\left(\frac{\sin y(X_3)}{\sqrt{3}}\right) - \frac{\pi}{6},  \qquad\;\label{fNLX3}\\
\hspace{-0.6in}
y(X_3)&=& \cos^{-1}\left(\frac{\sqrt{2}\,X_3}{\sqrt{1-X_3^{2}}}\right).
\eea
\esub
%==========================
From the above we see that for all  regions $\mu_4\in[\tfrac{1}{4},1]$ we have 
$F_3^{(N=4)(1,2,3)Num}(r_4) = f_{NL}(X_3)$ if $X_3\in [\tfrac{1}{3},\tfrac{1}{\sqrt{3}}]~\equiv~X_3^{(I)}$,
 and 
$F_3^{(N=4)(1,2)Num}(r_4) = \tfrac{\pi}{6}\,(2\,X_3-1)$ if $X_3\in [\tfrac{1}{\sqrt{3}},1]\equiv X_3^{(II)}$, 
(since for $F_3^{(N=4)(3)Num}(r_4)$ there simply is no region $X_3^{(II)}$).
While the numerator $F_3^{(N=4)(1,2,3)Num}(r_4)$  is essentially the same in each $\mu^{(1,2,3)}_4$ region,
what makes the CDF different in each region is the different normalizing denominators
 $F_3^{(N=4)(1,2,3)Denom}(r_4)$ in each of the three $\mu^{(1,2,3)}_4$ regions, i.e.
 $F_3^{(N=4)(1,2,3)}(r_4)=F_3^{(N=4)(1,2,3)Num}(r_4)/F_3^{(N=4)(1,2,3)Denom}(r_4)$.
 
% Note that for $X_3\ge \tfrac{1}{\sqrt{3}}$ in $\mu_4^{(i=1,2)}$
%the CDF $F_3^{(N=4)(1,2)}(X_3; r_4)$ is linear in $X_3$.
%In all three   $\mu_4^{(i)}$ regions $F_3^{(N=4)(i)}(X_3; r_4)$
% is nonlinear, characterized by $f_{NL}(X_3)$, for 
%$X_3\in X_3^{(I)} = [\tfrac{1}{3},\tfrac{1}{\sqrt{3}}]$.

It is also interesting to note that in region $\mu_4^{(1)}$
$F_3^{(N=4)(1)Denom}(r_4) = \tfrac{\pi}{6}\equiv\tfrac{4\pi}{4!}$.
This just indicates that region $\mu_4^{(1)}$ describes the ``in-sphere" which
can be completely inscribed inside the tetrahedron simplex $\Delta_3$, for which
the WC represents a fractional area of $1/(N=4)!$.

%====================================
\subsubsection{CDF for $r_4:\; F_4^{(N=4)}(r_4)$}
We now compute the radial CDF for $\rmu:\; F_4^{(N=4)}(\rmu)$ via ($r_4\equiv\rmu$)
\bsub
\bea{CDF:r4:formal}
 F_4^{(N=4)}(r_4)&=&
 \frac
 {
\int_{0}^{r_4\le\rbarmax=\frac{1}{2\sqrt{3}}} dr'_4 \,r^{'2}_4 \,F^{(N=4)Denom}(\Xbar_3)
 }
 {
 \int_{0}^{\rbarmax=\frac{1}{2\sqrt{3} }} dr'_4\, r^{'2}_4\, \,F^{(N=4)Denom}(\Xbar_3)
 },    \qquad\;\\
&\equiv&
  \frac
 {
F_4^{(N=4)Num}(r_4)
 }
 {
 F_4^{(N=4)Denom}
 }
\eea
\esub
in each of the three purity regions (Region 1, Region 2, Region 3) given respectively by
$\mu^{(1)}_4~\in~[\tfrac{1}{4},\tfrac{1}{3}]\leftrightarrow r^{(1)}_4\in[0,\tfrac{1}{2 \sqrt{3}}]$, \;
$\mu^{(2)}_4~\in~[\tfrac{1}{3},\tfrac{1}{2}]\leftrightarrow r^{(2)}_4\in[2 \sqrt{3},\tfrac{1}{2}]$, \;
$\mu^{(3)}_4~\in~[\tfrac{1}{2},1]\leftrightarrow r^{(3)}_4\in[\tfrac{1}{2},\tfrac{\sqrt{3}}{2}]$,
using the  $F_3^{(N=4)(i=\{i,2,3\})Denom}(r_4)$ from 
\Eq{F3N4:r4:Denom:by:region:1}, 
\Eq{F3N4:r4:Denom:by:region:2}, and 
\Eq{F3N4:r4:Denom:by:region:3}.
Note that the normalizing denominator $F_4^{(N=4)Denom}$ in \Eq{CDF:r4:formal} is a constant, while
the numerator $F_4^{(N=4)Num}(r_4)$ is $r_4$-dependent.

We proceed by first computing the indefinite integrals
$F_4^{(N=4)(I,\, II,\, III)Num}(r_4)$
required for $F_4^{(N=4)(i)Num}(r_4)$ for each Region~$i$:
\bsub
\bea{F4r4Num:by:m4:region}
% Region 1
\hspace{-0.5in}
\trm{\underline{Region 1}:\; }  
\mu^{(1)}_4~\in~[\tfrac{1}{4},\tfrac{1}{3}] &\leftrightarrow& r^{(1)}_4\in[0,\tfrac{1}{2 \sqrt{3}}], \no
\hspace{-0.5in}
%\left(F_3^{(N=4)(1)Denom}(r_4):\; \Eq{F3N4:r4:Denom:by:region:1}\right):\;  
F_4^{(N=4)(I)Num}(r_4) &=& \int^{r_4} dr_4\, r^{2}_4 \,\frac{\pi}{6} = \frac{\pi}{6}\,\frac{r^3_4}{3},\qquad
\label{F4r4Num:by:m4:region:1} \\
%%%
&{}&\no
%%%
% Region 2
%\hspace{-0.25in}
\trm{\underline{Region 2}:\; }  
\mu^{(2)}_4~\in~[\tfrac{1}{3},\tfrac{1}{2}] &\leftrightarrow& r^{(2)}_4\in[\tfrac{1}{2 \sqrt{3}},\tfrac{1}{2}],  \no
%
%\hspace{-0.25in}
%\left(F_3^{(N=4)(2)Denom}(r_4):\; \Eq{F3N4:r4:Denom:by:region:2}\right):\; 
%&{}& 
F_4^{(N=4)(II)Num}(r_4) &=& \int^{r_4} dr_4\, r^{2}_4 \,\frac{\pi}{6} \left(2\,\Xbar_3-1\right), \no
&=& \frac{\pi}{36}\left(\sqrt{3}-2\,r_4 \right) r^2_4, \label{F4r4Num:by:m4:region:2} %\\
\eea
\bea{}
%%%
%&{}&\no
%%%
% Region 3
\hspace{-0.5in}
\trm{\underline{Region 3}:\; }  
\mu^{(3)}_4~\in~[\tfrac{1}{2},1] &\leftrightarrow& r^{(3)}_4\in[\tfrac{1}{2},\tfrac{\sqrt{3}}{2 }], \no
\hspace{-0.5in}
%\left(F_3^{(N=4)(3)Denom}(r_4):\; \Eq{F3N4:r4:Denom:by:region:3}\right):\; 
%&{}& 
F_4^{(N=4)(III)Num}(r_4) &=& \int^{r_4} dr_4\, r^{2}_4 \,f_{NL}\big(\Xbar_3(r_4)\big), \no
&\equiv& g_{NL}(r_4)                                 
\label{F4r4Num:by:m4:region:3},  
\eea
\esub
where
\bea{gNLr4}
\hspace{-0.3in}
g_{NL}(r_4)  &=&
\frac{1}{432}
\left[
3\,\sqrt{8\,r_4^2-2} + \sqrt{3}\,\pi \left(+ 12\,r_4^4+ 8\,\sqrt{3}\, r^3_4 -1\right)
\right. \no
\hspace{-0.3in}
&-&
\sqrt{3}\,(36\,r^2_4)\sec^{-1}\left(\sqrt{6\,r_4^2-\tfrac{1}{2}}\right) 
\no
\hspace{-0.3in}
&+& \left. 
144 \, r_4^3\, \sin^{-1}\left( \sqrt{\frac{4\,r_4^2-1}{12\,r_4^2-1}}\right) 
\right],
\eea
with $f_{NL}\big(\Xbar_3(r_4)\big)$ defined in \Eq{F3N4:r4:Denom:by:region:3} 
with $\Xbar_3(r_4)=\tfrac{1}{2\sqrt{3}\,r_4}$.
Note: It is surprising that these integrals can be computed analytically in closed form.

With the above integrals we can now compute the normalizing constant denominator $F_4^{(N=4)Denom}$:
\bsub
\bea{F4Denom:calc}
\hspace{-0.5in}
F_4^{(N=4)Denom} &=&  
\left( F_4^{(N=4)(I)Num}(\tfrac{1}{2\sqrt{3}})-F_4^{(N=4)(I)Num}(0) \right), \no
&+& \left( F_4^{(N=4)(II)Num}(\tfrac{1}{2})-F_4^{(N=4)(II)Num}(\tfrac{1}{2\sqrt{3}}) \right) \no
&+& \left( F_4^{(N=4)(III)Num}(\tfrac{\sqrt{3}}{2})-F_4^{(N=4)(III)Num}(\tfrac{1}{2})\right),\qquad\; \\
&=& \frac{1}{72}.
\eea
\esub
%====================================
Using the above results we then have 
\bsub
%=================
\bea{F4r4:by:m4:region}
% Region 1
\hspace{-0.5in}
\trm{\underline{Region 1}:\; }  &{}&
\mu^{(1)}_4~\in~[\tfrac{1}{4},\tfrac{1}{3}] \leftrightarrow r^{(1)}_4 \in [0,\tfrac{1}{2 \sqrt{3}}], \no
\hspace{-0.5in}
F_4^{(N=4)(1)}(r_4) &=& 
\frac
{
\left( F_4^{(N=4)(I)Num}(r_4\le\tfrac{1}{2\sqrt{3}})-F_4^{(N=4)(I)Num}(0) \right)
}
{
F_4^{(N=4)Denom}
}, \no
&=& 4\,\pi\,r_4^3, \label{F4r4:by:m4:region:1} %\\
\eea
%%%%
%\hspace{-0.5in}
%&{}& \no
%%%%
\bea{}
% Region 2
\hspace{-0.5in}
\trm{\underline{Region 2}:\; }   &{}&
\mu^{(2)}_4~\in~[\tfrac{1}{3},\tfrac{1}{2}] \leftrightarrow r^{(2)}_4 \in [\tfrac{1}{2 \sqrt{3}},\tfrac{1}{2}], \no
\hspace{-0.65in}
F_4^{(N=4)(2)}(r_4) &=& 
\frac
{
F_4^{(N=4)(1)}(\tfrac{1}{2\sqrt{3}})
%\left( F_4^{(N=4)(I)Num}(\tfrac{1}{2\sqrt{3}})-F_4^{(N=4)(I)Num}(0) \right) 
%+ 
%\left( F_4^{(N=4)(II)Num}(r_4\le\tfrac{1}{2})-F_4^{(N=4)(II)Num}(\tfrac{1}{2\sqrt{3}}) \right)
}
{
F_4^{(N=4)Denom}
} \no
&+& 
\frac
{
\left( F_4^{(N=4)(II)Num}(r_4\le\tfrac{1}{2})-F_4^{(N=4)(II)Num}(\tfrac{1}{2\sqrt{3}}) \right)
}
{
F_4^{(N=4)Denom}
}
\no
\hspace{-0.5in}
&=& 2\,\sqrt{3}\,\pi\,r_4^2 -4\,\pi\,r_4^3 - \frac{\pi}{6\,\sqrt{3}}
\label{F4r4:by:m4:region:2} %\\
\eea
%%%%%
%&{}& \no
%%%%%
\bea{}
% Region 3
\hspace{-0.6in}
\trm{\underline{Region 3}:\; }   &{}&
\mu^{(3)}_4~\in~[\tfrac{1}{2},1] \leftrightarrow r^{(3)}_4 \in [\tfrac{1}{2},\tfrac{\sqrt{3}}{2}], \no
\hspace{-0.6in}
&{}& F_4^{(N=4)(3)}(r_4) = 
\frac
{
%\left( F_4^{(N=4)(I)Num}(\tfrac{1}{2\sqrt{3}})-F_4^{(N=4)(I)Num}(0) \right) 
F_4^{(N=4)(2)}(\tfrac{1}{2}) 
%+ 
%\left( F_4^{(N=4)(I)Num}(\tfrac{1}{2})-F_4^{(N=4)(I)Num}(\tfrac{1}{2\sqrt{3}}) \right)
}
{
F_4^{(N=4)Denom}
}, \qquad 
\no
\hspace{-0.6in}
&{}&+
\frac
{
\left( F_4^{(N=4)(III)Num}(r_4\le\tfrac{\sqrt{3}}{2})-F_4^{(N=4)(III)Num}(\tfrac{1}{2}) \right)
}
{
F_4^{(N=4)Denom}
}, \qquad\;
\eea
\bea{}
\hspace{-0.15in}
&{}& F_4^{(N=4)(3)}(r_4)= 2\,\sqrt{3}\,\pi\,r_4^2 + 4\,\pi\,r_4^3 - \frac{\pi}{6\,\sqrt{3}}
+ \sqrt{2\,r_4^2-\tfrac{1}{2}} 
\no
\hspace{-0.15in}
&+& \frac{(36\,r_4^2-1)}{2\,\sqrt{3}}\,
\sec\left(\sqrt{6\,r_4^2-\tfrac{1}{2}}\right)
+ 24\,r_4^3\,\sin^{-1}\left( \sqrt{\frac{4\,r_4^2-1}{12\,r_4^2-1}}\right).\qquad\;
\label{F4r4:by:m4:region:2} 
\eea
%=================
\esub
%===============================

%===============================
\subsection{Plots of radial CDFs}
In this section we show plots of the radial CDFs for $N=\{2,3,4\}$ from the analytic formulas derived in the previous section, and scatter plots of the distribution of the diagonal density matrices
on $S^{(1)}$ and  $S^{(2)}$ for $N=3$ and $N=4$, respectively.

\subsubsection{Composite plot of the radial CDF  for $N\in\{2, 3, 4\}$}
\begin{figure}[h]
\begin{tabular}{cc}
{\hspace{0.5em}} & \includegraphics[width=3.5in,height=2.25in]{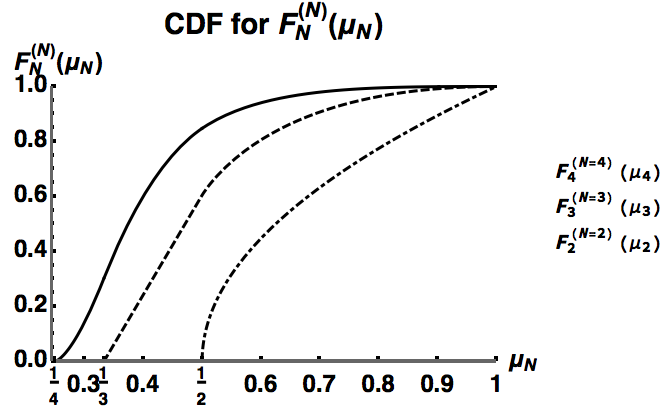} %&
%{\hspace{0.5em}} & \includegraphics[width=3.5in,height=2.25in]{CDF_FNkNrN_6Mar2022} %&
%\includegraphics[width=3.0in,height=1.5in]{alphaplus_wfdivw0_2_beta_0p1}
\end{tabular}
\caption{Radial CDFs  $F_N^{(N)}(\mu_N)$ for $N\in\{2,3,4\}$.
}\label{fig:CDF:FNkNrN:for:N:2:3:4}
\end{figure}

In \Fig{fig:CDF:FNkNrN:for:N:2:3:4}
we show  the radial CDFs for $N=\{2,3,4\}$ in a single composite plot.
%=========================================================
% from locate_NumRecC_23Feb2022.nb in Downloads
%=========================================================
%=========================================================
Note that as $N$ increases, the curves become increasingly flat for higher purities, i.e.
$F_N^{(N)}(\mu_N)~\sim~1-\epsilon$ for $\mu_N\to\mu^{(N)}_N\in[\tfrac{1}{2},1]$. At $N>4$ this increase becomes more severe, and as discussed earlier, in the case of $N=4$, posses numerical challenges for the interpolation of the CDFs and for the numerical inversion of the inverse CDFs (and not just for $r_N$, but also for higher angles $\varphi_k\sim N$, i.e. $k$ near $N$). This is exacerbated by the fact that the integration measure for a given $N$ contains a factor of $\sin^{(N-3)}\varphi_N$ (and $\sin^{(N-3-k)}\varphi_k$ for lower angles) which become increasingly peaked at $\varphi_k\sim\pi/2$ or $X_k=\cos\varphi_k\sim 1$. We discuss an approximation scheme for larger $N$ based on this fact in a later section. 

In \Tbl{tbl:purity:by:region} we show the percentage of purity by region obtained from the radial CDFs. 
The last two columns, illustrates the increasing flatness of $F_N^{(N)}(\mu_N)$ with increasing $N\in\{2,3,4\}$, which shows the increasing rarity of obtaining near maximal purity states by random sampling. As an example, for $N=4$ 
one would expect 
% PMA: orig
%$\sim 5\, (7)$ samples in the region $\mu_4\in[0.99,1]\, ([0.95,1]$)  per every $10^{7}\, (10^{5})$ trials 
% James' renomalized to same number of samples
$\sim 5 (660)$ samples in the region $\mu_{4}\in[0.99,1]\, ([0.95,1])$ per every $10^{7}$ trials 
of density matrices generated from conventional uniform sampling (via the Haar measure)  of random unitary matrices.
In comparison, in the plots generated in the next section, $10^3$ density matrices with exactly $\mu_4\equiv 0.99$ were generated in approximately $2$~secs (in \tit{Mathematica}, with no optimization).
%============================
% Set Latex for Everyone, p116
% PMA version
%============================
%\medskip
%\begin{table}[h]
%%\hspace{-0.5in}\hspace{-0.5em}
%%\begin{center}
%\begin{tabular}{|c|c|c|c|c|c|} \hline
%\multicolumn{6}{|c|}{\bf Percentage of Purity by Region} \\ \hline
%\multicolumn{1}{|c|}{} &
%\multicolumn{1}{|c|}{$\mu\in[\tfrac{1}{4},\tfrac{1}{3}]$\;} &
%\multicolumn{1}{|c|}{$\mu\in[\tfrac{1}{3},\tfrac{1}{2}]$\;}  &
%\multicolumn{1}{|c|}{$\mu\in[\tfrac{1}{2},1]$\;}                 &
%\multicolumn{1}{|c|}{$\mu\in[0.95,1]$\;}                           &
%\multicolumn{1}{|c|}{$\mu\in[0.99,1]$\;}  
%\\ [0.25em] \hline\hline
%$\,N=2$\, & - & - & 100\,\% & 5.13\% &1.01\,\%\\ \hline
%$\,N=3$\, & - & 60.46\,\% & 39.54\,\%& 0.20\,\% & \;$7.5\times 10^{-3}$\,\%\;  \\ \hline
%$\,N=4$\, & 30.23\,\% & 54.54\,\% & 15.23\,\% & \;$6.6\times 10^{-3}$\,\%\; &  \;$5.1\times 10^{-5}$\,\%\; \\ \hline
%\end{tabular}
%%\end{center}
%\caption{Percentage of Purity by Region for 
%(columns 2-4) $N\in\{2, 3, 4\}$, and
%(columns 5-6)  $\mu\in[0.95,1]$ and $\mu\in[0.99,1]$, respectively, illustrating the increasing rarity of obtaining near maximal purity states from uniform random sampling via the Haar measure.
%}\label{tbl:purity:by:region}
%\end{table}
%%============================
% James' table version: 10May2022
%============================
\begin{table}[h]
\begin{centering}
\begin{tabular}{|c|c|c|c|c|c|}
\hline
\multicolumn{6}{|c|}{$\textbf{$\%$ of states by Purity $\mu$ and dimension $N$ }$}\\
\hline
N & $\mu\!\in\![\frac{1}{4},\frac{1}{3}]$ & $\mu\!\in\![\frac{1}{3},\frac{1}{2}]$ & $\mu\!\in\![\frac{1}{2},1]$ & $\mu\!\in\![0.95,1]$ & $\mu\!\in\![0.99,1]$\\
[0.25em] 
\hline
\hline
2 & -  & - & 100 & 5.13 & 1.01 \\ 
\hline
3 & - & 60.46 & 39.54 & 0.20 & $7.5\times 10^{-3}$ \\
\hline
4 & 30.23 & 54.54 & 15.23 & $6.6\times 10^{-3}$ & $5.1\times 10^{-5}$\\
\hline
\end{tabular}
\end{centering}
\caption{Percentage of Purity by Region for 
(columns 2-4) $N\in\{2, 3, 4\}$, and
(columns 5-6)  $\mu\in[0.95,1]$ and $\mu\in[0.99,1]$, respectively, illustrating the increasing rarity of obtaining near maximal purity states from uniform random sampling via the Haar measure.
}\label{tbl:purity:by:region}
\end{table}
%=========================================================

\begin{figure}[h]
\begin{tabular}{c}
\includegraphics[width=3.0in,height=2.25in]{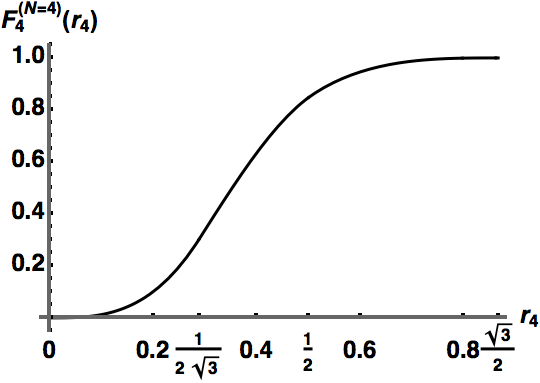} \\
%\includegraphics[width=3.0in,height=2.25in]{CDF_FN4r4_5Mar2022} \\
%\hspace{0.25in}
\includegraphics[width=3.0in,height=2.25in]{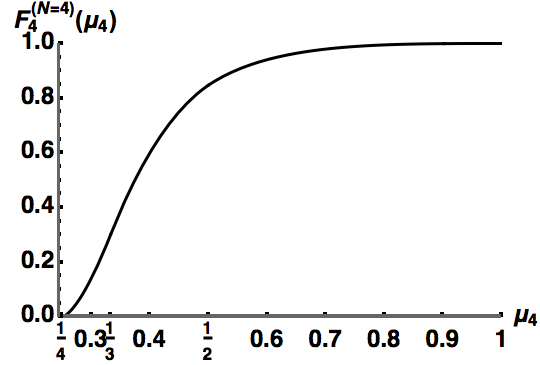}
\end{tabular}
\caption{
(top)  radial  CDF for $\rmu:\, F_4^{(N=4)}(r_4)$,
(bottom) purity CDF for $\mu_4:\, F_4^{(N=4)}\big(r_4(\mu_4)\big)$
where $\rmu = \sqrt{\mu_4-1/4}$.
}\label{fig:CDF:FN4:r4:and:mu4}
\end{figure}

\begin{figure}[h]
\begin{tabular}{cc}
\hspace{2em} & \includegraphics[width=3.5in,height=2.25in]{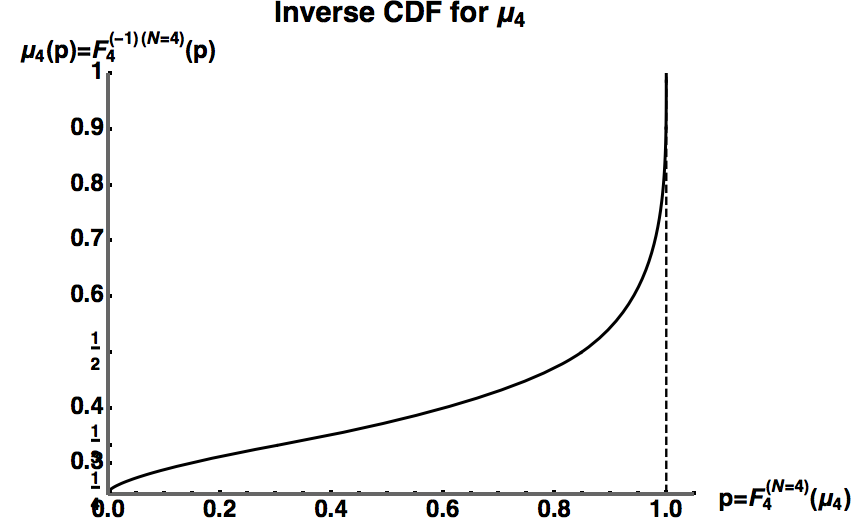}
%\hspace{2em} & \includegraphics[width=3.5in,height=2.25in]{CDFinv_FN4mu4_v2_5Mar2022}
\end{tabular}
\caption{
Inverse radial CDF for $\mu_4:\, F^{(N=4)}\big(r_4(\mu_4)\big)$.
}\label{fig:invCDF:FN4:mu4}
\end{figure}

In \Fig{fig:CDF:FN4:r4:and:mu4} 
%=========================================================
% from locate_NumRecC_23Feb2022.nb in Downloads
%=========================================================
%============================
we plot the CDF for (top) $F_4^{(N=4)}(r_4)$ and (bottom) $F_4^{(N=4)}\big(r_4(\mu_4)\big)$
where the later is obtained from the former by the variable substitution 
$\rmu = \sqrt{\mu_4-1/4}$. The three regions $r_4^{(i=\{1,2,3\})}$ and 
 $\mu_4^{(i=\{1,2,3\})}$ are indicated on the abscissa.
In \Fig{fig:invCDF:FN4:mu4} we plot the inverse CDF for $\mu_4$,
%=========================================================
% from locate_NumRecC_23Feb2022.nb in Downloads
%=========================================================
%============================
which simply switches the abscissa and ordinate in the bottom plot of \Fig{fig:CDF:FN4:r4:and:mu4}.
However, this figure illustrates two things. First, to uniformly sample a value of $\mu_4$ according to 
the CDF $F_4^{(N=4)}(\mu_4)$ we treat
$p=F_4^{(N=4)}(\mu_4)\in[0,1]$ as a uniform deviate (abscissa in \Fig{fig:invCDF:FN4:mu4})
which then determines a value of $\mu_4(p)$ (ordinate in  \Fig{fig:invCDF:FN4:mu4}).
Second, we see that for $p\sim 1-\epsilon$, the slope of $\mu_4(p)$ is nearly vertical (infinite), and
interpolation of the values $\{p, \mu_4(p) \}$ becomes increasingly inaccurate.
This is equivalent to the statement that in \Fig{fig:CDF:FN4:r4:and:mu4}
$F_4^{(N=4)}(\mu_4)\sim 1-\epsilon$ for  $\mu_4\sim1-\delta$, i.e. values of $F_4^{(N=4)}(\mu_4)$
are infinitesimally close to $1$ as $\mu_4\to 1$, and thus becomes computationally hard to represent and distinguish since the addition (subtraction) of a large number $1$ with an infinitesimal number $\epsilon$ becomes increasingly inaccurate in floating point arithmetic on a computer. Care must be taken in the interpolation of
pairs of values  $\{\mu_4, p(\mu_4) \}$ and it numerical inversion  $\{p, \mu_4(p) \}$.
Both the former and latter become significant numerical issues as we numerically compute the CDF integrals for 
$\{X_k,\, F_k^{(N)}(X_k; X_{k+1})\}$ for $N>4$, and then perform the numerical interpolation (or root finding procedure) of its inverse CDF
of pairs $\{p=F_k^{(N)}(X_k; X_{k+1}), X_k(p)\}$.

%==================================================================
\subsubsection{$N=3$: $1$-spheres $S^{(1)}$ in simplex $\Delta_2$ for fixed $\mu_3$}
\begin{figure}[h]
%\begin{tabular}{cc}
\includegraphics[width=3.0in,height=3.0in]{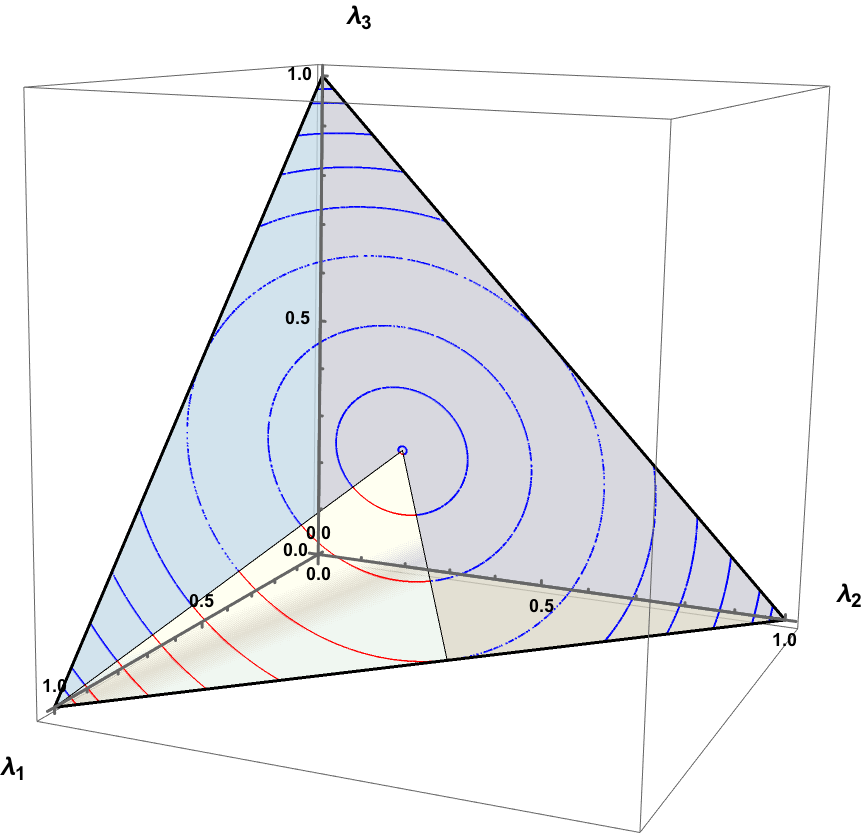}
%\includegraphics[width=3.0in,height=3.0in]{fig_S1_circles_of_constant_mu3_WC_and_2Simplex_14Mar2022}
%\end{tabular}
\caption{
Uniform sample of 1-spheres $S^{(1)}$ (blue) of fixed 
$\mu_3=\{0.3334, 0.35, 0.40, 0.50, 0.60, 0.70, 0.80, 0.90, 0.95, 0.99\}$
outward from the center ($\mu_3=1/3$) of 2-simplex $\Delta_2$    to the vertices ($\mu_3=1$).
The white triangle in the lower left corner is the WC ($1/3!$ of $\Delta_2$) with partial circular arcs (red) 
uniformly sampled from $F_2^{N=3}(\mu_3)$ in \Eq{FN3k2varphi2:r3}.
}\label{fig:N3:circles:of:fixed:mu3}
\end{figure}

In \Fig{fig:N3:circles:of:fixed:mu3} we plot the inverse CDF for $F_2^{(N=3)}(\varphi_2; r_3)$
 from \Eq{FN3k2varphi2:r3}. That is given a uniform deviate $p~=~F_2^{(N=3)}(\varphi_2; r_3)\in [0,1]$ for 
 a fixed value of the purity-radius $r_3=\sqrt{\mu_3-1/3}$, we compute $\varphi_2(p; r_3)$, which then uniformly samples $\varphi_2$
 over $F_2^{(N=3)}$. Since $F_2^{(N=3)}$ is linear in $\varphi_2$, this can be analytically performed (trivially).
 We then compute $\vec{p}_e^{\,(N=3)}~=~\big(\tfrac{1}{\sqrt{3}},\, \vec{r}^{\,(N=3)}(r_3, \varphi_2)\big)$, which is the diagonal eigenvalue vector $\vec{\lambda}_E^{\,(N=3)}= (\lambda_1, \lambda_2, \lambda_3)$ transformed to the $e$-basis 
 (where the first component is always $\tfrac{1}{\sqrt{3}}$),
 and plot the two-vector $\vec{r}^{\,(N=3)}~\equiv ~(r_3\,\cos\varphi_2, r_3\,\sin\varphi_2)$.
 The latter is a 2D vector in plane of the equilateral triangle that is the 
 $N=3$ eigenvalue  simplex $\Delta_2$, and is centered on the MMS 
 ($r_3=0\leftrightarrow \mu_3=\tfrac{1}{3})$ at the triangle's center. 
 The direction $\varphi_2=0$ points to the $N=2$ MMS  $\half(1,1,0)$  
 embedded in $N=3$ WC $\tilde{\Delta}_2$ (white triangle in lower left corner of \Fig{fig:N3:circles:of:fixed:mu3}), 
 while $\varphi_2=\tfrac{\pi}{3}$ points to the WC pure state $(1,0,0)$.
%=========================================================
% from locate_NumRecC_23Feb2022.nb in Downloads
%=========================================================
%============================
Each (blue) concentric circle represents diagonal density matrices of fixed radius $r_3=\sqrt{\mu_3-1/3}$.
The center of the triangle $r_3=0$ is the MMS of purity $\mu_3=\tfrac{1}{3}$. The largest inscribed circle is
at the ``in-radius" $r_3=\tfrac{1}{\sqrt{6}}=r_3^*$ at $\mu_3=\tfrac{1}{2}$. For $r_3>r_3^*$ the concentric circles lie outside the equilateral triangle and therefore only intersect the triangle in three segments near the vertices. The minimum angle is 
no longer 0, and is given by $\cos^{-1}\left( \tfrac{1}{\sqrt{3\cdot 2}\,r_3}\right)$ as in \Eq{X2:varphi2:limits:line3}, while the maximum angle in the WC is still $\pi/3$.

The above is a generic feature for higher dimensions. Namely, for a dimension $N$ the eigenvalue simplex $\Delta_{N-1}$ is an equilateral simplex (hyper-triangle) of dimension $N-1$. The ``in-radius" for the largest inscribed sphere $S^{(N-1)}$ occurs at 
$r_N^*~=~\sqrt{\tfrac{1}{N-1} - \tfrac{1}{N}} \leftrightarrow \mu_N = \tfrac{1}{N-1}$. For $r_N>r_N^*$ the spheres $S^{(N-1)}$ lie outside the  simplex $\Delta_{N-1}$, and therefore intersect it near its pure state vertices $\vec{e}_i = (0,\ldots,1_i,\ldots, 0)$.
This is the origin of the complicated, nested angle bound on $\varphi_k$ for general $N$.
We'll see this explicitly for the 2-spheres $S^{(2)}$ in the tetrahedron eigenvalue simplex for $N=4$ in the next section.

%==================================================================
\subsubsection{$N=4$: $2$-spheres $S^{(2)}$ in simplex $\Delta_3$ for fixed $\mu_4$}
\begin{figure*}[ht]
\begin{tabular}{ccc}
\includegraphics[width=2.55in,height=2.5in]{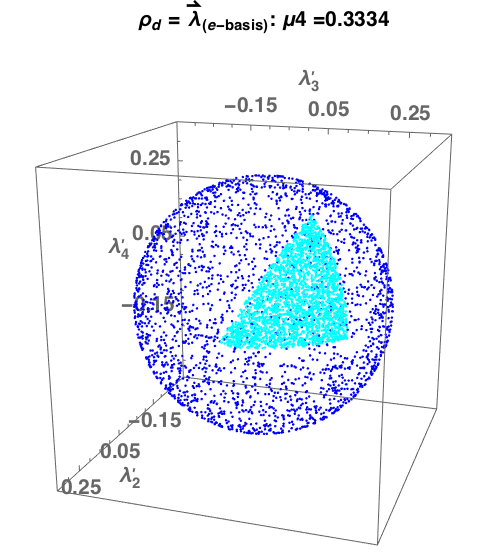} & \hspace{0.5in} &
\includegraphics[width=2.55in,height=2.5in]{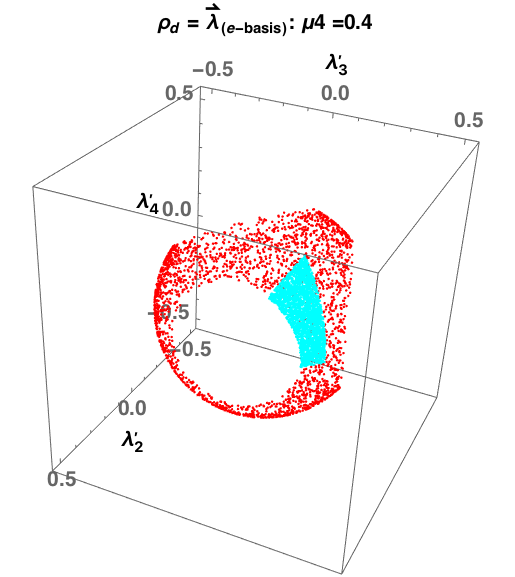} \\
\includegraphics[width=2.55in,height=2.5in]{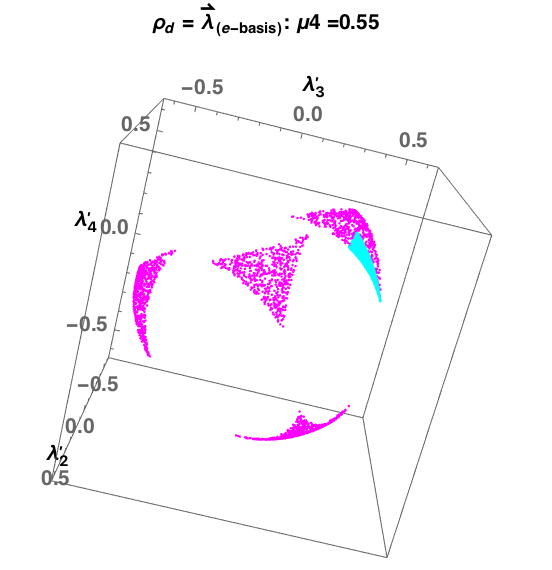}& \hspace{0.5in} &
\includegraphics[width=2.55in,height=2.5in]{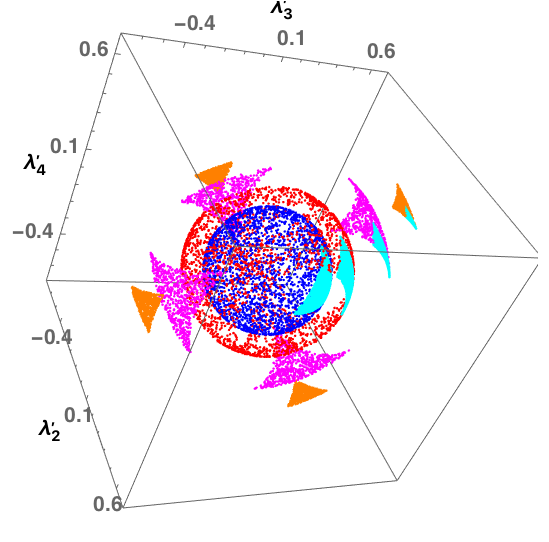}
\end{tabular}
\caption{
Diagonal density matrices 
%$\vec{r}^{\,(N=4)}(\varphi_2,\, \varphi_3,\,r_4)$
$\vec{r}^{\,(N=4)}(r_4\, \varphi_3,\, \varphi_2)$
 in the tetrahedral eigenvalue simplex $\Delta_3$ for $N=4$ for 
(top left, blue) $\mu_4=0.3334$,
(top right, red) $\mu_4=0.40$,
(bottom left, magenta) $\mu_4=0.55$,
(bottom right, blue, red, magenta, orange) $\mu_4=\{0.3334, 0.40, 0.55, 0.75\}$.
The cyan patches are the WC $\tilde{\Delta}_3$ where the eigenvalues
are arranged in decreasing order.
}\label{fig:N4:mu4:0p3334:0p40:0p55:0p75}
\end{figure*}

%In \Fig{fig:N4:mu4:0p3334} we plot  
In \Fig{fig:N4:mu4:0p3334:0p40:0p55:0p75}(top left) we plot
%=======================================================================
%================================================
% Try putting N=4 S^2 spheres in a figure*
% from locate_NumRecC_23Feb2022.nb in Downloads
%================================================
%=======================================================================
%%=========================================================
%% from locate_NumRecC_23Feb2022.nb in Downloads
%%=========================================================
%\begin{figure}[h]
%%\begin{tabular}{cc}
%\includegraphics[width=2.75in,height=2.75in]{fig_2-Spheres_N4_mu4_0p3334_15Mar2022_cropped}
%%\end{tabular}
%\caption{
%(blue) Diagonal density matrices $\vec{r}^{\,(N=4)}(r_4, \varphi_3, \varphi_2)$
% in the tetrahedral eigenvalue simplex $\Delta_3$ for $N=4$ for 
%$\mu_4~=~0.3334$. The cyan patch is the WC $\tilde{\Delta}_3$ where the eigenvalues
%are arranged in decreasing order.
%}\label{fig:N4:mu4:0p3334}
%\end{figure}
%%============================
 $\vec{p}_e^{\,(N=4)}=\big(\tfrac{1}{\sqrt{4}},\, \vec{r}^{\,(N=4)}(r_4, \varphi_3, \varphi_2)\big)$, 
 which is the diagonal eigenvalue vector $\vec{\lambda}_E^{\,(N=4)}= (\lambda_1, \lambda_2, \lambda_3, \lambda_4)$ transformed to the $e$-basis 
 (where the first component is always $\tfrac{1}{\sqrt{4}}$),
 and plot the three-vector 
 $\vec{r}^{\,(N=4)}~\equiv ~(r_4\,\cos\varphi_3,\, r_4\,\sin\varphi_3\cos\varphi_2,\, r_4\,\sin\varphi_3\sin\varphi_2)$ 
 for $\mu_4 = 0.3334\approx\tfrac{1}{3}$. This is the ``in-sphere," i.e. the largest $S^{2}$ that can be inscribed within the tetrahedron $\Delta_3$.
%
%In \Fig{fig:N4:mu4:0p40} we plot  
In \Fig{fig:N4:mu4:0p3334:0p40:0p55:0p75}(top right)
%%=========================================================
%% from locate_NumRecC_23Feb2022.nb in Downloads
%%=========================================================
%\begin{figure}[h]
%%\begin{tabular}{cc}
%\includegraphics[width=2.75in,height=2.75in]{fig_2-Spheres_N4_mu4_0p40_v3_15Mar2022_cropped}
%%\end{tabular}
%\caption{(red) Same as \Fig{fig:N4:mu4:0p3334} but now with $\mu_4~=~0.40$.
%The cyan patch is the WC $\tilde{\Delta}_3$ where the eigenvalues
%are arranged in decreasing order.
%}\label{fig:N4:mu4:0p40}
%\end{figure}
%%============================
the ``clipping" of the sphere $S^2$ is due its intersection with the triangular face of the 
eigenvalue simplex tetrahedron $\Delta_3$, since $\mu_4=0.40>\mu_4^* = \tfrac{1}{N-1} = \tfrac{1}{3}$ .
This clipping is even more pronounced in  \Fig{fig:N4:mu4:0p3334:0p40:0p55:0p75}(bottom left)
%\Fig{fig:N4:mu4:0p55}
%%=========================================================
%% from locate_NumRecC_23Feb2022.nb in Downloads
%%=========================================================
%\begin{figure}[h]
%%\begin{tabular}{cc}
%\includegraphics[width=2.75in,height=2.75in]{fig_2-Spheres_N4_mu4_0p55_15Mar2022_cropped}
%%\end{tabular}
%\caption{(magenta) Same as \Fig{fig:N4:mu4:0p3334} but now with $\mu_4~=~0.55$.
%The cyan patch is the WC $\tilde{\Delta}_3$ where the eigenvalues
%are arranged in decreasing order.
%}\label{fig:N4:mu4:0p55}
%\end{figure}
%%============================
and occurs more closely to the vertices of $\Delta_3$.
Finally, in \Fig{fig:N4:mu4:0p3334:0p40:0p55:0p75}(bottom right)
%\Fig{fig:N4:mu4:0p3334:0p40:0p55:0p75}
%%=========================================================
%% from locate_NumRecC_23Feb2022.nb in Downloads
%%=========================================================
%\begin{figure}[h]
%%\begin{tabular}{cc}
%\includegraphics[width=2.75in,height=2.75in]{fig_2-Spheres_N4_mu4_0p3334_0p40_0p55_0p75_15Mar2022_cropped}
%%\end{tabular}
%\caption{ Same as \Fig{fig:N4:mu4:0p3334} but now with 
%(blue, red, magenta, orange) $\mu_4=\{0.3334, 0.40, 0.55, 0.75\}$.
%The cyan patches are the WC $\tilde{\Delta}_3$ where the eigenvalues
%are arranged in decreasing order.
%}\label{fig:N4:mu4:0p3334:0p40:0p55:0p75}
%\end{figure}
%%============================
shows the spheres and clipped-spheres for $\mu_4=\{0.3334, 0.40, 0.55, 0.75\}$.
%
%These images were created by uniformly sampling from the marginal CDFs in the following ``top-down" sense.
To create these images by uniformly sampling from the marginal CDFs, we work ``top-down" in the following sense.

%\flushleft{\underline{Procedure for uniformly sample CDFs at fixed $\mu_{N=4}$:}}
\medskip
\underline{Procedure for uniformly sample CDFs at fixed $\mu_{N=4}$:}
%\begin{quote}
%\hspace{-0.25in}
\begin{itemize}
%\item[{}]
%\item[{}]  \underline{Procedure for uniformly sample CDFs at fixed $\mu_4$:}
 \item[(a)] Given a chosen purity $\tfrac{1}{4}\le\mu_4\le 1 \Leftrightarrow 0\le r_4 \le \sqrt{1-1/4}=\tfrac{\sqrt{3}}{2}$, 
\item[(b)] Given an $r_4$, pick an $X_3=\cos\varphi_3$ such that 
               $\tfrac{1}{3}\le X_3 \le \Xbarmax_3(r_4) 
   = \Min\left[\Xbar_3\equiv\tfrac{1}{2\sqrt{3}\,r_4}, \,1\right]$, 
   \newline \trm{where}\; $\Xbar_3=1 \;\trm{at}\; r_4=\tfrac{1}{\sqrt{4\,3}} = 
   \tfrac{1}{2\sqrt{3}} \leftrightarrow \mu_4 = \tfrac{1}{3}$,
 %,  
 %
 \item[(c)] Given an $X_3$, pick an $X_2=\cos\varphi_2$ such that
               $\tfrac{1}{2}\le X_2 \le \Xbarmax_2(X_3) 
             = \Min\left[\Xbar_2\equiv\frac{\sqrt{2} X_3}{\sqrt{1-X^2_3}}, \,1\right]$ 
             %\newline 
              \trm{where}\; $\Xbar_2~=~1 \;\trm{at}\; X_2=\tfrac{1}{\sqrt{3}}\equiv \Xstar_2$,
              \newline or
              $\Min\left[\varphibar_2\equiv\cos^{-1}\left(\Xbar_2\equiv\frac{\sqrt{2} X_3}{\sqrt{1-X^2_3}}\right),\,1\right]\le \varphibarmin_2\le \cos^{-1}(\tfrac{1}{2})~=~\tfrac{\pi}{3}$.
\end{itemize}
%\end{quote}
%
Note that the value of the chosen $r_4$ determines the ``break points" for $\Xbarmax_3(r_4)$ given by
the value at 
$\Xbar_3\equiv\tfrac{1}{2\sqrt{3}\,r_4} = \{\tfrac{1}{3},\tfrac{1}{\sqrt{3}},1,\infty\}$ 
at 
$ r_4=\{\tfrac{\sqrt{3}}{2}, \tfrac{1}{2}, \tfrac{1}{2\,\sqrt{3}},0\}\leftrightarrow $
$\mu_4  = \{1,\tfrac{1}{2},\tfrac{1}{3}, \tfrac{1}{4}\}$.
%= \tfrac{1}{N-i}|_{i\in\{0:N\}}       
We will see that later on, that for a given $N$ 
the previous determined break points for   $N'<N$ still apply (i.e. hold).  

\begin{figure}[!ht]
%\begin{tabular}{cc}
\includegraphics[width=2.75in,height=1.75in]{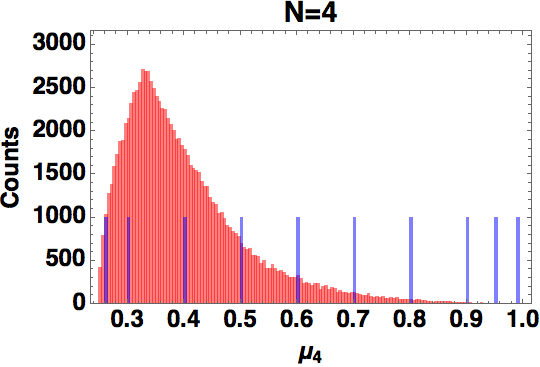}
%\includegraphics[width=2.75in,height=1.75in]{fig_histogram_of_mu4s_16Mar2022}
%\end{tabular}
\caption{
Histogram of purity values for $N=4$ generated by
(blue vertical bars) uniform sampling of diagonal density matrices of fixed purity ($10^3$) 
for $\mu_4=\{0.26, 0.30, 0.40, 0.50,$ $0.60, 0.70, 0.80, 0.90, 0.95, 0.99\}$,
%($10^3$ samples for each fixed value of the purity),
(red) uniform sampling ($10^5$ total samples, bin width $0.005$) of random unitary matrices $U', U$ such that 
$\rho_{diag}$ is the absolute square of a random row of  $U'$, and
$\rho=U\,\rho_{diag}\,U^\dag$ with $\mu_4 = \Tr[\rho^2]$.
}\label{fig:histogram:N4:mu4s}
\end{figure}

In \Fig{fig:histogram:N4:mu4s} we plot
%=========================================================
% from locate_NumRecC_23Feb2022.nb in Downloads
%=========================================================
%============================
a histogram of the purity for $N=4$ generated by two methods: 
(i) our uniform sampling of diagonal density matrices $\rho_{diag}$ at fixed purity $\mu_4$, and
(ii) the conventional method of uniformly sampling unitary matrices according to the Haar measure.
In (ii) $\rho_{diag}$ is the absolute square of a random row of  $U'$, and
$\rho=U\,\rho_{diag}\,U^\dag$ with $\mu_4 = \Tr[\rho^2]$.
In (i) we compute $10^3$  samples per fixed value of the purity $\mu_4$, and obtained $10^3$ identical purity values per chosen $\mu_4$ (see thin vertical bars).
In (ii) we computed a total of $10^5$ uniformly sampled unitary matrices, and used a bin width of $0.005$ for the histogram of purity values (red). We note that while the unitary matrices are sampled uniformly in (ii), the purity is sampled non-uniformly, being weighted towards lower values of the purity, especially in the Region 1 and 2, $\mu_4\in[\tfrac{1}{4},\tfrac{1}{2}]$.
As indicated in Table I, we expect on the order of $7$ samples per $10^5$ to be in the purity range $\mu_4\in[0.95, 1]$, which is borne out in the histogram in \Fig{fig:histogram:N4:mu4s}.

For general $N$ with variables $(X_2,\, X_3,\, \ldots, X_{N-1},\, r_N)$,
the bounds for the angles $X_k=\cos\varphi_k\, (X_2=\varphi_2)$ were outlined in 
\Eq{range:of:xs:N:1}-\Eq{range:of:xs:N:3},  
\Eq{X_k:constraint:1}-\Eq{X_k:constraint:2}, and
\Eq{XN:constraint:line1}-\Eq{XN:constraint:line2}. The procedure for uniformly sampling 
diagonal density matrices for fixed $\mu_N$ follows similarly to the case of $N=4$ above.
Given a fixed value of the purity $\mu_N$, one uniformly samples $F_{N-1}^{(N)}(X_{N-1};\, r_N)$
for the highest angle $X_{N-1}$ whose $r_4$-dependent bound is given by \Eq{XN:constraint:line1}.
For the lowest and middle angles $X_k$ for $k\in\{N-2,\dots,2\}$ one then successively uniformly samples, 
in descending $k$-order,
$F_{k}^{(N)}(X_{k};\, X_{k+1})$ whose $X_{k+1}$-dependent bound is given in \Eq{X_k:constraint:2}.
Once the diagonal density matrix $\rho_N^{(diag)}$ of fixed purity $\mu_N$ is uniformly sampled by the above procedure,
the general density $\rho_N$ can be obtained, as usual, from a uniformly (Haar) sampled unitary matrix $U$ via
$\rho_N = U\, \rho_N^{(diag)}\, U^\dag$, with identical eigenvalue spectrum and purity.
%================================================

%=======================================================================
\section{Applications}\label{sec:Applications}
\subsection{Entanglement in $\boldsymbol{N=4}$}\label{subsec:Ent:N4}
In this section we apply the $N=4$ analytic expressions for the CDFs $F_{k=\{2,3\}}^{(N=4)}$ to investigate well known measures of entanglement of 2-qubit systems, but now as functions of fixed values of the purity $\mu_4(\rho_{ab})$ of the composite systems, as well of the purity $\mu_2(\rho_{a})$ of a subsystem. As the canonical entanglement measure for 2-qubits we use the concurrence, $\mathcal{C}$ \cite{Wootters:1998,Wootters:2001} as well as the (more generalizable) logarithmic negativity ($LN$) \cite{Peres:1996, Horodecki:1997, Agarwal:2013}, and the quantum discord (QD) \cite{Zurek:2002,Luo:2008, Rulli_Sarandy:2011}. We will also explore the classical correlations $\mathcal{J}$, defined in relation to the quantum discord $\mathcal{Q}$, as the alternative definition of the mutual information between subsystems $A$ and $B$ conditioned on measurements on subsystem $B$ 
\cite{Henderson_Vedral:2001, Vedral:2003, Luo:2008, QDNote}, 
%=========================================================
% from locate_NumRecC_23Feb2022.nb in Downloads
%=========================================================
\begin{figure}[ht]
%\begin{tabular}{cc}
\hspace{-0.65in}
\includegraphics[width=4in,height=2.25in]{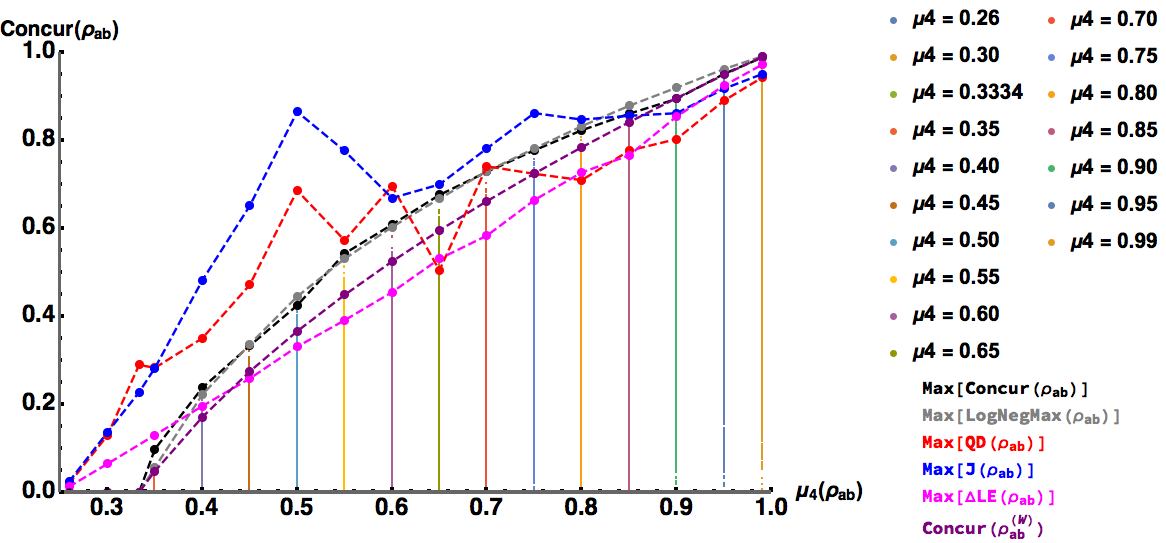}
%\includegraphics[width=4in,height=2.25in]{ConcurrLogNegDeltaLERaw_2500pts_QDJ_500pts_ConcurWerner2Qubit_vs_mu4_20Mar2022}
%\end{tabular}
\caption{
Plot of various entanglement measures (points with dashed curves) for 2-qubits $(N=4)$ vs $\mu_4$:
(black) $\Max[\mathcal{C}(\rho_{ab})]$ (concurrence),
(gray) $\Max[LN(\rho_{ab})]$ (logarithmic negativity),
(red) $\Max[\mathcal{Q}(\rho_{ab})]$ (quantum discord),
(blue) $\Max[\mathcal{J}(\rho_{ab})]$ (classical correlations),
(magenta) $\Max[\Delta LE(\rho_{ab})]$ ($\Delta$ Linear Entropy), and
(purple) $\mathcal{C}(\rho^{(Werner)}_{ab})$, (concurrence for 2-qubit Werner state).
The vertical points at fixed
$\mu_4=\{0.26, 0.30, 0.3334, 0.35, 0.40, 0.45, 0.50, 0.55, 0.60, 0.65, 0.70,$ $0.75, 0.80, 0.85, 0.90, 0.95, 0.99\}$
are the concurrences of $2500$ randomly sampled 2-qubit density matrices at fixed purity.
}\label{fig:EntMeasures:All:N4:mu4s}
\end{figure}
%============================
as well as a ``baseline" entanglement witness ($\Delta LE$) based on the difference in purity between the composite system $ab$ and its subsystem $a$, attempting to generalize the linear entropy (LE). %from pure to mixed states.

In \Fig{fig:EntMeasures:All:N4:mu4s} we plot the 
%measure of entanglement 
functions of the density matrix discussed above
for fixed discrete values of the purity 
$\mu_4~=~\{0.26, 0.30, 
0.3334, 0.35, 0.40, 0.45, 0.50, 0.55, 0.60,$ $0.65, 0.70, 0.75, 0.80,$
$0.85, 0.90, 0.95, 0.99\}$ (colored vertical ``lines" of points).
As is well-known \cite{Zyczkowski:1998, Zyczkowski_2ndEd:2020}, we see that separability occurs in the
region $\mu_4^{(1)}\in[\tfrac{1}{4},\tfrac{1}{3}]$ where the concurrence $\mathcal{C}$ goes to zero. 
At each fixed values of $\mu_4$, $2500$ composite density matrices $\rho_{ab}$ were generated and their concurrences computed, and values plotted vertically. The black-dashed curved represents 
$\underset{\{\rho_{ab}\}}{\trm{max}}[\mathcal{C}(\rho_{ab}|_{\mu_4})]$, that is, the maximum concurrence over the $2500$ $\rho_{ab}$ sampled at the same fixed value of the composite purity $\mu_4$.
The dashed-gray curve in  \Fig{fig:EntMeasures:All:N4:mu4s}
represents $\underset{\{\rho_{ab}\}}{\trm{max}}[LN(\rho_{ab}|_{\mu_4})]$, the corresponding maximum, now using the logarithmic negativity $LN$ \cite{LNnote} and is essentially the same as that for the concurrence, although the latter appears statistically slightly larger than the former at purities above $\mu_4\gtrsim 0.75$.

The dashed-purple curve in \Fig{fig:EntMeasures:All:N4:mu4s} represents the concurrence for the 2-qubit Werner state 
$\rho^{(W)}_{ab} = p\,\ket{\Phi^{(+)}}\bra{\Phi^{(+)}} + \tfrac{(1-p)}{4} \mathbb{I}$, where 
$\ket{\Phi^{(+)}} = \tfrac{1}{\sqrt{2}}\left(\ket{00}_{ab} + \ket{11}_{ab}\right)$ is the maximally entangled (symmetric) Bell state, and $\tfrac{1}{4}\,\mathbb{I}$ is the 2-qubit maximally mixed state (MMS).
A straightforward calculation of $\mu_N=\Tr[(\rho^{(W,N)}_{ab})^2]$ 
for a Werner state with a $d$-dimensional Bell state 
$\ket{\Phi_N^{(+)}} = \tfrac{1}{\sqrt{d}}\sum_{n=0}^{d-1} \ket{n, n}_{ab}$ (see \App{app:Werner:state})
reveals that
$p(\mu_N)=\sqrt{\tfrac{d^2\,\mu_N-1}{d^2-1}}$ with a sudden death of entanglement (as measured by the $LN$) at 
a critical value of $p_N^* = \tfrac{1}{d+1}$ corresponding to $\mu_N^* = \tfrac{2}{d\,(d+1)}$, as is borne out in
\Fig{fig:EntMeasures:All:N4:mu4s} for the case of $d=2$. 
%It is interesting that while 
Note that while 
$\mathcal{C}(\rho^{(W)}_{ab})$ qualitatively follows the same curve as 
$\underset{\{\rho_{ab}\}}{\trm{max}}[\mathcal{C}(\rho_{ab}|_{\mu_4})]$, the former is strictly less than the latter, except at the endpoints $\mu_4=1/4$ and $\mu_4=1$.

The red-dashed and blue-dashed curves in 
\Fig{fig:EntMeasures:All:N4:mu4s} plot 
$\underset{\{\rho_{ab}\}}{\trm{max}}[\mathcal{Q}(\rho_{ab}|_{\mu_4})]$
and
$\underset{\{\rho_{ab}\}}{\trm{max}}[\mathcal{J}(\rho_{ab}|_{\mu_4})]$, respectively.
The jaggedness in the curves are due to the finite sampling, now at $500$ density matrices per fixed
$\mu_4$ values (arising from the additional computational complexity due to computing the supremum over all measurements when calculating $\mathcal{Q}$ and $\mathcal{J}$ \cite{QDNote}). However, they do show instances where $\mathcal{Q}>\mathcal{J}$, as noted by Luo \cite{Luo:2008} in the special case of the 2-qubit Werner state, but now for random $\rho_{ab}$. As is well known,  $\mathcal{Q}>0$ (and  $\mathcal{J}>0$) for all  $\mu_4\in(\tfrac{1}{4},1]$ and therefore is non-zero in the no-entanglement  region $\mu_4\in[\tfrac{1}{4},\tfrac{1}{3}]$.
Instead, $\mathcal{Q}>0$ measures the quantum information-disturbance such that a classical state is defined by the fact that any projective measurement on $\rho_{ab}$ leaves it in the state before the measurement.

Finally, the dashed-magenta curve is a plot of a ``baseline" entanglement witness
$\underset{\{\rho_{ab}\}}{\trm{max}}[\Delta LE(\rho_{ab}|_{\mu_4})]$ for $N=4$,
where we have defined
for arbitrary dimension $N$
\bea{Delta:LE}
\Delta LE &\overset{\trm{def}}{=}& 
\frac{
\left( \mu(\rho^{(N)}_{ab}) -\tfrac{1}{N} \right) - 
\left( \mu(\rho^{(N_a)}_{a}) -\tfrac{1}{N_a} \right)
}
{
\left(1-\tfrac{1}{N}\right)
},
\no
\trm{where}\; N_a&=&\trm{dim}(\rho_a),\quad  \rho_a = \Tr_b[\rho_{ab}].
\eea
The reasoning behind \Eq{Delta:LE} is that if the purity of the subsystem $\rho_a$ is less than that of the composite state $\rho_{ab}$, i.e. $\mu(\rho^{(N_a)}_{a}) < \mu(\rho^{(N)}_{ab})$,
then the system is more mixed, and hence must contain entanglement.
However, since $\mu(\rho^{(N)}_{ab})\in[\tfrac{1}{N},1]$ and
$\mu(\rho^{(N_a)}_{a})\in[\tfrac{1}{N_a},1]$ we should measure these purities  relative to their respective MMS.
The denominator in \Eq{Delta:LE} simply normalizes $\Delta LE$ to unity for a maximally entangled state with
$\mu(\rho^{(N)}_{ab})=1$ and correspondingly $\mu(\rho^{(N_a)}_{a})=\tfrac{1}{N_a}$. We also have 
$\Delta LE(\rho_{ab})=0$ when $\rho_{ab} = \mathbb{I}/N$ is the $N$-dimensional MMS.
The dashed-magenta curve for $\underset{\{\rho_{ab}\}}{\trm{max}}[\Delta LE(\rho_{ab}|_{\mu_4})]$  
in \Fig{fig:EntMeasures:All:N4:mu4s} is a linear interpolation \cite{DeltaLE:note} of the purity
$\mu_4$ in the range $[\tfrac{1}{4},1]$  and acts as a lower bound to 
$\underset{\{\rho_{ab}\}}{\trm{max}}[\mathcal{C}(\rho_{ab}|_{\mu_4})]$ in the approximate region
$\mu_4 \approx [0.37,1]$, where the intersection of the two curves occurs at $\mu_4\approx 0.37>\tfrac{1}{3}$.
$\underset{\{\rho_{ab}\}}{\trm{Max}}[\Delta LE(\rho_{ab}|_{\mu_4})]$ is non-zero in the complement region
$\mu_4 \approx [\tfrac{1}{4}, 0.37]$, and in a sense behaves more like the quantum discord 
$\underset{\{\rho_{ab}\}}{\trm{max}}[\mathcal{Q}(\rho_{ab}|_{\mu_4})]$ in this region (although the two measure different things). 
Another way to view 
$\underset{\{\rho_{ab}\}}{\trm{max}}[\Delta LE(\rho_{ab}|_{\mu_4})]$ is that for all 
$\mu_4 \in [\tfrac{1}{4},1]$, it acts as a witness (proper lower bound) to both  
$\underset{\{\rho_{ab}\}}{\trm{max}}[\mathcal{Q}(\rho_{ab}|_{\mu_4})]$
and $\underset{\{\rho_{ab}\}}{\trm{max}}[\mathcal{J}(\rho_{ab}|_{\mu_4})]$.

Of course, the results of \Fig{fig:EntMeasures:All:N4:mu4s} could have also been obtained from the conventional method of  density matrix generation via sampling random unitaries with respect to the Haar measure.  
However, considering the distribution of purities $\mu_4$ given in the histogram in \Fig{fig:histogram:N4:mu4s},
it would have been much more difficult and time consuming to obtain the corresponding 
results for higher purity values in \Fig{fig:EntMeasures:All:N4:mu4s}.

\begin{figure}[ht]
%\begin{tabular}{cc}
%\hspace{-0.625in}
%\hspace{-0.5in}
\includegraphics[width=3.5in,height=2.25in]{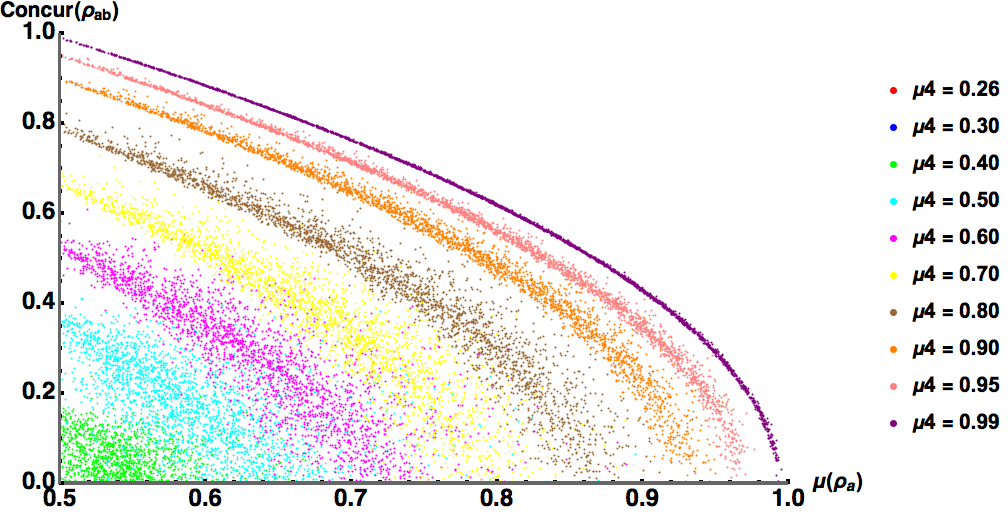}
%\includegraphics[width=3.5in,height=2.25in]{Concurrence_vs_mu2rhoa_2500pts_21Mar2022}
%\end{tabular}
\caption{
Concurrence $\mathcal{C}(\rho_{ab})$ vs purity $\mu_2(\rho_a)$ of the subsystem $a$ for 
fixed composite purity values (colored points)
$\mu_4(\rho_{ab})~=~\{0.26, 0.30, 0.40, 0.50, 0.60, 0.70, 0.80, 0.90, 0.95, 0.99\}$.
For each fixed value of the purity $\mu_4(\rho_{ab})$,  $2500$ density matrices (same colored points) were generated.
}\label{fig:Concurrence:vs:mu2:rhoa}
\end{figure}

Another informative approach to examining the above data is to plot the concurrence $\mathcal{C}(\rho_{ab})$ vs 
$\mu_2(\rho_a)$ the purity of the reduced density matrix of (say) subsystem $a$.
In \Fig{fig:Concurrence:vs:mu2:rhoa} we plot $2500$ generated density matrices for each fixed value  
$\mu_4(\rho_{ab})$ (different colored points) of the composite system $\rho_{ab}$ vs the purity 
$\mu_2(\rho_a)$ of the reduced density matrix $\rho_a$.
%=========================================================
% from locate_NumRecC_23Feb2022.nb in Downloads
%=========================================================
%============================
We see that as the subsystem purity $\mu_2(\rho_a)$ increases the concurrence approaches zero, indicating a separable composite system, while at low values of $\mu_2(\rho_a)$ the concurrence approaches its 
maximum $\mu_4(\rho_{ab})$-dependent value. As the purity $\mu_4(\rho_{ab})$ of the composite system increases toward unity, the (colored) bands of the $2500$ density matrices becomes narrower, and more pronounced. 
\medskip

\begin{figure*}[ht]
\begin{tabular}{c}
%\hspace{-0.5in}
\includegraphics[width=3.5in,height=2.25in]{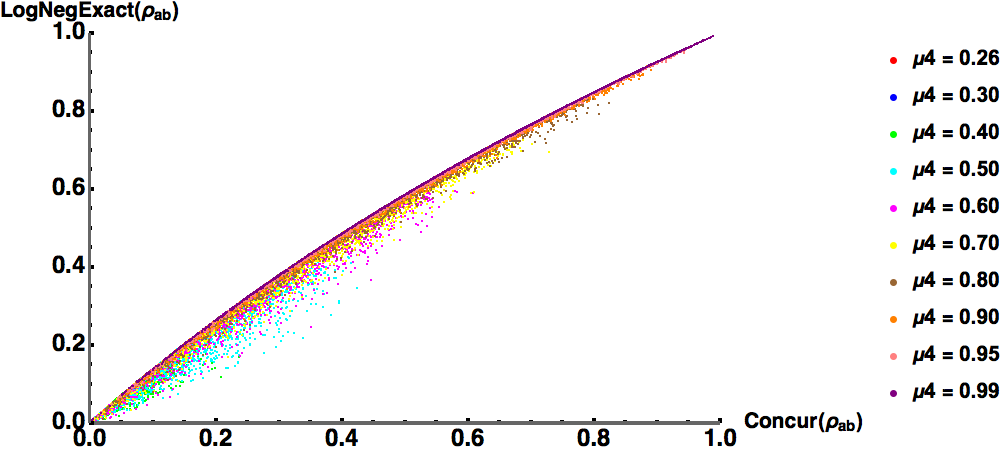} \\
{} \\
\includegraphics[width=3.5in,height=2.25in]{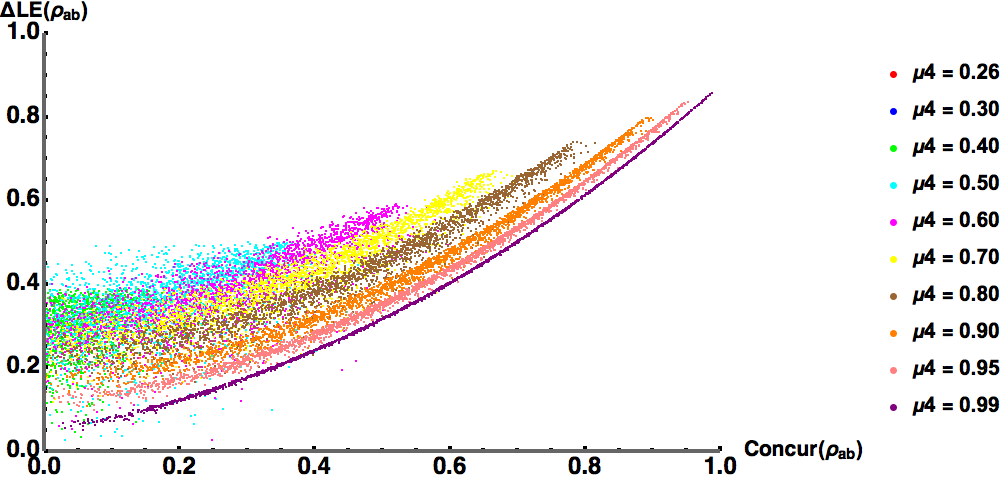} \\
{}\\
\includegraphics[width=3.5in,height=2.25in]{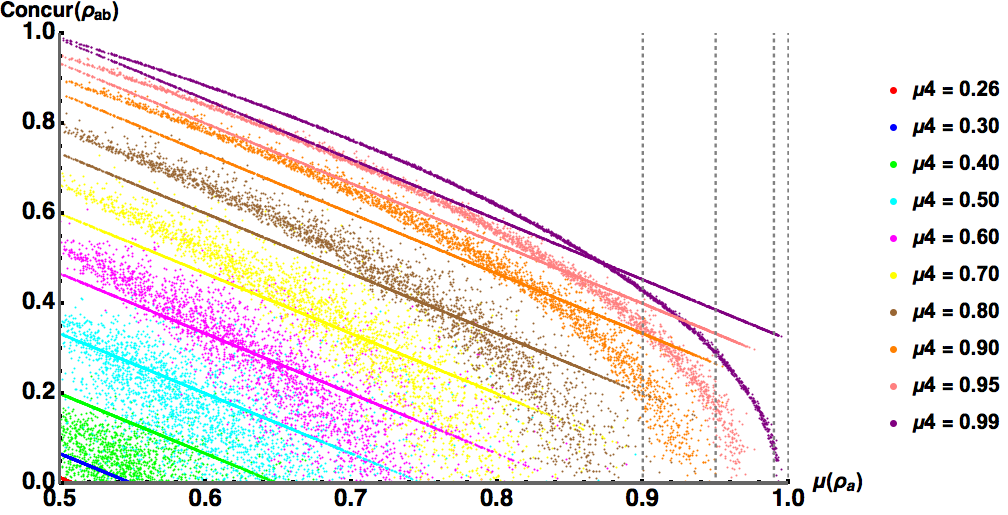} 
%
%\includegraphics[width=3.5in,height=2.25in]{LogNeg_vs_Concurrence_2500pts_21Mar2022} \\
%{} \\
%\includegraphics[width=3.5in,height=2.25in]{DeltaLE_vs_Concurrence_2500pts_21Mar2022} \\
%{}\\
%\includegraphics[width=3.5in,height=2.25in]{Concurrence_DeltaLE_vs_mu2rhoa_2500pts_21Mar2022} 
\end{tabular}
\caption{
(top) Logarithmic negativity $LN(\rho_{ab})$, and
(middle) $\Delta LE(\rho_{ab})$
vs the 
concurrence $\mathcal{C}(\rho_{ab})$. 
For each fixed value of the composite purity 
$\mu_4(\rho_{ab})=\{0.26, 0.30, 0.40, 0.50, 0.60, 0.70, 0.80,$ $ 0.90, 0.95, 0.99\}$,
$2500$ density matrices (same colored points)
were generated.
(bottom) Same as \Fig{fig:Concurrence:vs:mu2:rhoa} with  superimposed (line-like) scatter plots of $\Delta LE(\rho_{ab})$  vs $\mu_2(\rho_a)$ at the same colored fixed $\mu_4(\rho_{ab})$ values used to plot the concurrence.
The vertical dashed-gray lines are visual makers for $\mu_4 = (0.90, 0.95, 0.99, 1.0)$.
}\label{fig:LogNegvsConcurrence}
\end{figure*}

In \Fig{fig:LogNegvsConcurrence}(top) we plot (top) the logarithmic negativity $LN(\rho_{ab})$, and
(middle) $\Delta LE(\rho_{ab})$ \Eq{Delta:LE}
vs the concurrence $\mathcal{C}(\rho_{ab})$ for the same composite purity values $\mu_4(\rho_{ab})$
as in \Fig{fig:Concurrence:vs:mu2:rhoa} (again, $2500$ density matrices per fixed $\mu_4(\rho_{ab})$ (color) value).
The tight clustering of all points in \Fig{fig:LogNegvsConcurrence}(top) for all values of $\mu_4(\rho_{ab})$ indicates that the logarithmic negativity is a very close approximation to
a one-to-one function
 the concurrence (though it is well known that the former does not detect bound entanglement).
The much broader widths in \Fig{fig:LogNegvsConcurrence}(middle) indicates that the $\Delta LE(\rho_{ab})$ from \Eq{Delta:LE} is much less accurate approximation to the concurrence than $LN(\rho_{ab})$ (top), though still exhibiting some structural dependence on the purities.
%\medskip
%=========================================================
% from locate_NumRecC_23Feb2022.nb in Downloads
%=========================================================
%============================
%%% page 20 if invoking \clearpage\newpage
\clearpage
\newpage
%============================
%

\begin{figure}[ht]
%\begin{tabular}{c}
%\hspace{-0.65in}
%%% REDO THIS GRAPH! %%%
\includegraphics[width=3.5in,height=2.25in]{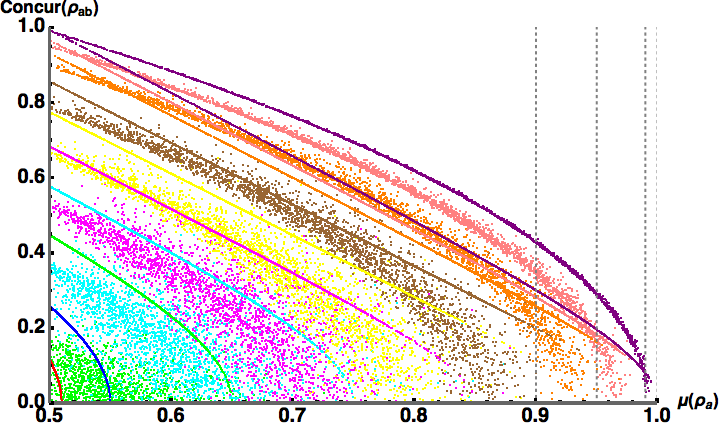} 
%\includegraphics[width=3.5in,height=2.25in]{DeltaLEModified_vs_Concurrence_2500pts_v2_21Mar2022} 
%\end{tabular}
\caption{
Same as \Fig{fig:LogNegvsConcurrence}(bottom) but now with
$\Delta LE'(\rho_{ab})$ as defined in \Eq{DeltaLE:modified}
vs the 
concurrence $\mathcal{C}(\rho_{ab})$. 
}\label{fig:DeltaLEModifiedvsConcurrence}
\end{figure}

Finally,  \Fig{fig:LogNegvsConcurrence}(bottom) is the same as  \Fig{fig:Concurrence:vs:mu2:rhoa}, but now with  superimposed (line-like) scatter plots of $\Delta LE(\rho_{ab})$  vs $\mu_2(\rho_a)$ at the same  fixed 
$\mu_4(\rho_{ab})$ (colored) values used to plot the concurrence $\mathcal{C}(\rho_{ab})$.
Each line-like plot of $\Delta LE(\rho_{ab})$ does a reasonably good job at passing through the same colored band corresponding to $\mathcal{C}(\rho_{ab})$ for the same value of $\mu_4(\rho_{ab})$. This suggests that something else is needed to attempt to introduce the necessary concavity to $\Delta LE(\rho_{ab})$ in order to ``bend" it more into the shape of the $\mathcal{C}(\rho_{ab})$. We require something that does not disturb the endpoint values, which suggests a new definition of a $\Delta LE'(\rho_{ab})$ that is also an explicit function of $\mu_2(\rho_{a})$.
In \Fig{fig:DeltaLEModifiedvsConcurrence} we plot the same figure as in  \Fig{fig:LogNegvsConcurrence}(bottom)
%=========================================================
% from locate_NumRecC_23Feb2022.nb in Downloads
%=========================================================
%============================
but now with 
\be{DeltaLE:modified}
\Delta LE'(\rho_{ab}) \overset{\trm{def}}{=} \sqrt{2\,\big(1-\mu_2(\rho_{a})\big)\,\Delta LE(\rho_{ab})},
\ee
where the factor $\sqrt{2\,\big(1-\mu_4(\rho_{ab})\big)}$ was chosen since it is $0$ at $\mu_2(\rho_{ab})=1$ 
and unity at $\mu_2(\rho_{ab})=\tfrac{1}{2}$, as an attempt to introduce the desired concavity.
While \Eq{DeltaLE:modified} is admittedly an ad hoc attempt, the slight concavity introduced 
in \Fig{fig:DeltaLEModifiedvsConcurrence}
indicates that it is a step in the right direction, suggesting other possibilities 
such as introducing other polynomial powers
$\left[ 2\,\big(1-\mu_4(\rho_{ab})\big) \right]^{1/q}$ with $q> 2$ (with appropriate scaling at the  
$\mu_2(\rho_{ab})$ endpoints).
This will be investigated further in future research.
%=======================================================================

%=======================================================================
\subsection{Induced Measures 
%from a pure state of composite dimension $\boldsymbol{N K}$ for$\boldsymbol{N=\{2,3,4\}}$
}\label{subsec:Induced:Measures}
%=======================================================================
With the formulas developed above, we now have a geometric representation (in the WC) of the eigenvalues  $\{\lambda_i\}$ in dimension $N$. Transforming back to the $E$-basis we have the following forms 
$\vec{\lambda}_N$  for the eigenvalues for $N=2$ and $N=3$
\bsub
\be{lambdavecs:N2:N3}
\hspace{-0.225in}
%N=2
\left(\begin{array}{c}\lambda_1 \\ \lambda_2\end{array}\right)_{N=2} = 
\left(\begin{array}{c}
\frac{1+\sqrt{2}\,r_2}{2} \\ 
\frac{1-\sqrt{2}\,r_2}{2}
\end{array}\right), \;
%
% N=3
\left(\begin{array}{c}\lambda_1 \\ \lambda_2\\ \lambda_3\end{array}\right)_{N=3} = 
\left(\begin{array}{c}
\frac{2+r_3\,\cos\varphi_2 + \sqrt{3}\,r_3\,\sin\varphi_2}{6} \\ 
\frac{2+r_3\,\cos\varphi_2 + \sqrt{3}\,r_3\,\sin\varphi_2}{6} \\
\frac{1-r_3\,\cos\varphi_2}{3}
\end{array}\right), 
\ee
and for $N=4$
\be{lambdavecs:N4}
\hspace{-0.2in}
% N=4
\left(\begin{array}{c}\lambda_1 \\ \lambda_2\\ \lambda_3\\ \lambda_4\end{array}\right)_{N=4} = 
\left(\begin{array}{c}
\frac{3+ 2 \sqrt{3}\, r_4\cos\varphi_3 + 2 \sqrt{6}\, r_4\sin\varphi_3\cos\varphi_2 + 
        6 \sqrt{2}\, r_4\sin\varphi_3\sin\varphi_2 }{12} \\ 
\frac{3+ 2 \sqrt{3}\, r_4\cos\varphi_3 + 2 \sqrt{6}\, r_4\sin\varphi_3\cos\varphi_2 - 
        6 \sqrt{2}\, r_4\sin\varphi_3\sin\varphi_2 }{12}  \\
\frac{3+ 2 \sqrt{3}\, r_4\cos\varphi_3 - 4 \sqrt{6}\, r_4\sin\varphi_3\cos\varphi_2 }{12} \\
\frac{1- 2 \sqrt{3} r_4\,\cos\varphi_3}{4}
\end{array}\right), 
\ee
\esub
each with unit trace. These formulas can be easily generated for arbitrary $N$.

\subsubsection{The probability distribution of the eigenvalues of $\rho_N$}
With the above formulas, one can then explore metrics and measures on the 
space of density matrices $\mathcal{M}^{(N)}$ of dimension $N$. 
%
%=============. from N4_diagonal_density_matrix_generation_18Feb2021.tex ==================
 For example
 (see sections 15.3 and 15.4, pp410-414 of \cite{Zyczkowski_2ndEd:2020} for further details),
  $P_{HS}(\lambda_1,\ldots,\lambda_N)$  is the probability distribution of eigenvalues of 
  $\rho=U\,\rho_d\,\,U^\dag$
 when \tit{averaged over all unitaries $U\in SU(N )$} with respect to the Haar measure (uniform measure on  sphere $S^{N^2-1}$),
where $\rho_d  \equiv \Lambda =\trm{diagonal}(\lambda_1,\ldots,\lambda_N) \leftrightarrow \vec{\lambda}_N$.
 This probability distribution of eigenvalues is given by
 \bsub
 \bea{GQS:15.35}
P_{HS}(\lambda_1,\ldots,\lambda_N) &=& C_N^{HS} 
\delta\left( 1 - \sum_{j=1}^N \lambda_j \right)\,
\prod_{j<k}^{N} (\lambda_j-\lambda_k)^2,\qquad\;  \label{GQS:15.35:line1} \\
 C_N^{HS}  &=& \dfrac{\G(N K)}{\prod_{k=1}^{N}\G(k)\,\G(k+1)}, \label{GQS:15.35:line2}\\
 \int_0^1\,d\lambda_N\,&\cdots&\int_0^1\,d\lambda_1\,P_{HS}(\lambda_1,\ldots,\lambda_N) = 1.
 \label{GQS:15.35:line3}
 \eea
 \esub
 Here, the delta function $\delta\left( 1 - \sum_{j=1}^N \lambda_j \right)$ enforces the unit trace condition on the ddm.
 The second term $\prod_{j<k}^{N} (\lambda_j-\lambda_k)^2$ arises from the introduction of the 
 unitary $U$ in $\rho=U\,\rho_d\,\,U^\dag$. The Hilbert-Schmidt (HS) metric is then defined by $ds^2_{HS} = \tfrac{1}{2}\,\Tr[(d\rho)^2]$, where the factor of $\tfrac{1}{2}$ is used to make closed geodesics on $\mathbb{C}P^N\sim S^{N^2-1}$ have length $\pi$.
 Writing $d\rho~=~U \left[ d\Lambda + U^\dag dU \Lambda - \Lambda U^\dag dU\right] U^\dag$ (employing $d(U\,dU^\dag)=0$), one has that in the spectral representation   
 $d\rho_{ij} = d\lambda_i\,\delta_{ij} + (\lambda_j-\lambda_i)\,(U^\dag\,dU)_{ij}$. 
 The first (diagonal) term is the classical-like Fisher-Rao metric (of the eigenvalue probability vector), while
 the second term introduces the off-diagonal differences of eigenvalues $ (\lambda_j-\lambda_i)$. 
 (Note:  writing $U=e^{-i\,H}$, one then has 
 $(U^\dag\,dU)_{ij} = -i\,H_{ij}$). Thus the HS metric becomes (taking ``one'' factor from  each squared term of $d\rho$)
 $ds^2_{HS} = \tfrac{1}{2} \sum_{i=1}^N (d\lambda_i)^2 + \sum_{i<j}^N (\lambda_i-\lambda_j)^2\,|(U^\dag\,dU)_{ij}|^2$.
 This leads to the HS volume element (taking into account a normalization factor respecting the delta function, i.e that $\lambda_N \overset{\trm{def}}{=} 1-\sum_{i=1}^{N-1} \lambda_i$) 
 \bsub
 \bea{GQS2:15.34}
 dV_{HS} &=& \sqrt{\dfrac{N}{2^{N-1}}}\,
 \prod_{j=1}^{N-1} d\lambda_j\,
 \prod_{j<k}^{1\cdots N} (\lambda_i-\lambda_j)^2\, \no
 &\times&  \left|
 \prod_{j<k}^{1\cdots N} \trm{Re}(U^\dag\,dU)_{jk} \, \trm{Im}(U^\dag\,dU)_{jk} 
 \right|, \label{GQS2:15.34:line1} \\
 &\equiv& P_{HS}(\vec{\lambda}) \times Q(\bold{F}^N). \label{GQS2:15.34:line2}
 \eea
 \esub
The above shows that $P(\rho)=P_{HS}(\vec{\lambda}) \times Q(\bold{F}^N)$ is a \tit{product metric}  
with $P_{HS}(\vec{\lambda})$ depending on the eigenvalues of $\rho$ acting as the \tit{radial} portion, and 
$Q(\bold{F}^N)$ depending on the eigenvectors of $\rho$ (i.e the column vectors of $U$) acting as the \tit{angular} portion, which one can roughly think of as $d\Omega_{\mathbb{C}^N=S^{N^2-1}}$. 
(Technically, $\bold{F}^N$ is  \tit{Flag Manifold} \cite{Zyczkowski_2ndEd:2020, Flag:Manifold:Note}).
Thus, $P_{HS}(\vec{\lambda})$ in \Eq{GQS:15.35} is just the radial portion of 
 $dV_{HS}$ in \Eq{GQS2:15.34}, depending only on the eigenvalues of $\rho$.
 In many cases one is \tit{not} interested in the ``angular" distribution of the eigenvalues $\vec{\lambda}$ and so one integrates over this angular portion. This is what $P_{HS}(\vec{\lambda})$ in \Eq{GQS:15.35}  represents, i.e. a random 
 $\rho_N=U\,\rho_d\,U^\dag$ averaged over all unitaries $U$ with respect to the Haar measure.
 
 \subsubsection{$\rho_N$ considered as the reduced density matrix of a randomly chosen pure state of composite dimension $N K$}
 From a fiber-bundle perspective, one can also think of $\mathcal{M}^{(N)}$ as the base space in which the 
 $\rho_N$ live. Above each $\rho_N$ (labeled by their eigenvalues $\vec{\lambda}_N$) is a fiber containing all the purifications of the density matrix, each of which maps down from bundle space to $\rho_N$ in the base space. These pure states $\ket{\Psi}_{N,K}$ lie in a composite Hilbert space
  $\H_{N,K} = \H_N \otimes \H_K$, where  $ \H_N$ is the \tit{system} subspace of dimension $N$, and  $\H_K$ is the \tit{ancilla (reservoir)} subspace of dimension $K$, which we will take as $K\ge N$. We can conveniently write  
$\ket{\Psi}_{N,K} = \sum_{n,k} A_{nk}\ket{n}_N\otimes\ket{k}_K$  for an $N\times K$ rectangular complex matrix $A$ with matrix elements
 % $A_{nk} = {}_N\bra{n}\,\otimes\,{}_K\bra{k} \Psi_{N,K}\rangle$,
 $A_{nk} = {}_{N,K}\bra{n,k}\Psi_{N,K}\rangle$,
with orthogonal bases $\ket{n}_N\in\H_{N}$ and $\ket{k}_K\in\H_{K}$, respectively.
The only restriction placed upon $A$ is that $||\Psi_{N,K} ||^2 = \Tr[A\,A^\dag]=1$.
The corresponding density matrix $\rho_{NK} =\ket{\Psi}_{N,K}\bra{\Psi_{N,K}}$ is
represented by a 4-indexed object $(\rho_{NK})_{n k, n' k'}$ which upon partial tracing over 
$\H_K$ yields the $N\times N$ reduced density matrix $\rho_N=\Tr_K[A\,A^\dag]$. One can then define a family of measures in the space of mixed states $\mathcal{M}^{(N)}$ labeled by the size $K$ of the reservoir.

Here we will consider the measures 
$P^{\trm{trace}}_{N,K}(\lambda_1,\ldots,\lambda_N)$ 
 induced by the above partial tracing over the reservoir. 
 Since we consider pure states $\ket{\Psi}_{N,K}$ that are drawn randomly (with respect to the Fubini-Study  metric), the induced measure $P^{\trm{trace}}_{N,K}$ also has the product form of \Eq{GQS2:15.34:line2}. 
 Hence, the distribution of eigenvectors is determined by the Haar measure on $U(N)$. 
 The radial distribution of eigenvalues can be derived and is given by 
(see section Eq.(15.59), p417 in \cite{Zyczkowski_2ndEd:2020})
\bsub
\bea{PNK:GQS:15.59}
P^{\trm{trace}}_{N,K\ge N}(\lambda_1,&\ldots&,\lambda_N) = C_N^{HS} 
\delta\left( 1 - \sum_{j=1}^N \lambda_j \right)\, \no
&\times&
\prod_{j<k}^{N} (\lambda_j-\lambda_k)^2\, 
\prod_{i}^{N} (\lambda_i)^{K-N}, \label{PNK:GQS:15.59:line1} \\
 C_{N,K}  &=& \dfrac{\G(N K)}{\prod_{j=0}^{N-1}\G(K-j)\,\G(K-j+1)}, \label{PNK:GQS:15.59:line2} \\
 \int_0^1\,d\lambda_N\,&\cdots&\int_0^1\,d\lambda_1\,P^{\trm{trace}}_{N,K}(\lambda_1,\ldots,\lambda_N) = 1. \qquad \label{PNK:GQS:15.59:line2} 
 \eea
 \esub
 The net effect of tracing over the reservoir system of dimension $K$ is to modify the 
 previous metric $P_{HS}$ \Eq{GQS:15.35} with the
 extra product terms $\prod_{i}^{N} (\lambda_i)^{K-N}$ (here written for $K\ge N$) in \Eq{PNK:GQS:15.59}.
 Note that for $K\to N \Rightarrow P^{\trm{trace}}_{N, N} = P_{HS}$.
 In other words,  for a composite bipartite system composed of two subsystems of equal dimension $N$, tracing out over one of the systems produces the distribution of eigenvalues given by $P_{HS}$.

While in the past many authors  have been been interested in 
 $P^{\trm{trace}}_{N,K}(\lambda_1,\ldots,\lambda_N)$ and have used it to derive the average purity, entropy, etc... (see Chapter 15 of \cite{Zyczkowski_2ndEd:2020} and references therein) of systems, a new feature of this present work is that we are able to produce \tit{exact analytic expressions} for the joint probability distribution $P^{\trm{trace}}_{NK}$ in terms of the geometric coordinates $(\varphi_2, \varphi_3,\dots,\varphi_N, r_N)$ for an arbitrary dimension $N$ by using the expressions $\vec{\lambda}_N$ as in \Eq{lambdavecs:N2:N3}, \Eq{lambdavecs:N4}, and their (easy to produce) generalizations. By integrating over the angular variables, given their boundary regions of the previous sections, we are able to produce marginal probability and cumulative distribution functions in terms of the radius $r_N$ or purity $\mu_N$. In some cases, as we shall show below, we can actually perform the integrations analytically and obtain 
 closed-formed expressions.
In general, these analytic equations quickly become unwieldy for general $N, K$, and one must resort to numerical integration. 
% 
% for the probability and cumulative probability distributions \tit{as a function of the purity} (i.e. \tit{not} just producing formulas for general $N,K$ for averages - which is also of strong interest). While in principle we can derive such analytic formulas in general, producing the probability and cumulative distribution involves integration over sub-purities until only the highest purity $\mu_N$. 
% 
% Because such formulas become \tit{piecewise} in each \tit{Region-k} $\tfrac{1}{k}\le\mu_N\le\tfrac{1}{k-1}$ for $k\in\{N,N-1,\ldots,2\}$, it becomes increasingly cumbersome to derive such formulas for \tit{general}  $N,K$. 

 Thus, in the following we will show formulas and plots for the case of $N=2, K=2^{k\in\{1,2,3,4\}}$, though we generate a formula for arbitrary $K$. This system is a qubit in a  reservoir of (1,2,3,4) qubits.
 The case $N=2$ is not surprising, since the independent eigenvalue $\lambda_1$ (with $\lambda_2 \equiv 1-\lambda_1)$ of a qubit ($N=2$) is completely characterized by its purity $\mu_2$, i.e $\lambda_1 = \tfrac{1}{2} \left( 1+ \sqrt{\mu_2-1/2}\right)$. 
 The first non-trivial case occurs for $N=3$ which depends on the coordinates $(r_3,\varphi_2)$ and thus we investigate the explicit case 
 $N=3, K=\{3,6,9\}$. Here the system is a qutrit, and the reservoir consists of  1-qutrit, a qubit-qutrit, and 2-qutrits, respectively.
 Lastly, we investigate the case of $N=4, K=4$ in which the system is a ``ququad" 
 ($\mathbb{C}^4$, of which a system of 2 qubits is a subset), 
and the reservoir is of the same size, $K=4$. 
%(this could also be considered as pure state of 4 qubits for which we trace out two of them). 
 Here we integrate (the non-trivial task) out $(X_3=\cos\varphi_3,\varphi_2)$ 
 and produce probability distributions and cumulative probability distributions as a function of 
 $\mu_4$ ($r_4$) alone.

%=======================================================================
\subsubsection{$\boldmath{N=2}$}\label{subsubsec:N2}
\begin{figure}[ht]
\begin{tabular}{c}
%\hspace{-4em}
%\includegraphics[width=6.0in,height=2.25in]{PN2K2_4_8_16_8Mar2021} \\
%\includegraphics[width=6.0in,height=2.25in]{FPN2K2_4_8_16_8Mar2021} 
\hspace{-0.5in}
\includegraphics[width=3.25in,height=1.75in]{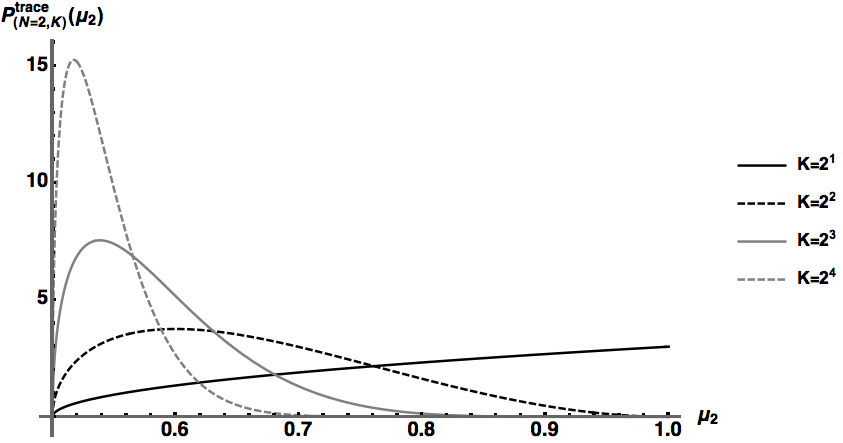} \\
\hspace{-0.5in}
\includegraphics[width=3.25in,height=1.75in]{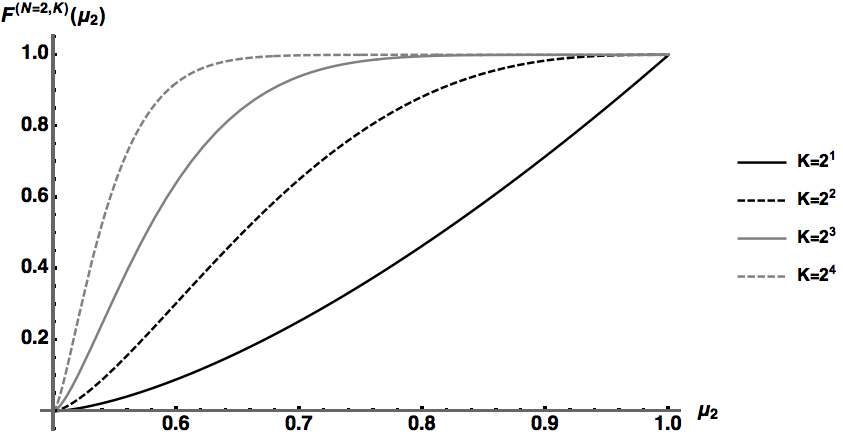} 
%
%\hspace{-0.5in}
%\includegraphics[width=3.25in,height=1.75in]{fig_PtraceN2K2_4_8_16_mu2_2Apr2022} \\
%\hspace{-0.5in}
%\includegraphics[width=3.25in,height=1.75in]{fig_CDFtraceN2K2_4_8_16_mu2_2Apr2022} 
\end{tabular}
\caption{(top)  Probability $P^{\trm{trace}}_{N=2,K} (\mu_2)$,
(bottom) Cumulative probability distribution function $F^{\trm{trace}}_{N=2,K} (\mu_2)$
for $N=2$, and $K=2^{k\in\{1,2,3,4\}}$.
}\label{fig:PN2K}
\end{figure}

%=======================================================================
The case of $N=2$ is straightforward since any $2\times 2$ matrix is completely determined by two invariants, its trace and determinant. Comparing 
$ \pNvecE{2}(x_2)$ from \Eq{N2:param} with $\vec{\lambda}^{(N=2)} = (\lambda_1,\lambda_2=1-\lambda_1)^T$ in either the E-basis or e-basis, on finds that 
$\lambda_1 = (1+x_2)/2 =  \left(1+\sqrt{2\,\mu_2-1}\right)/2$. Using \Eq{PNK:GQS:15.59} for the case of $N=2$ and carrying out the Jacobian transformations
$J_{\lambda,  x}\,J_{x, \mu} = \det\left(\tfrac{d\lambda_1}{dx_2}\right)\, \det\left(\tfrac{d x_2}{d\mu_2}\right)$ to convert
$d\lambda_1 = J_{\lambda, x}\,dx_2 = J_{\lambda, x}\,J_{x, \mu}\,d\mu_2$ one derives for arbitrary $K\ge N$
%=================
\bsub
\bea{PN2K}
\hspace{-0.4in}
P^{\trm{trace}}_{N=2,K} (\mu_2) &=& \dfrac{(1-\mu_2)^{K-2}\,(2\mu_2-1)^{1/2}\,\G(2 K)}{2^{K-1}\,\G(K)\,\G(K-1)}, \\
\hspace{-0.4in}
F^{\trm{trace}}_{N=2,K} (\mu_2) &=& \int_{1/2}^{\mu_2}\,d\mu'_2\,P^{\trm{trace}}_{N=2,K} (\mu'_2), \no 
\hspace{-0.4in}
&=& 
\dfrac{
 (2\,\mu_2-1)^{3/2}\,{}_{2}F_1(\tfrac{3}{2},2-K, \tfrac{5}{2},2\mu_2-1)\,\G(2 K)
}
{
3\;2^{2 K-3}\,\G(K)\,\G(K-1)
},\qquad\;\;
\eea
\esub
%=================
where ${}_{2}F_1(a,b; c; z)=\sum_{k=0}^\infty \tfrac{(a)_k\,(b)_k}{(c)_k}\,\tfrac{z^k}{k!}$ is the hypergeometric function,
with Pochhammer symbol $(a)_k \equiv a\,(a+1)\,\ldots\,(a+k-1) = \tfrac{\G(a+k)}{\G(a)}$.
These are plotted in \Fig{fig:PN2K} for $K=2^{k\in\{1,2,3,4\}}$
%======================================
% from locate_NumRecC_F6N6mu6_24Mar2022
% compute on 2Apr2022
%=======================================
%============================
The general trend of these plots is that as the dimension $K$ of the reservoir system  increases, the probability distribution $P^{\trm{trace}}_{N=2,K} (\mu_2)$
peaks further and further to the left, approaching $\mu_2\to1/2$, i.e. the MMS (where it approaches an almost delta function-like behavior). This is also reflected in the cumulative probability distribution $F^{\trm{trace}}_{N=2,K} (\mu_2)$ which saturates to unity for smaller values of $\mu_2$ as $K$ increases.
%============================
\begin{figure}[ht]
\begin{tabular}{c}
%\hspace{-4em}
%\includegraphics[width=3.5in,height=2.25in]{PS2N2K2_4_8_16_8Mar2021} \\
%\includegraphics[width=3.5in,height=2.25in]{PS2N2K2_4_8_16_32_64_8Mar2021} 
\includegraphics[width=3.25in,height=1.75in]{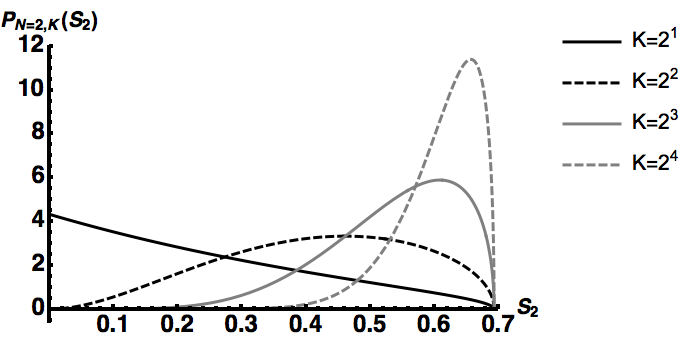} \\
\includegraphics[width=3.25in,height=1.75in]{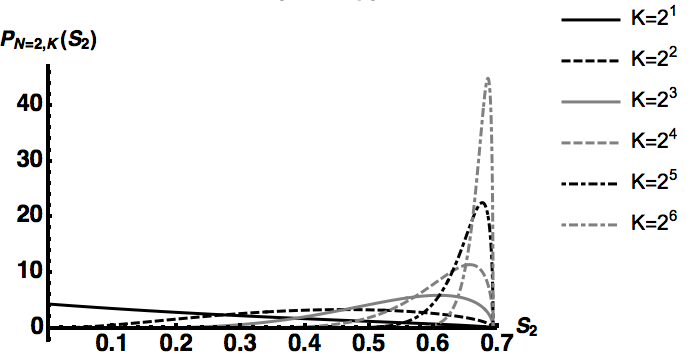} 
\end{tabular}
\caption{ Probability $P^{\trm{trace}}_{N=2,K} (S^{(N)}_2)$,
as a function of the R\'{e}nyi entropy $S^{(N)}_2 = -\ln(\mu_N)$
for $N=2$, and 
(top) $K=2^{k\in\{1,2,3,4\}}$, and
(bottom) $K=2^{k\in\{1,2,3,4,5,6\}}$
Note: $S^{(N)}_2 = S_2(\rho_N) = -\ln \left( \sum_{i=1}^N \lambda_i^2\right)~\in~\{0,\ln(N)\} \leftrightarrow \{\trm{pure},\trm{MMS}\}$.
}\label{fig:PS2N2K}
\end{figure}
%============================

We can use  $P^{\trm{trace}}_{N=2,K} (\mu_2)$ to obtain the probability distribution as a function of the 
R\'{e}nyi entropy 
 $S^{(N)}_2 = S_2(\rho_N) = -\ln \left( \sum_{i=1}^N \lambda_i^2\right)~\in~\{0,\ln(N)\} \leftrightarrow \{\trm{pure},\trm{MMS}\}$
 which takes it minimum value $0$ on pure states and its maximum value $\ln(N)$  ($\ln(2)\approx 0.693$) on the MMS.
Here we use the fact that $P(S_2)\,dS_2=P(\mu_2)\,d\mu_2 \Rightarrow P(S_2) = P(\mu_2\to e^{-S_2})\,\left|\tfrac{d\mu_2}{dS_2}\right|$ to obtain
\be{PS2N2K}
P^{\trm{trace}}_{N=2,K} (S_2) = 
\dfrac{(1-e^{-S_2})^{K-2}\,(2e^{-S_2}-1)^{1/2}\,\G(2 K)}{2^{K-1}\,\G(K)\,\G(K-1)}\,e^{-S_2}.
\ee
This is plotted in \Fig{fig:PS2N2K}(top) for  $K=2^{k\in\{1,2,3,4\}}$ and also  $K=2^{k\in\{1,2,3,4,5,6\}}$ on the (bottom), in order to exhibit the spikey behavior near the MMS for very large~$K$.
%=======================================================================

%=======================================================================
\subsubsection{$\boldmath{N=3}$}\label{subsubsec:N3}
%=======================================================================
For the case of $N=3$, we substitute \Eq{lambdavecs:N2:N3}(right)  into 
the unnormalized portion 
$\tilde{P}^{\trm{trace}}_{N,K\ge N}(\lambda_1,\ldots,\lambda_N) \overset{\trm{def}}{=} \prod_{j<k}^{N} (\lambda_j-\lambda_k)^2\, \prod_{i}^{N} (\lambda_i)^{K-N}$ of the probability distribution
in \Eq{PNK:GQS:15.59:line2} 
(and also include the Jacobian $J_{\lambda,\varphi}~\overset{\trm{def}}{=}~\trm{Det}\left[\left(\tfrac{\partial (\lambda_1, \lambda_2)}{\partial (r_3, \varphi_2)}\right)\right]$ of the transformation between the 
$(\lambda_1,\lambda_2)$ and $(r_3,\varphi_2)$ coordinates)
to obtain
\bsub
\bea{tildePtrace:N3:K3:Kge4}
%\hspace{-0.65in}
\tilde{P}^{\trm{trace}}_{N=3,K=3} &=& \frac{r_3^6}{2^4\,3^3}\,\sin^2(3\,\varphi_2), \\
%
%\hspace{-0.65in}
%\tilde{P}^{\trm{trace}}_{N=3,K=6} &=& \frac{r_3^6}{2^{10}\,3^{12}}\,
%(4 - 3\,r_3^2-r_3^3\,\cos(3\,\varphi_2))^3
%\sin^2(3\,\varphi_2),\qquad \\
%
%\hspace{-0.65in}
\tilde{P}^{\trm{trace}}_{N=3,K\ge4} &=& \frac{r_3^6}{2^{2(K-1)}\,3^{3(K-2)}}\,\sin^2(3\,\varphi_2) \no
&\times& (4 - 3\,r_3^2-r_3^3\,\cos(3\,\varphi_2))^{K-3}.
\qquad \;
\eea
\esub
We then compute
\bsub
\bea{}
\hspace{-0.5in}
P^{\trm{trace}}_{N=3,K}(r_3) &=& 
\frac
{
\int_{\varphibarmin_2(r_2)}^{\pi/3} d\varphi_2\, \tilde{P}^{\trm{trace}}_{N=3,K}(r_3,\varphi_2)
}
{
\int_{0}^{r_3^{max}=\sqrt{2/3}}\,dr'_3\int_{\varphibarmin_2(r'_3)}^{\pi/3}\,\tilde{P}^{\trm{trace}}_{N=3,K}(r'_3,\varphi_2)
}, \qquad\\
\hspace{-0.5in}
\varphibarmin_2(r_3) &=& \cos^{-1}\left(\Min\left[\frac{1}{\sqrt{3\cdot 2}\,r_3},\,1\right] \right), \\
\hspace{-0.5in}
P^{\trm{trace}}_{N=3,K}(\mu_3) &=& P^{\trm{trace}}_{N=3,K}(r_3(\mu_3)) \, \frac{1}{2\,\sqrt{\mu_3-1/3}},
\eea
\esub
where in the last line we have used the fundamental law of probability that
$P(\mu_3)\,d\mu_3 = P(r_3)\,dr_3\Rightarrow P(\mu_3)\,J_{r_3,\mu_3}$ 
where here the Jacobian of the transformation is simply 
$J_{r_3,\mu_3} = \tfrac{dr_3}{d\mu_3} =  \frac{1}{2\,\sqrt{\mu_3-1/3}}$, using
$r_3 = \sqrt{\mu_3-1/3}$. In the $r_3$ variable, we can then compute the CDF
$F^{\trm{trace}}_{N=3,K}(r_3)$~as
\be{}
F^{\trm{trace}}_{N=3,K}(r_3) = 
\int_{0}^{r_3\le r_3^{max}} dr'_3\, P^{\trm{trace}}_{N=3,K}(r'_3,\varphi_2).
\ee
%============================
\begin{figure}[ht]
\begin{tabular}{c}
%\hspace{-4em}
%\includegraphics[width=3.5in,height=2.25in]{PS2N2K2_4_8_16_8Mar2021} \\
%\includegraphics[width=3.5in,height=2.25in]{PS2N2K2_4_8_16_32_64_8Mar2021} 
\includegraphics[width=3.0in,height=1.75in]{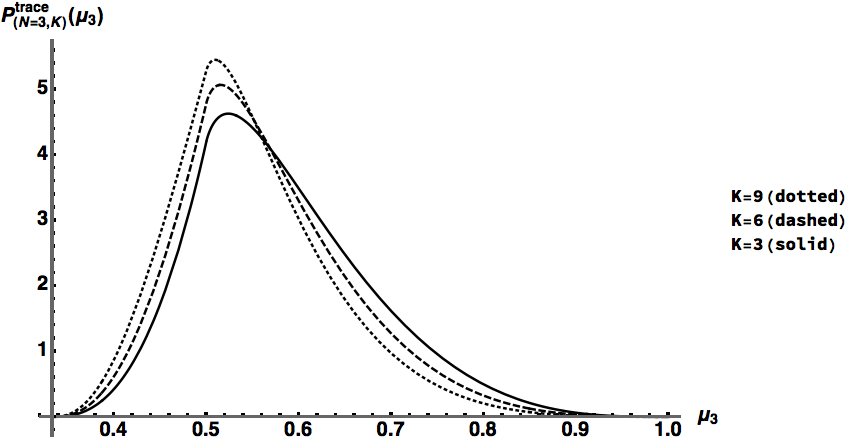} \\
\includegraphics[width=3.0in,height=1.75in]{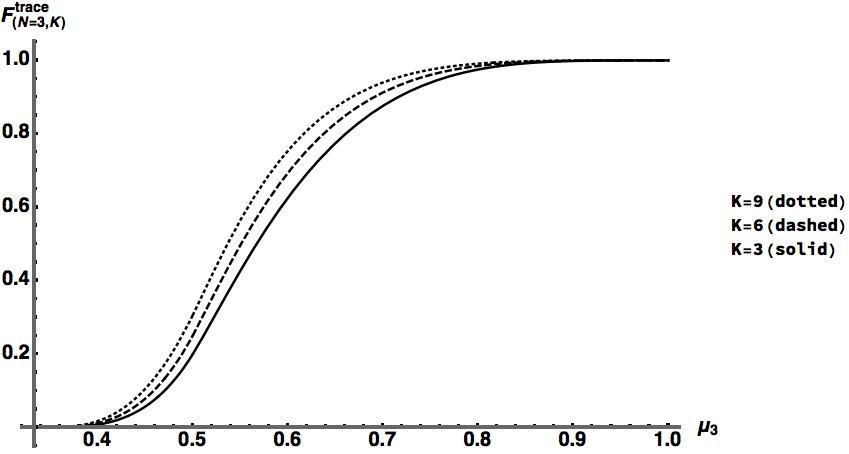} 
\end{tabular}
\caption{
(top) $P^{\trm{trace}}_{N=3,K}(\mu_3)$ and  
(bottom) $F^{\trm{trace}}_{N=3,K}(\mu_3)$
for $K~=~\{3, 6, 9\}$.
}\label{fig:P:F:N3K369}
\end{figure}
%============================
To plot $F^{\trm{trace}}_{N=3,K}(r_3)$ as a function of $\mu_3$ we can simply substitute in 
$r_3=\sqrt{\mu_3-1/3}$.
Remarkably, all the above integrals can be performed analytically in closed form, though they are somewhat wieldy to write down.
In \Fig{fig:P:F:N3K369} we plot $P^{\trm{trace}}_{N=3,K}(r_3)$ and  $F^{\trm{trace}}_{N=3,K}(r_3)$
for $K=\{3, 6, 9\}$. The general trend is the same as for $N=2$ case, namely that as $K$ increases, 
$P^{\trm{trace}}_{N=3,K}(r_3)$ moves leftward, centered on lower values of $\mu_3$, and at the same time narrows, and increases in height (eventually delta-function spiking at the MMS at $\mu_3=1/3$ as 
$K\to\infty$). This feature is reflected in the CDF $F^{\trm{trace}}_{N=3,K}(r_3)$ which also moves leftward , and saturates earlier near unity value at ever smaller values of $\mu_3$ as $K$ increases \cite{Renyi:entropy:note}.

%=======================================================================
\subsubsection{$\boldmath{N=4}$}\label{subsubsec:N4}
\begin{figure}[ht]
%\begin{tabular}{c}
%\hspace{-4em}
%\includegraphics[width=3.5in,height=2.25in]{P_FP_N4K4_8Mar2021} 
\includegraphics[width=3.0in,height=1.75in]{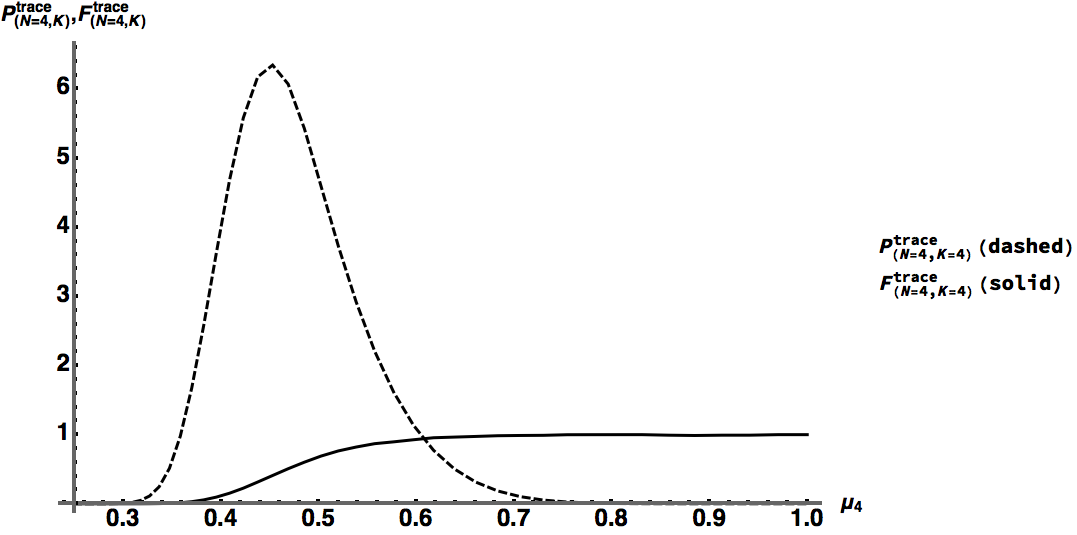} 
%\includegraphics[width=3.0in,height=1.75in]{fig_PandCDFtraceN4K4mu4_2Apr2022} 
%\end{tabular}
\caption{Composite plot of
$P^{\trm{trace}}_{N=4,K=4}(\mu_4)$ and  
$F^{\trm{trace}}_{N=4,K=4}(\mu_4)$.
}\label{fig:P:F:N4K4}
\end{figure}

The case of $N=4$ follows analogously to the case of $N=3$ above, with some minor modification of the limits. For $N=4$ the variables are now $(r_4, X_3=\cos\varphi_3, \varphi_2)$.
The expression for $\tilde{P}^{\trm{trace}}_{N=4,K=4}(r_4)$ (including the Jacobian of the transformation between  the $(\lambda_1, \lambda_2, \lambda_3)$ and the $(r_4, X_3=\cos\varphi_3, \varphi_2)$ coordinates) is given by
%=============
\bea{}
%\hspace{-0.25in}
&{}&\tilde{P}^{\trm{trace}}_{N=4,K=4}(r_4) =
\frac{r_4^{14}}{2^6\,3^3}\,\sin^2(\varphi2)\,\sin^7(\varphi_3)\,\big(1+2\cos^2(2\varphi_2)\big)^2 \no
\hspace{-0.25in}
&{}&
\times
\left(
21\,\cos(\varphi_3) + 11\,\cos(3\,\varphi_3) -\sqrt{8}\,\cos(3\,\varphi_2)\,\sin^2(\varphi_3)
\right)^2. \qquad
\eea
%=============
As we integrate out both $X_3$ and $\varphi_2$ to obtain the marginal distributions for
 $P^{\trm{trace}}_{N=4,K}(r_4)$ and  
$F^{\trm{trace}}_{N=4,K}(r_4)$ we must remember to now use
$\varphibarmin_2(X_3)=\cos^{-1}\left( \Min\left[\tfrac{\sqrt{2}\,X_3}{\sqrt{1-X_3^2}},\,1 \right]\right)$,
$\Xbarmax_3 =\Min\left[\tfrac{1}{\sqrt{4\cdot 3}\,r_4},\,1 \right]$ and $X_3^{(min)} =1/3$.
These integrals cannot be performed in closed form analytically and so must be computed numerically. 
In \Fig{fig:P:F:N4K4} we illustrate the $N=4$  
case for $K=4$.
%============================
%%============================
As in the $N=3$ case, we see that $P^{\trm{trace}}_{N=4,K=4}(\mu_4)$ is already becoming
narrower and left-shifted, and $F^{\trm{trace}}_{N=4,K=4}(\mu_4)$ is plateauing to unity
at much lower values of the purity $\mu_4$ than for the corresponding case for $N=3$ in 
\Fig{fig:P:F:N3K369}.
%=======================================================================

%================================================================================
\subsection{Complementary-Quantum Correlation (CQC) relation: $\boldsymbol{N=4}$}\label{subsec:CQC}
%================================================================================
As a final application we re-examine the complementary-quantum correlation (CQC) relation conjectured by Schneeloch \tit{et al} \cite{Scheeloch:2014} which  lower bounds the quantum mutual information (QMI) of a bipartite system $AB$ of dimension $M\otimes N$ by the sum of the classical mutual informations (CMI) obtained from pairs of mutually unbiased measurements on the subsystems.
As usual, the QMI is given by $I(A:B) = S(A)+S(B)-S(AB)$ where $S$ is the von Neumann entropy 
$S(\rho)=-\Tr[\rho\log\rho]$. For the CMI, the authors considered the post-measurement-QMI 
obtained by local projective measurements of observables $\hat{Q}^A$ and $\hat{Q}^B$ on $A$ and $B$, respectively. The post-measurement-QMI is equal to the CMI obtained from the joint probability distribution of measurement outcomes $P(q_i^A, q_j^B)$, where $i$ and $j$ run over all measurement outcomes.
The CMI for the measurements $\{\hat{Q}^A,\hat{Q}^B\}$
 is then given by $H(\hat{Q}^A:\hat{Q}^B) = H(\hat{Q}^A)+ H(\hat{Q}^B)- H(\hat{Q}^A,\hat{Q}^B)$,
%%============================
%\begin{figure}[ht]
%\begin{tabular}{cc}
%\hspace{-0.15in}
%\includegraphics[width=2in,height=1.25in]{QMI_vs_CMI_mu4_0p30_Npts2500_13Apr2022} & 
%\hspace{-0.15in}
%\includegraphics[width=2in,height=1.25in]{QMI_vs_CMI_mu4_0p50_Npts2500_13Apr2022} \\
%%
%\hspace{-0.15in}
%\includegraphics[width=2in,height=1.25in]{QMI_vs_CMI_mu4_0p70_Npts2500_13Apr2022} & 
%\hspace{-0.15in}
%\includegraphics[width=2in,height=1.25in]{QMI_vs_CMI_mu4_0p99_Npts2500_13Apr2022} 
%\end{tabular}
%\caption{Composite plot of
%$P^{\trm{trace}}_{N=4,K=4}(\mu_4)$ and  
%$F^{\trm{trace}}_{N=4,K=4}(\mu_4)$.
%}\label{fig:P:F:N4K4}
%\end{figure}
%%%============================
%
%============================
\begin{figure}[ht]
\begin{tabular}{ccc}
\hspace{-0.75in}
%=======================================================================
\includegraphics[width=1.35in,height=1.0in]{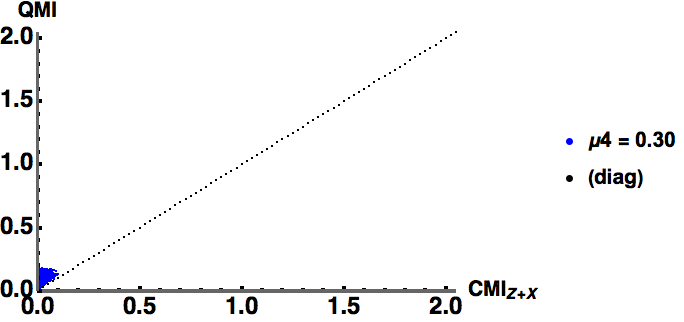} & 
%\hspace{-0.25in}
\includegraphics[width=1.35in,height=1.0in]{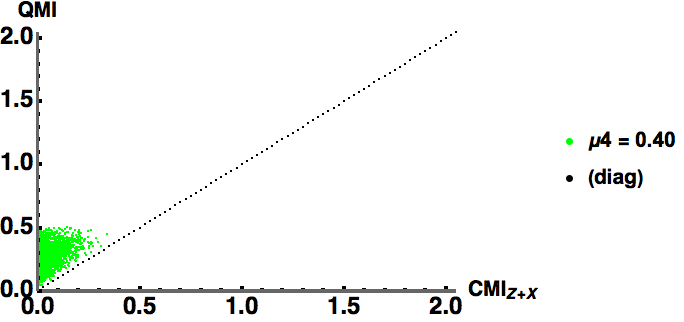} &
%\hspace{-0.45in}
\includegraphics[width=1.35in,height=1.0in]{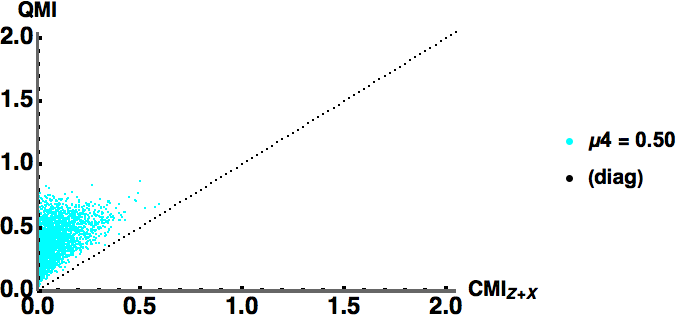} \\
\hspace{-0.75in}
\includegraphics[width=1.35in,height=1.0in]{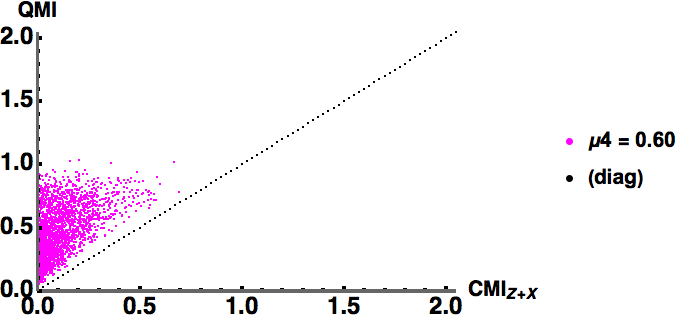} & 
%\hspace{-0.60in}
\includegraphics[width=1.35in,height=1.0in]{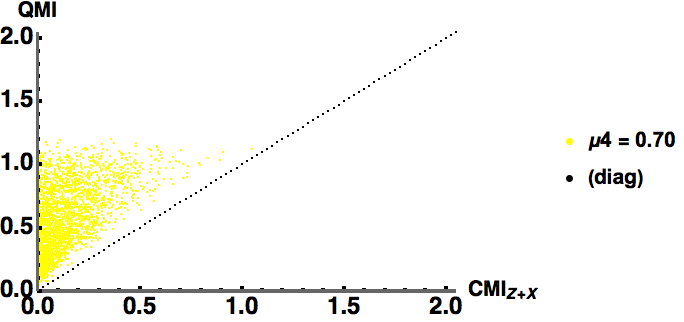} &
%\hspace{-0.45in}
\includegraphics[width=1.35in,height=1.0in]{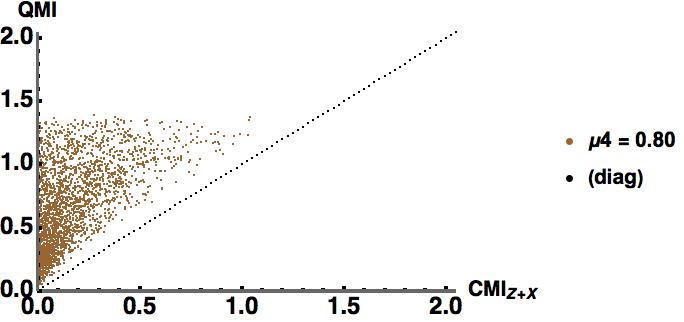} \\
\hspace{-0.75in}
\includegraphics[width=1.35in,height=1.0in]{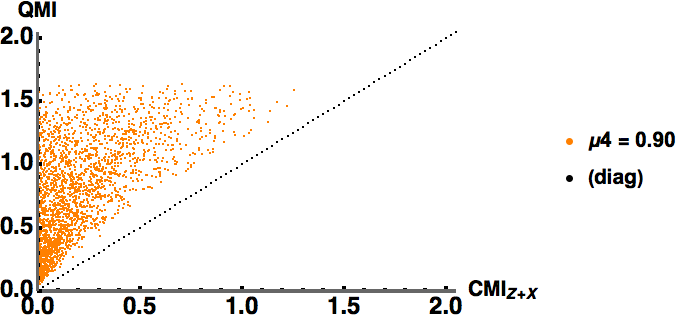} & 
%\hspace{-0.60in}
\includegraphics[width=1.35in,height=1.0in]{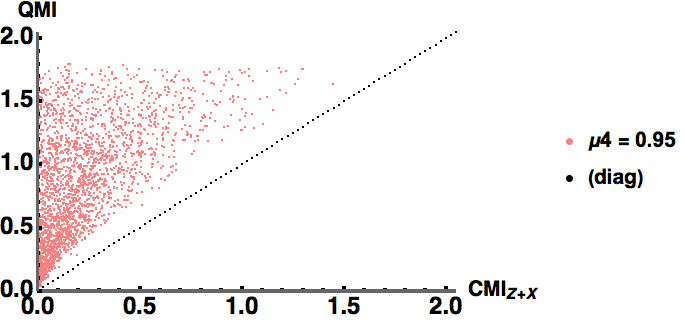} &
%\hspace{-0.45in}
\includegraphics[width=1.35in,height=1.0in]{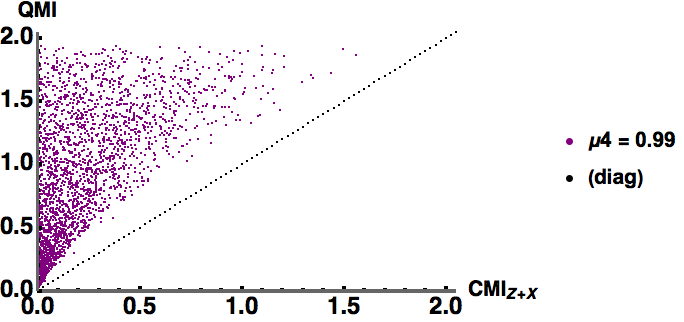} 
%=======================================================================
%\includegraphics[width=1.35in,height=1.0in]{QMI_vs_CMI_mu4_0p30_Npts2500_13Apr2022} & 
%%\hspace{-0.25in}
%\includegraphics[width=1.35in,height=1.0in]{QMI_vs_CMI_mu4_0p40_Npts2500_13Apr2022} &
%%\hspace{-0.45in}
%\includegraphics[width=1.35in,height=1.0in]{QMI_vs_CMI_mu4_0p50_Npts2500_13Apr2022} \\
%%
%\hspace{-0.75in}
%\includegraphics[width=1.35in,height=1.0in]{QMI_vs_CMI_mu4_0p60_Npts2500_13Apr2022} & 
%%\hspace{-0.60in}
%\includegraphics[width=1.35in,height=1.0in]{QMI_vs_CMI_mu4_0p70_Npts2500_13Apr2022} &
%%\hspace{-0.45in}
%\includegraphics[width=1.35in,height=1.0in]{QMI_vs_CMI_mu4_0p80_Npts2500_13Apr2022} \\
%%
%\hspace{-0.75in}
%\includegraphics[width=1.35in,height=1.0in]{QMI_vs_CMI_mu4_0p90_Npts2500_13Apr2022} & 
%%\hspace{-0.60in}
%\includegraphics[width=1.35in,height=1.0in]{QMI_vs_CMI_mu4_0p95_Npts2500_13Apr2022} &
%%\hspace{-0.45in}
%\includegraphics[width=1.35in,height=1.0in]{QMI_vs_CMI_mu4_0p99_Npts2500_13Apr2022} 
%=======================================================================
\end{tabular}
\caption{Pairs of points $(CMI_{Z+X}, QMI)$ for $2500$ density matrices, each at a fixed value of 
purity $\mu_4 \equiv \Tr[\rho^2_{AB}]$, for $\mu_4\in\{0.26, 0.30, 0.40, 0.50, 0.60, 0.70, 0.80, 0.90, 0.95, 0.99\}$.
}\label{fig:QMI:vs:CMI:mu4s:2500pts}
\end{figure}
%%============================
where $H(\vec{p}) = -\sum_i p_i\log p_i$ is the classical Shannon entropy.

The CQC relation conjectured in \cite{Scheeloch:2014} was given by
 \be{CQC:relation}
 I(A:B)\ge H(\hat{Q}^A:\hat{Q}^B) + H(\hat{R}^A:\hat{R}^B), 
 \ee
where $\{\hat{R}^A,\hat{R}^B\}$ is another set of arbitrary measurement observables subject only to the constraint that they are mutually unbiased to the observables $\{\hat{Q}^A,\hat{Q}^B\}$ for each subsystem, respectively.
Numerical evidence on $k\otimes k$ systems for $k\in\{2,3,4\}$ showed that for ($10^7$) density matrices generated randomly from unitary matrices distributed according to the Haar measure, that pairs of points
$(CMI_{Z+X}, QMI)$ always lay within and filled the  triangle with vertices $\{(0,0), (0,k), (k,k)\}$.
Here, we have denoted $CMI_{Z+X}$ as the righthand side of \Eq{CQC:relation} for
$\hat{Q}\to \hat{Z}$ and $\hat{R}\to \hat{X}$, where $\{\hat{X},\hat{Y},\hat{Z}\}$ are the usual single qubit Pauli spin operators. In the following we examine the case of $k=2$ ($N=4$, two-qubits), but now from the perspective of sampling density matrices at fixed purity values $\mu_4 \equiv \Tr[\rho^2_{AB}]$.

In \Fig{fig:QMI:vs:CMI:mu4s:2500pts} we plot pairs of points 
 $(CMI_{Z+X}, QMI)$ for $2500$ density matrices, each at a fixed value of 
purity $\mu_4 \equiv \Tr[\rho^2_{AB}]$, for $\mu_4\in\{0.26, 0.30, 0.40, 0.50, 0.60, 0.70, 0.80, 0.90, 0.95, 0.99\}$.
%============================
\begin{figure}[ht]
\begin{tabular}{ccc}
\hspace{-0.2in}
\includegraphics[width=1.35in,height=1.0in]{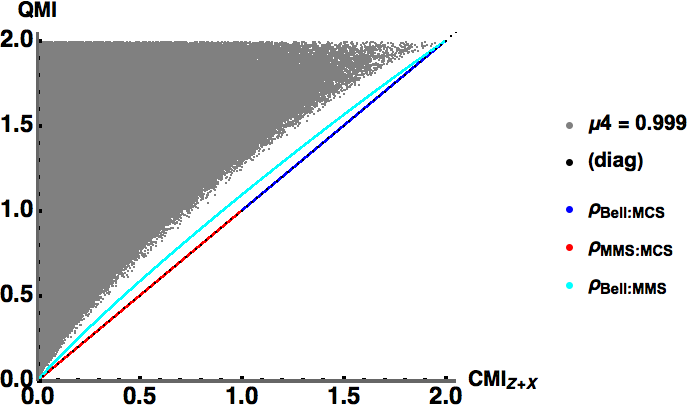} & 
%\hspace{-0.25in}
\includegraphics[width=1.35in,height=1.0in]{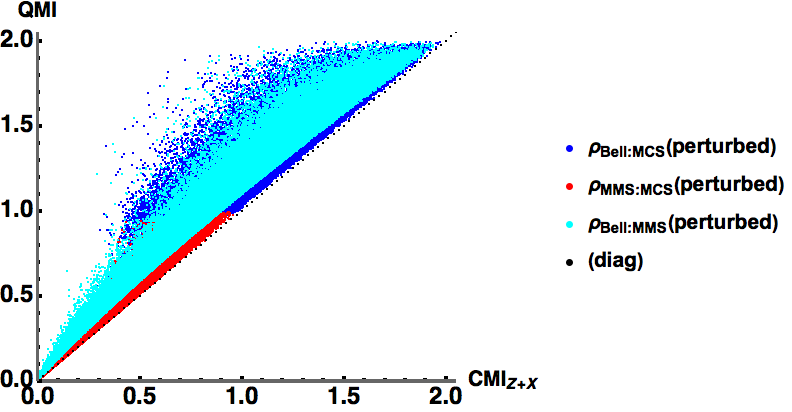} &
%\hspace{-0.45in}
\includegraphics[width=1.35in,height=1.0in]{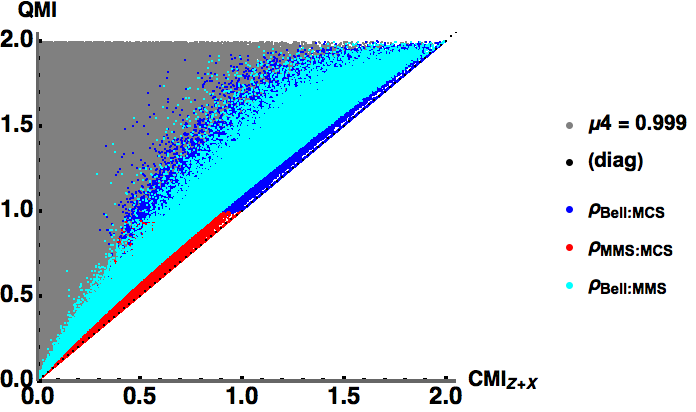} \\
%
%\includegraphics[width=1.35in,height=1.0in]{CQC_mu4_0p999_1Mpts_and_BellMMSMCS_unPerturbed_250Kpts_eps_0p15_13Apr2022} & 
%%\hspace{-0.25in}
%\includegraphics[width=1.35in,height=1.0in]{CQC_BellMMSMCS_U-Perturbed_250Kpts_eps_0p15_only_with_legends_13Apr2022} &
%%\hspace{-0.45in}
%\includegraphics[width=1.35in,height=1.0in]{CQC_mu4_0p999_BellMMSMCS_U-Perturbed_250Kpts_eps_0p15_13Apr2022} \\
%
\end{tabular}
\caption{
Pairs of points $(CMI_{Z+X}, QMI)$ for 
(left, gray) $10^6$~density matrices at $\mu_4=0.999$,
and $10^4$ convex combinations of density matrices: 
(left, blue) Bell and maximally correlated states (MCS) $\rho_{Bell:MCS}$,
(left, red) maximally mixed state and MCS $\rho_{MMS:MCS}$, and
(left, cyan) Bell and MMS $\rho_{Bell:MMS}$.
(center) $2.5\times 10^4$ perturbations $U\,\rho\,U^\dag$ of each (left) convex combinations $\rho$, 
with $U$ near the identity matrix.
(right) Composite plot of left and middle plots.
}\label{fig:QMI:vs:CMI:mu4:0p999:1Mpts:Uperturbed:250Kpts}
\end{figure}
%=======================================================================
We see that for each fixed value of $\mu_4$, a scaled down triangle similar to the full triangle for $N=4$,
with vertices $\{(0,0), (0,2), (2,2)\}$,
 is populated, growing larger with larger values of the purity.
%============================
\begin{figure}[ht]
\begin{tabular}{c}
\hspace{0.10in}
\includegraphics[width=3.75in,height=2.0in]{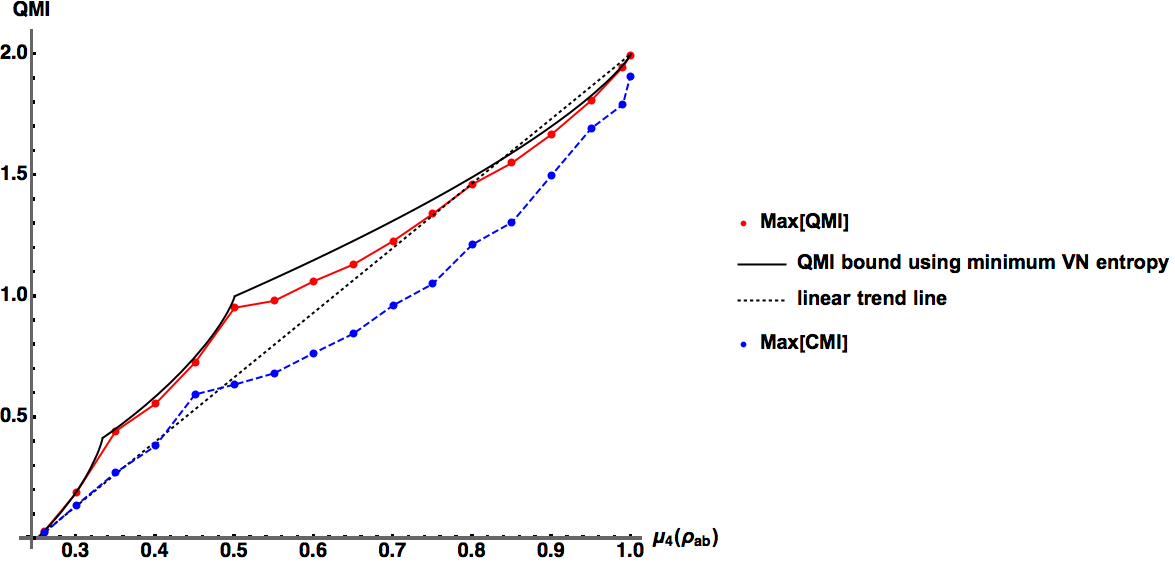} 
%\includegraphics[width=3.75in,height=2.0in]{MaxQMI_vsmu4_Npts25K_with_SminBoundOnly_CMI_18Apr2022} 
%\hspace{-0.25in}
%\includegraphics[width=1.35in,height=1.0in]{CQC_BellMMSMCS_U-Perturbed_250Kpts_eps_0p15_only_with_legends_13Apr2022} &
%\hspace{-0.45in}
%\includegraphics[width=1.35in,height=1.0in]{CQC_mu4_0p999_BellMMSMCS_U-Perturbed_250Kpts_eps_0p15_13Apr2022} \\
%
\end{tabular}
\caption{(red) $QMI \,(\equiv\Max[QMI]$) value corresponding to 
(blue) $\Max[CMI_{Z+X}]$  from $25,000$ density matrices, 
each at a fixed value of purity $\mu_4 \equiv \Tr[\rho^2_{AB}]$, for 
$\mu_4~\in~\{0.26, 0.30, 0.40, 0.50, 0.60, 0.70, 0.80, 0.90, 0.95, 0.99\}$.
(black, solid) QMI bound using minimum von Neumann entropy (as described in text);
(black, dashed) linear trend line.
}\label{fig:MaxQMI:MaxCMI:vs:mu4}
\end{figure}
%============================
In \Fig{fig:QMI:vs:CMI:mu4:0p999:1Mpts:Uperturbed:250Kpts} the gray points in the (left) plot represents $10^6$ random density matrices all at fixed purity $\mu_4=0.999$ (note: for density matrices generated   from unitaries uniformly sampled from the Haar measure, this would have required $\sim\mathcal{O}(10^{13})$ samples to obtain the same number of states with $\mu_4=0.999$). We observe that the $N=4$ triangle with vertices $\{(0,0), (0,2), (2,2)\}$ is almost, though not fully populated by states with  $\mu_4=0.999$, missing a (white) crescent shaped region near the diagonal
\cite{filling:out:the:triangle:note}.
Thus, in the (left) figure, we also plot various convex combinations of the maximally entangled Bell state
$\ket{Bell}\bra{Bell}$ (with $\ket{Bell} = \tfrac{1}{\sqrt{N}}\,\sum_{i}\ket{ii}_{ab}$), with
the maximally correlated state (MCS, left, blue) given by $\rho_{MCS} = \sum_{i,i}\ket{i,i}_{ab}\bra{i,j}$,
and with the maximally mixed state (MMS, left cyan) given by $\rho_{MMS} = \sum_{i,j}\ket{i,j}_{ab}\bra{i,j}$.
The (left red) are convex combinations of the MMS with the MCS.
We see that 
(blue) $\rho_{Bell:MCS} = p\,\ket{Bell}\bra{Bell}+(1-p)\,\rho_{MCS}$ and
(red) $\rho_{MMS:MCS} = p\,\rho_{MMS}+(1-p)\,\rho_{MCS}$ fill the upper and lower ends of the diagonal
of the triangle, while 
(cyan) $\rho_{MMS:MCS} = p\,\ket{MMS}\bra{Bell}+(1-p)\,\rho_{MCS}$ is a curved upper bound to the diagonal, lying outside the gray sampled points.

Thus, to fill in the white crescent shaped region, we perturb the previous mixed states by
$\rho\to U\,\rho\,U^\dag$ with $U$ a unitary near the identity \cite{U-perturbed:note}, as shown in 
\Fig{fig:QMI:vs:CMI:mu4:0p999:1Mpts:Uperturbed:250Kpts}(middle) (using the same color coding as in the left plot).
In \Fig{fig:QMI:vs:CMI:mu4:0p999:1Mpts:Uperturbed:250Kpts}(right) we show the composite of the (left) and (middle) figures, which now almost completely fills the full $N=4$ triangle with  vertices $\{(0,0), (0,2), (2,2)\}$ using only the additional randomly sampled states with fixed purity (gray) $\mu_4=0.999$ (note: states of purity with 
$\mu_4=1-\epsilon$ with $\epsilon\to0$ would completely fill the triangle).

The apparent constant maximum value of QMI (for all CMI) in 
\Fig{fig:QMI:vs:CMI:mu4s:2500pts}
%\Fig{fig:QMI:vs:CMI:mu4:0p999:1Mpts:Uperturbed:250Kpts} 
as a function of purity $\mu_4$ suggests there exists a well defined purity dependent upper bound to the QMI.
In \Fig{fig:MaxQMI:MaxCMI:vs:mu4} the (red) points are 
the values of $QMI$ corresponding to $\Max[CMI_{Z+X}]$
%\Fig{fig:QMI:vs:CMI:mu4:0p999:1Mpts:Uperturbed:250Kpts}, 
%\Fig{fig:QMI:vs:CMI:mu4s:2500pts},
now sampled with $2.5 \times 10^4$ density matrices for each fixed value of the purity 
(additionally including $\mu_4~=~\{0.35, 0.45, 0.55, 0.65, 0.75, 0.85\}$).
These $QMI = \Max[QMI]$ values (a numerical approximation to the supremum of the QMI over all $\rho$ of each fixed $\mu_4$)
clearly outline three crescent shaped curves for each of the three purity regions 
$\mu_4^{(1)} = [\tfrac{1}{4},\tfrac{1}{3}]$, 
$\mu_4^{(2)} = [\tfrac{1}{3},\tfrac{1}{2}]$, and
$\mu_4^{(3)} = [\tfrac{1}{2},1]$.
The (black, solid) curve represents 
the upper bound based on the using the minimum entropy in each 
$\mu_4^{(i)}$ region, to be described shortly.
In fact, the black solid curve is actually the maximum value to the QMI given the joint purity $\mu_{4}$, since there are states saturating this bound
\cite{black:curve:note}.
Note that in regions $\mu_4^{(1)}$ and $\mu_4^{(2)}$ the red points nearly coincide with the black curve, but deviate from it in region $\mu_4^{(3)}$.
We conjecture that red points of $\Max[QMI]$ should actually lie on the upper bound black curve, and the deviation arises solely from finite sampling size. 
The (black, dashed) curve is simply a linear trend line (for visual guidance) between the points $(1/4,0)$ and $(1,2)$.
%The (blue) points are $\Max[CMI_{Z+X}]$ over the same data set for which the (red) points $\Max[QMI]$ were obtained (i.e. not the value of $CMI_{Z+X}$ for the $\Max[QMI]$ at each fixed $\mu_4$). 
The difference 
$\Max[QMI]-\Max[CMI]$ then gives the minimum ``gap" between the quantum and classical mutual information at each fixed purity value $\mu_4$.

The (black, solid) curve is derived as follows.
For a given purity, there is an exact lower limit to the von Neumann entropy because there is a minimum entropy probability distribution for constant purity \cite{Berry_Sanders:2003}. In \cite{Schneeloch:etal:CondEnt:2022}, we express those distributions in terms of the purity.
In particular, our lower bound to the joint entropy has the form:
\be{Smin}
\hspace{-0.15in}
S(AB)\geq \frac{-(1-g)\log\left(\frac{1-g}{1+\kappa}\right) -(\kappa +g)\log\left(\frac{\kappa+g}{\kappa(1+\kappa)}\right)}{1+\kappa}
\ee
where $g=\sqrt{\kappa(\mu_{4}(\kappa+1)-1)}$ and $\kappa = \texttt{Floor}[\tfrac{1}{\mu_4}]$, i.e. the value of 
$\tfrac{1}{\mu_4}$ rounded down to the nearest integer.
We obtain \Eq{Smin} by considering diagonal density matrices of the following form
$\rho_d\to\vec{p}_{min} = (p_0,\ldots,p_0,1-~\kappa\,p_0,0,0,\ldots,0)$, where $\kappa$ dictates the number of $p_0$ entries. Note that $p_0=1\Rightarrow\kappa=1$, a pure state, which has the minimum (zero) value of the von Neumann entropy. Therefore, $\vec{p}_{min}$ represents diagonal density matrices that deviate from the pure state as $p_0$ decreases from unity. We now fix $p_0$ by setting  $\vec{p}^{\,2}_{min} = \mu_4$ and solving for 
$p_0 = p_0(\mu_4)$ (choosing the root that gives $p_0=\mu_4$ if $\kappa\equiv\tfrac{1}{\mu_4}$).
We then simply compute $S_{min} = -p_0(\mu_4)\,\log_2 p_0(\mu_4) - (1-p_0(\mu_4))\,\log_2(1-p_0(\mu_4))$
with  $\kappa\to\texttt{Floor}[\tfrac{1}{\mu_4}]$
to obtain the (black) upper bound.

It is interesting to note that the (black) serrated curve yields a much tighter bound than a simple upper bound estimate obtained \cite{Schneeloch:etal:CondEnt:2022} by considering the maximum  of the von Neumann entropy $S_{max}$ derived  from considering 
$\vec{p}_{max} = (p_0,\ldots,p_0,1-(N-1)p_0)$, where all but one entry is given by $p_0$. Again we fix $p_0$ by
setting  $\vec{p}^{\,2}_{max} = \mu_4$, and solve for $p_0=p_0(\mu_4)$ (choosing the root such that
$p_0\le1-(N-1)p_0 $).
One then notes that the collision entropy defined by $S_2(\mu_4) = -\log_2(\mu_4)$ \cite{Renyi:entropy:note} is always less than or equal to the joint von Neumann entropy $S(A,B)$, so that $S_{max}\equiv S(A)+S(B)-S_2(\mu_4)\ge I(A:B)$.
The curve for $S_{max}$   would be a smooth concave arc in 
\Fig{fig:MaxQMI:MaxCMI:vs:mu4}  (not shown) with 
endpoints at points $(\tfrac{1}{4},0)$ and $(1,2)$ which also just touches the
$\Max[QMI]$ values at the cusp values at $\mu_4 = \tfrac{1}{3}$ and $\mu_4 = \tfrac{1}{2}$, i.e.
at all the four boundary points of the three $\mu_4^{(i)}$ regions.
As observed in \Fig{fig:MaxQMI:MaxCMI:vs:mu4}, the bound using the minimum entropy $S_{min}$ produces a much tighter bound (in fact, the maximum value of the QMI) than that using $S_{min}$.

%=======================================================================
%=======================================================================
\clearpage
\newpage
%=======================================================================
\section{$\boldsymbol{N=5, 6}$ and higher dimensions, plus numerical considerations}\label{sec:N5:and:beyond}
\begin{figure}[ht]
%\begin{tabular}{c}
%\hspace{-0.65in}
\includegraphics[width=3.5in,height=2.25in]{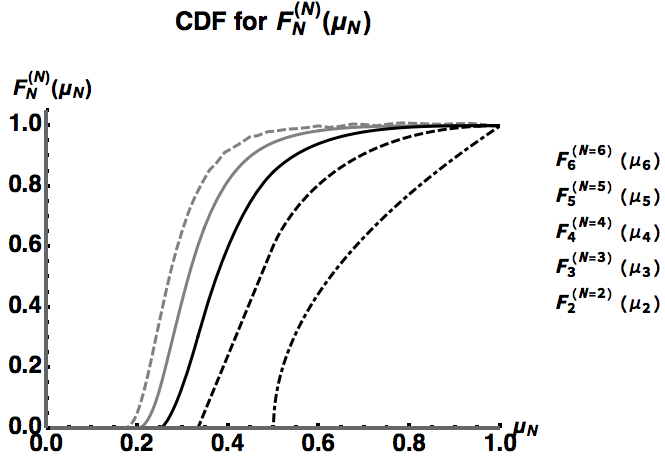} 
%\includegraphics[width=3.5in,height=2.25in]{fig_FNkNmuN_N_23456_25Mar2022}
%\end{tabular}
\caption{Radial CDFs  $F_N^{(N)}(\mu_N)$
Same as \Fig{fig:CDF:FNkNrN:for:N:2:3:4} but 
now additionally including $N=5$ (gray-solid) and $N=6$ (gray-dashed).
}\label{fig:CDF:FNkNrN:for:N:2:3:4:5:6}
\end{figure}

%In this section we consider extensions of the previous CDFs and sampling for $N>4$.
%
In \Fig{fig:CDF:FNkNrN:for:N:2:3:4:5:6} we plot the same radial CDFs $F_N^{(N)}(\mu_N)$ vs $\mu_N$ but now additionally including $N=5$ (gray-solid) and $N=6$ (gray-dashed). As $N$ increases we observe a  steeper 
%=========================================================
% from locate_NumRecC_23Feb2022.nb in Downloads and 
% locate_NumRecC_F6N6mu6_24Mar2022.nb 
% See also
% locate_NumRecC_F3N3mu3_F4N4mu4_F5N5mu5_just_plots_24Mar2022.nb, and
% locate_NumRecC_F3N3mu3_F4N4mu4_F5N5mu5_formulas_and_plots_24Mar2022.nb
%=========================================================
%============================
transition from low to high CDF values occurring at much lower purity values.
The more extended flat region of $F_N^{(N)}(\mu_N)\sim 1$ as $N$ increases presents challenges for both numerical accuracy in this region, as well as for numerically computing the inverse of the CDF for the purposes of sampling.

The curve for $F_5^{(N=5)}(\mu_5)$ in  \Fig{fig:CDF:FNkNrN:for:N:2:3:4:5:6} was computed as
in terms of $r_5=\sqrt{\mu_5-\tfrac{1}{5}}$ as 
\bsub
\bea{F5N5r5}
F_5^{(N=5)}(r_5) &=&
\frac
{
\int_{0}^{r_5\le r_5^{(max)}} \,dr'_5\, r^{'3}_5\,F_4^{(N=5)Denom}(r'_5)
}
{
\int_{0}^{r^{(max)}_5 = \sqrt{4/5}} \,dr'_5\, r^{'3}_5\,F_4^{(N=5)Denom}(r'_5)
}, \qquad\;  \label{F5N4r5:line1}  \\
&\equiv&
\frac{F_5^{(N=5)Num}(r_5)}{F_5^{(N=5)Denom}}
\label{F5N4r5:line2}
\eea
\esub
Here $F_5^{(N=5)Denom}$
is the $r_5$-independent normalization constant given by the denominator of 
\Eq{F5N4r5:line1}, which is just the numerator of \Eq{F5N4r5:line1} evaluated at the maximum 
purity value $\mu_5=1\leftrightarrow r_5=\sqrt{4/5}$.
In turn, the descending $F_{k=4,3,2}^{(N=5)Denom}$ terms are given by 
\bsub
\bea{F4N5r5}
\hspace{-0.25in}
F_4^{(N=5)Denom}(r_5) &=& 
\int_{1/4}^{\Xbarmax_4(r_5)=\Min\left[\tfrac{1}{5\cdot4\, r_5},\, 1\right]} dX_4\, \no
&{}&\hspace{2em} \times\sqrt{1-X_4^{2}}\, F_3^{(N=5)Denom}(X_4), \qquad\;\;
\eea
%%%%%%%%%
\bea{F3N5X4}
\hspace{-0.25in}
F_3^{(N=5)Denom}(X_4) &=& 
\int_{1/3}^{\Xbarmax_3(X_4)=\Min\left[\tfrac{\sqrt{5/3}X_4}{\sqrt{1-X_4^2}},\, 1\right]} dX_3\, \no
&{}&\hspace{2em} \times F_2^{(N=5)Denom}(X_3), \qquad
\eea
%%%%%%%%%
\be{F2N5X3}
%\hspace{-0.25in}
F_2^{(N=5)Denom}(X_3) =
\int^{\pi/3}_{\varphibarmax_2(X_3)=\cos^{-1}\left(\Min\left[\tfrac{\sqrt{2}X_3}{\sqrt{1-X_3^2}},\,1\right]\right)} 
\,d\varphi_2. \qquad
\ee
\esub
Now, for all $N$, $F_2^{(N)Denom}(X_3)$ is the expression given in \Eq{F2N4DenomX3} for 
$F_2^{(N=4)Denom}(X_3)$, since it is the lowest angle.
For $N\ge5$ 
$F_3^{(N=5)Denom}(X_4)$ is given by the expressions 
 in \Eq{F3N4:r4:Denom:by:region:1}-\Eq{F3N4:r4:Denom:by:region:3} for $F_3^{(N=4)Denom}(r_4)$
 with the substitution of 
 $\Xbarmax_3(r_4)=\Min\left[ \tfrac{1}{\sqrt{4\cdot 3\,r_4}}, 1\right]\to
\Xbarmax_3(X_4)= \Min\left[ \tfrac{\sqrt{5/3}\,X_4}{\sqrt{1-X_4^2}},1\right]$.
 This occurs since for $N=4$, $X_{3=N-1}$ is the highest angle. However, for $N\ge 5$, $X_3$ is now a bona fide middle angle. The highest angle for $N=5$ is now $X_{4=N-1}$ with 
 $\Xbarmax_4(r_5)=\Min\left[ \tfrac{1}{\sqrt{5\cdot 4\,r_5}}, 1\right]$.
 In addition, each $F_{k>2}^{(N)Denom}(X_{k+1})$ is integrated over 
 $(1-X_k^2)^{(k-3)/2}\,dX_k$ coming from the $\sin^{(k-2)}(\varphi_k)\,d\varphi_k$
 in the volume measure $dV_N$. The procedure in \Eq{F5N4r5:line1} through 
 \Eq{F2N5X3} generalizes to arbitrary $N$.
 
 Surprisingly, for all but one of the sub-integrations over $r_5$ in $F_5^{(N=5)Denom}(r_5)$,
 the numerator of  \Eq{F5N4r5:line1} can be carried out in closed from.
 These are given in \App{app:N5:analytics}.
 The holdout term arises from the subexpression $\tan^{-1}\Big[\sqrt{\tfrac{1-6\,X_4^2}{2(1-X_4^2)}}\Big]$ 
 that occurs in  $F_3^{(N=5)Denom}(X_4)$ in the region $X_4\in[\tfrac{1}{4},\tfrac{1}{\sqrt{6}}]$, which 
 when integrated over $\sqrt{1-X_4^2}$ to contribute to $F_4^{(N=5)Denom}(r_5)$, cannot be done in closed form. This latter integration must be performed numerically, and once again
 when integrated over $r^{'3}_5$ to contribute to the final result $F_5^{(N=5)Denom}(r_5)$.
 The gray-solid line in \Fig{fig:CDF:FNkNrN:for:N:2:3:4:5:6} for the CDF $F_5^{(N=5)}(r_5)$ was 
 computed using as many analytic formulas as possible, with numerical integration reserved for
 the terms just discussed.
 
 This reduction to mostly analytic expressions quickly fails at the $N\ge6$ level, primarily due to the volume
%==========================================================================
\begin{lstlisting}[mathescape,language=Mathematica,
caption={Nested integral Mathematica example: $N=6$},label=Mathematica:Code:1,frame=single]
F6N6r6Num[r6value_] := 
Block[{$\$$MaxExtraPrecision = 1000},
  NIntegrate[$r6^4\,(1-X5^2)\,\sqrt{1-X4^2}$, 
    {r6, 0, r6value        },
    {X5, $\tfrac{1}{5}$, $\Min\left[\tfrac{1}{\sqrt{6\times5}\;r6} , 1\right]\quad\;$  },
    {X4, $\tfrac{1}{4}$, $\Min\left[\sqrt{\tfrac{6}{4}}\tfrac{X5}{\sqrt{1-X5^2}}, 1\right]\,$},
    {X3, $\tfrac{1}{3}$, $\Min\left[\sqrt{\tfrac{5}{3}}\tfrac{X4}{\sqrt{1-X4^2}}, 1\right]$},
    {$\varphi2$, $\Min\left[\sqrt{\tfrac{4}{2}}\tfrac{X3}{\sqrt{1-X3^2}}, 1\right]$,$\tfrac{\pi}{3}$ },
  Method$\rightarrow$"Trapezoidal",
  WorkingPrecision$\rightarrow$40]
]
(*given $\mu6$, r6value = $\sqrt{\mu6-1/6}$; call as*)
F6N6Denom = F6N6r6Num[$\sqrt{4/5}$]; (* once *)
F6N6 = F6N6r6Num[r6value]/F6N6Denom
\end{lstlisting}
%==========================================================================
 integration factors of  $(1-X_k^2)^{(k-3)/2}$. Thus, in contrast to the involved analytical integration over various regions of 
 $\{\varphi_2, X_3, \ldots, X_{N-1}, r_N\}$ as was carried out for  $F_5^{(N=5)}(r_5)$, the gray-dashed curve in \Fig{fig:CDF:FNkNrN:for:N:2:3:4:5:6} for $F_6^{(N=6)}(r_6)$ was carried out by a straightforward nested integration over $\{\varphi_2, X_3, X_4, X_5, r_6\}$ for $N=6$, which is trivially coded up in just a few lines as
 shown in \Lst{Mathematica:Code:1}. This was the \tit{Mathematica} code used to produce
the gray-dashed curve for $N=6$ in \Fig{fig:CDF:FNkNrN:for:N:2:3:4:5:6}. 
Note that the same code can be executed in terms of angles $\varphi_k$ by replacing $X_k\to\cos\varphi_k$ in the integrand,
and replacing the integration limits  
{\tiny
$\Big\{X_k, \tfrac{1}{k}, \Min\Big[\sqrt{\tfrac{k+2}{k}}\tfrac{X_{k+1}}{\sqrt{1-X_{k+1}^2}}, 1\Big]\,\Big\}\to$
$\Big\{\varphi_k,  \trm{ArcSec}\left[\Max\left[\sqrt{\tfrac{k}{k+2}}\,\trm{Tan}(\varphi_{k+1})\right]\right]\,, \trm{ArcSec}[k]\Big\}$
} for the lowest and middle angles, and 
{\tiny
$\Big\{ X_{N-1}, \tfrac{1}{N-1}, \Max\left[\tfrac{1}{\sqrt{N(N-1)\,rN}}, 1\right]\Big\}\to$ 
$\Big\{ \varphi_{N-1}, \trm{ArcSec}\left[\Max\left[\tfrac{1}{\sqrt{N(N-1)\,rN}}, 1\right]\right],\trm{ArcSec}[N-1] \Big\}$
} for the highest angle.
However, we have found that the code in  \Lst{Mathematica:Code:1} in terms of the 
cosine variables $X_k=\cos\varphi_k$ converge much faster than the exact same code for the angles $\varphi_k$.

While in \Lst{Mathematica:Code:1} a modest attempt has been made to increase the precision of the nested numerical integrations, more attention must be made to this issue as revealed by the undulations at higher purity values (flat regions of the CDF) in the 
gray-dashed curve in \Fig{fig:CDF:FNkNrN:for:N:2:3:4:5:6} for $N=6$. This issue arises since in the region where
$F_N^{(N)}(\mu_N)\sim 1-\epsilon$, differences in values of the CDF are exponentially small, which compound the  inaccuracies when a small number $\epsilon_2\ll 1$ is added in the integration procedure to a prior large number 
$1-\epsilon_1\sim\mathcal{O}(1)$ to obtain the next $\mathcal{O}(1)$ value $1+ (\epsilon_2-\epsilon_1)$.
For $N=6$ it was found that a globally adaptive integration strategy produced a much smoother curve of 
$F_6^{(N=6)}(\mu_6)$  (not shown), especially for higher purity values, over that using a locally adaptive integration strategy (at least in \tit{Mathematica}). 
%Various Monte Carlo strategies were also tried, which produced results similar to $F_6^{(N=6)}(\mu_6)$ in \Fig{fig:CDF:FNkNrN:for:N:2:3:4:5:6}, though in much shorter execution times. The tradeoff is to find an integration strategy that produces acceptable levels of precision and accuracy in a tolerable amount of execution time.
%
These numerical issues will be addressed more fully in future work.
%\cite{N6:integration:note}. 

In order to create a uniform sampling procedure for a fixed purity $\mu_N$ (or $r_N$), all CDFs
$F_k^{(N)}(X_k; X_{k+1})$ for $k\in\{2,\ldots,N-1\}$ must be computed, which can be performed numerically analogous to  \Lst{Mathematica:Code:1} by truncating the nested integrals at level $k$, and appropriately adjusting the limits of the highest angle $k~=~N-1$. One then forms $F_k^{(N)Num}(X_k; X_{k+1})$, from which 
$F_k^{(N)Denom}(X_{k+1})$ is evaluated at $X_k~\to~\Xbarmax_k$ and
the CDF is formed by 
$F_k^{(N)}(X_k; X_{k+1})= F_k^{(N)Num}(X_k; X_{k+1})/F_k^{(N)Denom}(X_{k+1})$.

Even with careful attention paid to numerical accuracy and precision discussed above, the nested integration arising from the inherent dependency of the integration limits of one angle $X_k$  on the value of the next higher angle $X_{k+1}$ creates a ``curse-of-dimensionality" issue for large $N$. If $N_k$ points per integration region for angle $X_k$ are used, the total number of points used in the nested integration for a given $N$ is $N_{total} = \prod_{k=2}^{N-1}\,N_k\sim \left(N_{k_0}\right)^{N-2}$ for $N_k = N_{k_0},\; \forall\, k$.
Adaptive Monte Carlo integration was applied to the computation of $F_6^{(N=6)}(r_6)$ 
 producing qualitatively  similar results (but less accurate, for the number of MonteCarlo points used)  to the straightforward trapezoidal rule utilized in  \Lst{Mathematica:Code:1}. A trade-off between accuracy and execution time (number of function calls) arises for large values of $N$. Nevertheless, while decreased accuracy also effects the uniformity of the sampling of the diagonal density matrices, the matrices sampled will all have the same fixed, chosen value of the purity.
 
Lastly, the factors of $(\sin\varphi_k)^{k-2} \leftrightarrow (1-X_k^2)^{(k-3)/2}$ arising from the volume integration measure become sharply peaked around $\varphi_k\sim\pi/2 \leftrightarrow  X_k\sim1$, a well known fact for the volumes of $N$-spheres \cite{Zyczkowski_2ndEd:2020}. This implies that for large $N$,
$(1-X_k^2)^{\tfrac{(k-3)}{2}}\sim 1-\tfrac{(k-3)}{2}\,X_k^2\approx e^{-\tfrac{(k-3)}{2}\,X_k^2}$ is an ever increasingly accurate approximation such that one can additionally add the constraint to the limits of integration of $X_k$ that it be sampled from 
\be{Xk:largeN:approx}
%\hspace{-0.25in}
X_{k\gg3}\in
\left[
\sqrt{\tfrac{2}{k-3}},\,
\Max\left[\sqrt{\tfrac{2}{k-3}}, \Min\Big[\sqrt{\tfrac{k+2}{k}}\tfrac{X_{k+1}}{\sqrt{1-X_{k+1}^2}}, 1\Big]\,\right]
\right],
\ee
which implies that sampling of $X_k$ ceases when 
$\Xbarmax_k(X_{k+1})\equiv\sqrt{\tfrac{k+2}{k}}\tfrac{X_{k+1}}{\sqrt{1-X_{k+1}^2}}$ $< \sqrt{\tfrac{2}{k-3}}$. 
This approximation for larger values of $N$ would aid in sampling from the relevant non-neglible contributions to the integration over $X_k$.
%=======================================================================

\section{Summary and Conclusion}\label{sec:Discussion}
In this work we have presented a formulation to uniformly sample density matrices $\rho$  of fixed purity for 
arbitrary dimensions $N$ based on generating fixed purity random diagonal density matrices $\rho_{diag}$,
where $\rho=U\,\rho_{diag}\,U^\dag$, with $U$ a unitary matrix uniformly sampled from the Haar measure.
We have provided analytic formulas for the case of generating $\rho_{diag}$ in dimensions $N\in\{2, 3, 4\}$,
and analytical/numerical formulation for $N=5$, and provided simple implementable nested-integration numerical code for any dimension (particularly for $N\ge 6$). We used the analytic formulation for $N=4$, the case of two-qubits, to explore the relation of well known entanglement measures, and a ``baseline" entanglement witness, on the purity 
$\mu_4(\rho_{ab})$ of the two-qubit composite state, as well as on the purity of $\mu_2(\rho_{a})$ of the reduced single qubit system. In addition, we have used our method to sample density matrices at arbitrary constant purity to explore the complementary-quantum correlation (CQC) conjecture.
While the investigations for $N=4$ could in principle be carried out purely by uniformly sampling $U$ from the Haar measure (where $\rho_{diag}$ can be obtained as the absolute square of another randomly generated $U'$), we have shown that this latter sampling is heavily weighted toward lower values of the purity $\mu_4(\rho_{ab})$. This rarity of generating high purity states by the latter method is only exacerbated as the dimension increases, as we have shown by computing the radial cumulative probability distribution function (CDF) for $N\in\{2, 3, 4, 5, 6\}$.

Many entanglement measures (or witnesses) rely on constructions utilizing a reduced density matrix of the composite system (such as entropy-based methods), or on the manipulations of the composite density matrix itself (such as the eigenvalues of the partial transpose of the composite density matrix in the case of the logarithmic negativity). 
While clearly the eigenvalues $\{\lambda_i\}$ of the composite system $\rho$ are the same as its diagonal (spectral representation) $\rho_{diag}$, the reduced density matrices $\rho_{a} = \Tr[\rho_{ab}]$ (where here, both $a$ and $b$  each now represent a possible collection of subsystems) depend on both the eigenvalues $\{\lambda_i\}$ as well as the reduced matrix elements of the random $U$, since 
$d\rho = U\left[ d\rho_{diag} + U^{-1}\,dU\,\rho_{diag} - \rho_{diag}\,U^{-1}\,dU\right]\,U^{-1} \Rightarrow$
$(d\rho)_{ij} = d\lambda_{i}\,\delta_{ij} + (\lambda_j-\lambda_i)\,(U^{-1}\,dU)_{ij}$ \cite{Zyczkowski_2ndEd:2020}.
Thus, it is in general impossible to construct a uniform distribution of reduced density matrices of fixed purity 
$\mu(\rho_a)$ starting from a higher dimensional mixed composite system $\rho_{ab}$. 
However, what we have shown is the opposite, namely that we can construct composite mixed density matrices of fixed purity $\mu(\rho_{ab})$ and study the dependence of the entanglement measures (or witnesses) constructed from the subsequent further randomized sets of purities 
$\mu(\rho_a)$ of the lower dimensional reduced density matrices. This is especially important as we study effects of entanglement measures resulting from composite systems of high purity, where our method can act 
as an efficient means to generate a statistically relevant sample of random states to more accurately explore this purity regime.

Further, as in \Sec{subsec:Induced:Measures}, we can alternatively consider $\rho_N$ to be the reduced density matrix (averaged over all unitary equivalents) derived from a higher-dimensional purification with arbitrary reservoir dimension $K$. The formalism presented in this work then allows for a spherical 
polar description of the eigenvalues 
$(\lambda_1, \lambda_2, \ldots, \lambda_N)$ in terms of the variables 
$(\varphi_2, \varphi_3,\ldots, \varphi_{N-1}, r_N)$ describing these eigenvalues in the Weyl-Chamber.
For the joint probability distribution of the eigenvalues we can obtain closed formed analytic expressions in terms of the spherical polar variables for arbitrary $N$ and $K$.
By integrating out the angular variables $\{\varphi_k\}$ we can construct probability and cumulative probability distribution functions of the reduced state $\rho_N$ in terms of its purity $\mu_N$, which for certain lower dimensional cases can be expressed analytically.
 
Even at the lower purity regime, favored by the uniform (w.r.t to the Haar measure) $U$ approach, our method act can act as a surgical tool to more efficiently and precisely explore certain questions. It is well known that for the case of 
$N=4$ (two-qubits) and $N=6$ (qubit-qutrit) the boundary between separable and entangled states occurs at
composite purity $\mu_N = \tfrac{1}{N-1}$. For all other dimensions, separable states lie somewhere just outside the regime $\mu^{(1)}_N\in\{\tfrac{1}{N},\tfrac{1}{N-1}\}$ (for a proof, see \cite{Zyczkowski:1998}). Our method of generating composite density matrices of fixed purity can be used in a numerical search routine to more precisely pinpoint this transition boundary between separable and entangled states (say using the logarithmic negativity as the entanglement measure) than would be obtained by uniform (Haar) random $U$ approach (see also the work of Zyczkowski \cite{Zyczkowski:1999}). This will be explored in future work.

Lastly, the straightforward nested-numerical integration procedure presented in \Lst{Mathematica:Code:1} is trivially implementable for arbitrary dimensions $N$. However, the price to pay for this ease of coding manifests itself both in the ``curse of dimensionality" as well as in the accuracy of the results for various angular CDFs in the ``flat" regions near unity. Investigations of simple (i.e. less time consuming) integration schemes using high precision arithmetic, as well as the use of high precision local and globally adaptive Monte Carlo integrations schemes, will be explored in future work in order to investigate multipartite systems such as 
$N=6$ (qubit-qutrit), 
$N=8$ (three-qubits), 
$N=9$ (qutrit-qutrit),
$N=12$ (qubit-qubit-qutrit)
and $N=16$ (four-qubits) systems.

%=======================================================================
\vspace{-.25in}
\begin{acknowledgments}
\vspace{-1em}
PMA and JS acknowledge support of this work from
the Air Force Office of Scientific Research.
%PMA, AMS, and MLF would like to acknowledge support of this work from
%the Air Force Office of Scientific Research (AFOSR).
%PLK and EEH would like to acknowledge support for this work was provided by 
%the Air Force Research Laboratory (AFRL) Summer Faculty Fellowship Program (SFFP).
%The authors wish thank Paul Kwiat  for useful discussions and helpful suggestions.
Any opinions, findings and conclusions or recommendations
expressed in this material are those of the author(s) and do not
necessarily reflect the views of Air Force Research Laboratory.
The appearance of external hyperlinks does not constitute endorsement by the United States Department of Defense (DoD) of the linked websites, or the information, products, or services contained therein.  The DoD does not exercise any editorial, security, or other control over the information you may find at these locations.
\end{acknowledgments}

%=======================================================================
%=======================================================================
%=======================================================================
\clearpage
\newpage 
%=======================================================================
\appendix
%=======================================================================
\section{Werner States: Logarithmic Negativity $\boldsymbol{LN}$
and Variant of Linear Entropy $\boldsymbol{\Delta LE}$}\label{app:Werner:state}
\begin{figure}[h]
%\begin{tabular}{c}
\hspace{-0.5in}
\includegraphics[width=3.75in,height=2.05in]{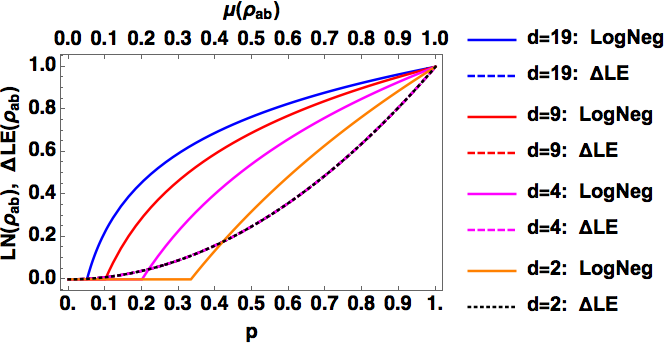} 
%\includegraphics[width=3.75in,height=2.05in]{fig_LogNegExact_DeltaLE_for_BellWernerStates_M1M3M8M18_27Mar2022} 
%\end{tabular}
\caption{Scaled $LN(\rho^{(W,\,d^2)}_{ab})$ (by $\log_2(d)$): for $d^2=(M+1)^2$ dimensional Werner states 
with probability (to be in the pure state) $0\le p\le~1$, for dimensions
(left to right)
$d=\{19, 9, 4, 2\}\leftrightarrow$\{blue, red, magenta, orange\}
for 
(solid) $LN(\rho_{ab})$, and
(dashed) $\Delta LE(\rho_{ab})$
Note: All $\Delta LE(\rho_{ab})$ (dashed) curves are identical, and fall on the black-dashed curve associated with $d=2$.
}\label{fig:rho:Werner:M1:M3:M8:M18}
\end{figure}

%\subsection{$\boldsymbol{LN^{(Aprpox)}}$ on Werner states }
 Because they are analytically tractable,
 it is informative to examine bipartite Werner states of dimension $d^2$ 
 (where $d=(M+1)$), i.e
 $\rho^{(W,\,d^2)}_{ab}~=~p  \ket{\Psi_{Bell}}_{ab}\bra{\Psi_{Bell}} + \tfrac{(1-p)}{d^2} I_{d^2\times d^2}$ with 
 $\ket{\Psi_{Bell}}_{ab}~=~\frac{1}{\sqrt{d}}\sum_{n=0}^{d-1}\,\ket{nn}_{ab}$.
 (Note: $ \ket{\Psi_{Bell}}$ is a $d^2\times 1$ vector, 
 so $\rho^{(W,\,d^2)}_{ab}$ is a $d^2\times d^2$ matrix).
 In \Fig{fig:rho:Werner:M1:M3:M8:M18} we show 
 the logarithmic negativity $LN(\rho_{ab})$,  
 for Werner states with (top) $M=1$ (two qubits) and $M=\{3, 8, 18\}$.
%============================
%============================
It is straightforward to compute that for the Werner state 
the negative eigenvalues of the partial transpose are given by 
$\lambda_{-} = \tfrac{1}{d^2}\,\big(1-(d+1)p \big)$ for $\tfrac{1}{d+1}\le p\le 1$
with multiplicity
$\binom{d}{2} = \tfrac{d(d-1)}{2}$.
Therefore, the negativity in this region is given by 
 $\N=\tfrac{1}{2} \tfrac{d-1}{d} \big((d+1)p-1\big)$ with 
 $LN = \log_2(1+2\,\N)$, 
 yielding $LN(p=\frac{1}{d+1})=0$ and $LN(p=1)=\log_2(d)$.
Thus, $\rho^{(W,\,d^2)}_{ab}$ is entangled ($LN~>~0$) for $\tfrac{1}{(d+1)} <p\le 1$ 
and separable ($LN\le0$) for $0\le p\le\tfrac{1}{(d+1)}$. 
The dimensions $d=(M+1) = \{2,4,9,19\}$ where chosen so that the 
sudden death of entanglement occurred at easily recognizable points
$p=\{\tfrac{1}{3}, \tfrac{1}{5}, \tfrac{1}{10}, \tfrac{1}{20}\}$ on the 
abscissa in \Fig{fig:rho:Werner:M1:M3:M8:M18}.

Finally, one can also show that the
purity $\mu(\rho)=\Tr[\rho^2]$ for the Werner states is given by
$\mu_{(W,\,d^2)}= \tfrac{1+(d^2-1)p}{d^2}$ corresponding to 
to a critical value $\mu^*_{(W,\,d^2)} = \tfrac{2}{d(d+1)}$ at the sudden death of entanglement
at $p= \tfrac{1}{d+1}$. This is plotted as the top abscissa in \Fig{fig:rho:Werner:M1:M3:M8:M18}.

In \Fig{fig:rho:Werner:M1:M3:M8:M18} we also plot $\Delta LE(\rho_{ab})$ as given by 
\be{Delta:LE:general:M}
\hspace{-0.25in}
\Delta LE(\rho_{ab}) = 
\Max
\left[
0, 
\dfrac
{
\left(\mu(\rho_{ab}) - \tfrac{1}{d^2} \right) -
\left(\mu(\rho_{a}) - \tfrac{1}{d} \right)
}
{
1 - \tfrac{1}{d^2}
}
\right]. 
\ee
Note that the $\Delta LE(\rho_{ab})$  curves all fall on the same (multi-colored dashed) curve.
This occurs because the $d$-dimensional Bell state has reduced density matrix $\rho_a$ equal to the MMS for all dimensions $d$. $\Delta LE(\rho_{ab})$ acts a better lower bound to $LN(\rho_{ab})$ the more $d$ increases.
%=======================================================================
%\clearpage
%\newpage 
%===============
\section{Uniform generation of random unitary and diagonal density matrices}\label{app:uniform:U:rho}
The following \tit{Mathematica} codes follow the procedure outlined in Mezzadri \cite{Mezzadri:2007} to generate a uniform random $n\times n$  unitary $U$ matrix (with respect to the Haar measure), and then subsequently generate a uniform random density matrix of dimension $n$.
%%%%%%%%%%%%%%
\begin{lstlisting}[mathescape,language=Mathematica,
caption={Mathematica code to generate random unitary $U$},label=Mathematica:Code:2,frame=single]
URandom[n_] := 
Module[{Z, Q, R, diagR, $\Lambda$}, 
 RG:=RandomVariate[NormalDistribution[]];
 
 Z = Table[$\tfrac{1}{\sqrt{2}}$(RG + I RG), {n}, {n})];
 {Q,R}=QRDecomposition[Z];
 
 (* Note: Z=Q.R=(Q $\Lambda$).($\Lambda^{-1}$R).*)
 (*Make R (hence Q) unique by forcing R*)
 (*to have positive diagonal entries*)
 
 diagR=Diagonal[R];
 
 (* diagonal matrix $\tfrac{R_{ii}}{|R_{ii}|}$*)
 (* Note: $\Lambda^{-1}$R makes diagonal entries *)
 (* of $R$ to be $|R_{ii}|$*)

$\Lambda$=DiagonalMatrix[diagR/Abs[diagR]]//Chop;
  
 (* return unique unitary matrix *)
 Q = Q . $\Lambda$ // Chop
]
\end{lstlisting}
%%%%%%%%%%%%%% 
%(* Note: Z=Q R= (Q \[CapitalLambda]) (\[CapitalLambda]^-1R). Make R \
%(and hence Q) unique  by forcing R to have positive diagonal entries *)

%%%%%%%%%%%%%%
\begin{lstlisting}[mathescape,language=Mathematica,
caption={Mathematica code to generate random $\rho$},label=Mathematica:Code:3,frame=single]
$\rho\trm{R}$andom[n_] := 
Module[{U,$\rho$d}, 
 RI:=RandomInteger[{1,n}];
 
 (*$\rho$d = $\rho_{diagonal}$: take a random row of  U*)
 U=URandom[n];
 $\rho$d = Abs[ U[[RI]] ]$^2$ // Chop;
 
 (*form $\rho = U . \rho\trm{d} . U^\dag $*)
  U=URandom[n];
  
 (*return random density matrix $\rho$ *)
 $\rho$ = U.$\rho$d.U$^\dag$ // Chop
]
\end{lstlisting}
%%%%%%%%%%%%%% 
%=======================================================================
\clearpage
\newpage
\bwt
\section{Analytic expressions involved in $\boldsymbol{N=5}$ radial CDF $\boldsymbol{F_5^{(N=5)}(r_5)}$ in  \Eq{F5N4r5:line1}}\label{app:N5:analytics}
%%%%%%%%%
%\bwt
%==================
\begin{lstlisting}[mathescape,language=Mathematica,
caption={Mathematica code for $F_5^{(N=5)}(r_5)$ (including high precision constants)},label=Mathematica:Code:4,frame=single]
F5N5r5[r5_] := 
Piecewise[
 {
  {F5N5Ir5,         $0\le r5\le \tfrac{1}{\sqrt{5*4}}$,
  {F5N5II1r5,      $\tfrac{1}{\sqrt{5*4}}\le r5\le \tfrac{\sqrt{2}}{\sqrt{5*3}}$,
  {F5N5II2r5,      $\tfrac{\sqrt{2}}{\sqrt{5*3}}\le r5\le \tfrac{\sqrt{3}}{\sqrt{5*2}}$,
  {F5N5II3r5[r5], $\tfrac{\sqrt{3}}{\sqrt{5*2}}\le r5\le \tfrac{\sqrt{4}}{\sqrt{5}}$
  }
]

where

F5N5Ir5 = 52.96585557488949` r5$^4$;

F5N5II1r5 = 0.13241463893722372` + 1287.9751667619084` * 
   (-0.0001028083791688112` - 0.06168502751048966` r5$^4$ + 
   (-0.00021816615649929114` + 0.004363323129985823` r5$^2$) * $\sqrt{-1 + 20\, r5^2}$ +
   $\sqrt{20 - \tfrac{1}{r5^2}}$ *(0.0001090830782496456` r5 + 0.0010908307824964558` r5$^3$) + 
   0.06544984694978735` r5$^4$ ArcCsc[4.47213595499958` r5]);
   
   
F5N5II2r5 = 0.603079 + 1287.98 (-0.00112542 + 0.0168991 r5$^2$ + 0.0245855 r5$^4$ + 
   (0.000218166 - 0.00436332 r5^2) $\sqrt{-1 + 20\, r5^2}$ + 
   $\sqrt{20 - \tfrac{1}{r5^2}}$ (-0.000109083 r5 - 0.00109083 r5^3) - 0.0654498 r5$^4$ ArcCsc[4.47214 r5])


F5N5II3r5[r5] :=
   (0.9443908789779503` + 1287.9751667619084` (-0.0022682775810888662`+
   0.016899077816556717` r5$^2$ + 0.03433940488566341` r5$^4$ + 
   0.00811898816047911` $\sqrt{-1. + 3.3333333333333335` r5^2}$ - 
    0.006051536478449089` $\sqrt{-1.8 + 6\, r5^2}$ - 
    0.0010416666666666658` $\sqrt{-3 + 10\, r5^2}$ - 
    5.421010862427522`*^-19 r5^2 $\sqrt{-3 + 10\, r5^2}$ + 
    0.00021816615649929117` $\sqrt{-1 + 20.` r5^2}$ - 
    0.004363323129985823` r5^2 $\sqrt{-1 + 20\, r5^2}$ + 
    (-0.004059494080239555` + 0.0451054897804395` r5$^4$) *
    ArcCot[$\sqrt{-1 + 3.3333333333333335`\, r5^2}$] + 
   (0.00544638283060418` - 0.06051536478449089` r5$^4$) ArcCot[$\sqrt{-1.8 + 6\, r5^2}$] - 
    0.2617993877991494` (($\sqrt{20 - 1/r5^2}$ (r5 + 10\, r5$^3$))/2400 + 
   1/4 r5$^4$ ArcCsc[2 $\sqrt{5}$ r5]) - 
   0.004059494080239555` ArcTan[$\sqrt{-1. + 3.3333333333333335` r5^2}$] + 
   0.004370554123324342` ArcTan[$\sqrt{-1.8 + 6\, r5^2]}$ + 
   1/$\sqrt{-3 + 10\, r5^2}$ (-0.0062499999999999995` +  0.020833333333333332` r5$^2$ - 
   0.01613743060919757` r5$^2$ $\sqrt{-3 + 10\, r5^2}$
   ArcSec[$\sqrt{-(4/5) + 6\, r5^2}$] - 0.0021516574145596756` $\sqrt{3 - 10 r5^2}$
   ArcTanh[$\sqrt{9/5 - 6\, r5^2}$]))) + 
   F5N5II3r5NumericalPortion[r5]
\end{lstlisting}
%==================
%\ewt
%%%%%
\clearpage
\newpage
%%%%%
where  % this (or some text) is needed to flush the page after \ewt (end wide text)
%%%%%%%%%%%%%%
\begin{lstlisting}[mathescape,language=Mathematica,
caption={Mathematica code for $F_5^{(N=5)}(r_5)$ (including high precision constants)},label=Mathematica:Code:5,frame=single]
$\gamma\trm{N}$[X4_] := NIntegrate[$\sqrt{1-X4\trm{prime}^2}$ ArcTan$\left[\sqrt{\dfrac{1-6 X4\trm{prime}^2}{2 (1-X4\trm{prime}^2)}}\, \right]$, {X4prime, 1/4, X4}]
 
$\trm{my}\gamma\trm{N}$[$\gamma\trm{N}$_, r5_?NumericQ] :=$\gamma\trm{N}\left[\dfrac{1}{\sqrt{5*4}\, r5}\right]$

F5N5II3r5NumericalPortion[r5_] := 
NIntegrate[r5prime$^3$ $\trm{my}\gamma\trm{N}$[$\gamma\trm{N}$,r5prime], {r5prime, $\tfrac{\sqrt{3}}{\sqrt{5*2}}$, r5}]/F5N5Denom

where

F5N5Denom = NIntegrate[(r5$^3$ F4N5r5Denom[r5]), {r5, 0, $\sqrt{\tfrac{4}{5}}$}]

with

F4N5r5Denom[r5_] :=
Piecewise[
{
   (* A: r5$^{(I)}$ *)
   {0.16449340621388117`, $0\le r5\le \tfrac{1}{\sqrt{5*4}}$},
   
   (* B: r5$^{(II.1)}$ *)
   {-0.24674011004195864` + (
     0.013089969389957469` $\sqrt{-1 + 20\, r5^2}$)/r5$^2$ + 
     0.2617993877991494` ArcCsc[2 $\sqrt{5}$ r5], $\tfrac{1}{\sqrt{5*4}}\le r5\le \tfrac{\sqrt{2}}{\sqrt{5*3}}$
  },
    
   (* C: r5$^{(II.2)}$ *)
   {  0.09834187056496882` + (
      0.03379815563311343` - 0.01308996938995747` $\sqrt{-1 + 20\, r5^2}$)/
      r5^2 - 0.2617993877991494` ArcCsc[2 $\sqrt{5}$ r5], $\tfrac{2}{\sqrt{5*3}}\le r5\le \tfrac{\sqrt{3}}{\sqrt{5*2}}$
  },
    
   (* D: r5$^{(II.3)}$ *)
   {  0.13735761954265363` + 
      1/(720 r5$^2$) (-3 $\pi$ $\sqrt{-1 + 20\, r5^2}$ + 
      15 r5$^2$ (5 $\sqrt{3}$ ArcCot[$\sqrt{-1 + (10\, r5^2)/3}$] -3 $\sqrt{15}$ ArcCot[$\sqrt{-\tfrac{9}{5} + 6\, r5^2}$] - 
      4 $\pi$ ArcCsc[2 Sqrt[5] r5]) + 2 $\sqrt{15}$ ($\pi$ - 3 ArcSec[$\sqrt{-\tfrac{4}{5} + 6\, r5^2}$]))+
      $\trm{my}\gamma\trm{N}$[$\gamma\trm{N}$, r5],  $\tfrac{3}{\sqrt{5*2}}\le r5\le \tfrac{\sqrt{4}}{\sqrt{5}}$
  }
}
  ]
\end{lstlisting}
%====================
\ewt
{\hspace{1in}} % needed (or text) to finish \ewt before \clearpage\newpage
%====================
\clearpage
\newpage
%=======================================================================
%=======================================================================

%============================
%--------------------------------------------
%\clearpage
%\newpage
\bibliography{alsing_composite_bibfile}

\begin{thebibliography}{32}
\bibitem [{CDF()}]{CDF:note}%
  \BibitemOpen
  \href@noop {} {}\bibinfo {note} {For a general probability density function
  (PDF) $p(x)$, the cumulative distribution function (CDF) is given by $F(x) =
  \int_{x_0}^{x}\,p(x')\,dx'$ such that $\textrm{Prob}(a\le x\le b) =
  \int_{a}^{b}\,p(x')\,dx' = \int_{x_0}^{b}\,p(x')\,dx' -
  \int_{x_0}^{a}\,p(x')\,dx' = F(b)-F(a)$.}\BibitemShut {Stop}%
\bibitem [{Tr:()}]{Tr:rho:cubed:note}%
  \BibitemOpen
  \href@noop {} {}\bibinfo {note} {Unfortunately, this fortuitous dependence of
  $x_k$ only on the two purities $(\mu_k,\, \mu_{k-1})$ does not occur for
  other traces of higher powers of $\rho$, i.e. $\Tr[\rho^q]$ for $q>2$, and so
  the enumeration of the range of the coordinates is intractable analytically,
  except for case of $q=2$.}\BibitemShut {Stop}%
\bibitem [{QDN()}]{QDNote}%
  \BibitemOpen
  \href@noop {} {}\bibinfo {note} {Recall that the quantum discord
  $\mathcal{Q}$ for a bipartite system $\rho_{ab}$ is defined as the difference
  between two expressions for the quantum mutual information:
  (i)~$\mathcal{I}(\rho_{ab}) = S(\rho_{a})+S(\rho_{b})-S(\rho_{ab})$ and
  (ii)~$\mathcal{J}(\rho_{ab}|\{\Pi^k_b\}) =
  S(\rho_{a})-S(\rho_{ab}|\{\Pi^k_b\})$ where $\{\Pi^k_b\}$ are a complete set
  of projective measurements on subsystem $b$. Here $S(\rho_{ab}|\{\Pi^k_B\})
  \overset{\trm{def}}{=} \sum_k p_k\,S(\rho^{(k)}_a)$ with $\rho^{(k)}_a =
  \tfrac{1}{p_k}(I_a\otimes \Pi^k_b)\rho_{ab}(I_a\otimes \Pi^k_b)$ and $p_k =
  \Tr[(I\otimes \Pi^k_b)\rho_{ab}(I\otimes \Pi^k_b)]$. The classical
  correlations are defined as $\mathcal{J}(\rho_{ab}) =
  \underset{\{\Pi^k_b\}}{\trm{sup}}\,\mathcal{J}(\rho_{ab}|\{\Pi^k_b\})$, and
  the quantum discord is defined as the difference $\mathcal{Q}(\rho_{ab}) =
  \mathcal{I}(\rho_{ab})-\mathcal{J}(\rho_{ab})$.}\BibitemShut {Stop}%
\bibitem [{LNn()}]{LNnote}%
  \BibitemOpen
  \href@noop {} {}\bibinfo {note} {The logarithmic negativity $LN$ is given by
  $LN(\rho_{ab}) = \log_2\left[1+2\,\mathcal{N}(\rho_{ab})\right]$, where the
  negativity $\mathcal{N}(\rho_{ab})$ is defined as the sum of the absolute
  values of the negative eigenvalues of the partial transpose (on either
  subsystem $a$ or $b$) of the composite state $\rho_{ab}$.}\BibitemShut
  {Stop}%
\bibitem [{Del()}]{DeltaLE:note}%
  \BibitemOpen
  \href@noop {} {}\bibinfo {note} {For the case of the Werner state used here
  with the $d$-dimensional Bell state, whose reduced density matrix is
  maximally mixed, the second term $\mu(\rho^{(N_a)}_{a}) -\tfrac{1}{N_a}$ in
  \Eq{Delta:LE} is exactly zero, and $\Delta LE$ is strictly a linear function
  of $ \mu(\rho^{(N)}_{ab})$. For other choices of the pure state component of
  the Werner state, this second term would not be zero, and would come into
  play.}\BibitemShut {Stop}%
\bibitem [{Ren()}]{Renyi:entropy:note}%
  \BibitemOpen
  \href@noop {} {}\bibinfo {note} {If desired, we could also use the
  probability distribution $P^{\trm{trace}}_{N=3,K} (\mu_3)$ to write
  $P^{\trm{trace}}_{N=3,K} (S_3)$ in terms fo the R\'{e}nyi entropy $S^{(N)}_3
  = S_3(\rho_N) = -\ln \left( \sum_{i=1}^N \lambda_i^3\right)$ as in
  \Fig{fig:PS2N2K}. The family of R\'{e}nyi entropies
  $S_{\alpha}(AB)=S_{\alpha}(\hat{\rho}_{AB})\equiv\frac{1}{1-\alpha}\log\Big(\text{Tr}[\hat{\rho}_{AB}^{\alpha}]\Big)$
  include the von Neuman entropy $\lim_{\alpha\rightarrow
  1}S_{\alpha}(AB)=-\text{Tr}[\hat{\rho}_{AB}\log(\hat{\rho}_{AB})]$, and the
  collision entropy $\lim_{\alpha\rightarrow
  2}S_{\alpha}(AB)=-\log\Big(\text{Tr}[\hat{\rho}_{AB}^{2}]\Big)=-\log\Big(\mu(\rho_{ab})\Big)$.}\BibitemShut
  {Stop}%
\bibitem [{fil()}]{filling:out:the:triangle:note}%
  \BibitemOpen
  \href@noop {} {}\bibinfo {note} {It is worth noting that one can prove that
  the entire triangle can be filled out (not counting possible violations). The
  hypotenuse represents mixtures of the maximally entangled state, maximally
  correlated mixed state, and the maximally mixed state. By changing the local
  basis upon which classical correlations are measured, one can trace out a
  flat horizontal line from each point on the diagonal to a point on the left
  side of the triangle, filling it entirely.}\BibitemShut {Stop}%
\bibitem [{U-p()}]{U-perturbed:note}%
  \BibitemOpen
  \href@noop {} {}\bibinfo {note} {We generate the unitary $U$ near the
  identity as follows. We create an $N\times N$ matrix with entries drawn from
  a normal distribution with unit variance, and scale the entries by a fixed
  chosen $\epsilon<1$ which sets the scale of the deviation of $U$ from the
  $N\times N$ identity matrix (in the plot we chose $\epsilon = 0.15$). Next,
  we replace the diagonal entries by unity, and Gram-Schmidt orthonormalize the
  matrix (\texttt{Orthogonalize} in \textit{Mathematica}) to produce the
  unitary $U$.}\BibitemShut {Stop}%
\bibitem [{bla()}]{black:curve:note}%
  \BibitemOpen
  \href@noop {} {}\bibinfo {note} {In particular, consider a mixture of four
  orthogonal Bell states with a total purity of $\mu_4$ but whose eigenvalues
  form a minimum entropy distribution at constant purity. For this set of mixed
  states, the marginal states are maximally mixed, and the quantum mutual
  information will be what is given by the black solid curve.}\BibitemShut
  {Stop}%
\end{thebibliography}
%-------------------------------------------

%-------------------------------------------

%============================
% Appendices
%============================
\end{document}